\UseRawInputEncoding  
\documentclass[aps,twocolumn,prx,longbilbliography,superscriptaddress,shopacs]{revtex4-2} 
\usepackage{graphicx}
\usepackage{bm}
\usepackage{gensymb}
\usepackage[none]{hyphenat}
\usepackage{epstopdf}
\usepackage{multirow}
\usepackage{float}
\usepackage{tabularx}
\usepackage{gensymb}
\usepackage{color}
\usepackage{siunitx}
\usepackage{xr}
\usepackage{amsmath}
\usepackage[colorlinks=true]{hyperref} 
\usepackage{placeins}
\usepackage{multirow}

\begin{document}

\title{Highly tunable magnetic phases in transition metal dichalcogenide Fe$_{1/3+\delta}$NbS$_2$}

\author{Shan Wu}
\email{shanwu@berkeley.edu}
\affiliation{Department of Physics, University of California, Berkeley, California, 94720, USA}
\affiliation{Material Sciences Division, Lawrence Berkeley National Lab, Berkeley, California, 94720, USA}

\author{Zhijun Xu}
\affiliation{NIST Center for Neutron Research, National Institute of Standards and Technology, Gaithersburg, Maryland 20899, USA}
\affiliation{Department of Materials Science and Engineering, University of Maryland, College Park, Maryland 20742, USA}

\author{Shannon C. Haley}
\affiliation{Department of Physics, University of California, Berkeley, California, 94720, USA}
\affiliation{Material Sciences Division, Lawrence Berkeley National Lab, Berkeley, California, 94720, USA}

\author{Sophie F. Weber}
\affiliation{Department of Physics, University of California, Berkeley, California, 94720, USA}
\affiliation{Material Sciences Division, Lawrence Berkeley National Lab, Berkeley, California, 94720, USA}

\author{Arany Acharya}
\affiliation{Department of Physics, University of California, Berkeley, California, 94720, USA}

\author{Eran Maniv}
\affiliation{Department of Physics, University of California, Berkeley, California, 94720, USA}
\affiliation{Material Sciences Division, Lawrence Berkeley National Lab, Berkeley, California, 94720, USA}

\author{Yiming Qiu}
\affiliation{NIST Center for Neutron Research, National Institute of Standards and Technology, Gaithersburg, Maryland 20899, USA}

\author{A. A. Aczel}
\affiliation{Neutron Scattering Division, Oak Ridge National Laboratory, Oak ridge, Tennessee 37831, USA}

\author{Jeffrey B. Neaton}
\affiliation{Department of Physics, University of California, Berkeley, California, 94720, USA}
\affiliation{Kavli Energy Nanosciences Institute at Berkeley, Berkeley, CA 94720}
\affiliation{Material Sciences Division, Lawrence Berkeley National Lab, Berkeley, California, 94720, USA}

\author{James G. Analytis}
\affiliation{Department of Physics, University of California, Berkeley, California, 94720, USA}
\affiliation{Material Sciences Division, Lawrence Berkeley National Lab, Berkeley, California, 94720, USA}

\author{Robert J. Birgeneau}
\email{robertjb@berkeley.edu}
\affiliation{Department of Physics, University of California, Berkeley, California, 94720, USA}
\affiliation{Material Sciences Division, Lawrence Berkeley National Lab, Berkeley, California, 94720, USA}

\date{\today}
\begin{abstract}
Layered transition metal dichalcogenides (TMDCs) host a plethora of interesting physical phenomena ranging from charge order to superconductivity. By introducing magnetic ions into 2H-NbS$_2$, the material forms a family of magnetic intercalated TMDCs T$_x$NbS$_2$ (T = 3d transition metal). Recently, Fe$_{1/3+\delta}$NbS$_2$  has been found to possess intriguing resistance switching and magnetic memory effects coupled to the N\'{e}el temperature of T$_N \sim 45$ K \cite{maniv2021_1,maniv2021_2}. We present comprehensive single crystal neutron diffraction measurements on under-intercalated ($\delta \sim -0.01$), stoichiometric, and over-intercalated ($\delta \sim 0.01$) samples. Magnetic defects are usually considered to suppress magnetic correlations and, concomitantly, transition temperatures. Instead, we observe highly tunable magnetic long-ranged states as the Fe concentration is varied from under-intercalated to over-intercalated, that is from Fe vacancies to Fe interstitials. The under- and over- intercalated samples reveal distinct antiferromagnetic stripe and zig-zag orders, associated with wave vectors $k_1$ = (0.5, 0, 0) and $k_2$ = (0.25, 0.5, 0), respectively.  The stoichiometric sample shows two successive magnetic phase transitions for these two wave vectors with an unusual rise-and-fall feature in the intensities connected to $k_1$. We ascribe this sensitive tunability to the competing next nearest neighbor exchange interactions and the oscillatory nature of the Ruderman-Kittel-Kasuya-Yosida (RKKY) mechanism. We discuss experimental observations that relate to the observed intriguing switching resistance behaviors. Our discovery of a magnetic defect tuning of the magnetic structure in bulk crystals Fe$_{1/3+\delta}$NbS$_2$  provides a possible new avenue to  implement controllable antiferromagnetic spintronic devices.
\end{abstract}

\maketitle

\section{Introduction}
Layered magnetic van der Waals (vdW) materials have recently attracted tremendous interest, resulting in rapid progress in fundamental studies of novel vdW physical phenomena together with promising potential for spintronic applications \cite{kenneth2018,jgp2016,pulickel2016,cheng2019}. The weak van der Waals bonds make single crystals readily cleavable thereby offering a new platform to study the evolution of the behavior from three dimensions (3D) down to the 2D limit. Moreover, the wide flexibility of 2D atomic samples allows for an efficient manipulation of magnetic states through external perturbations, such as strain, gating, proximity effect and pressure \cite{kin2019,gibertini2019,ting2019,tiancheng2019}. In bulk magnetic vdW crystals, usually high hydrostatic pressure \cite{matthew2021,hanies2018} or significant chemical substitution \cite{andrew2020,gil2018}  is utilized to modulate the magnetic state or the effective dimensionality via tuning of the interlayer exchange couplings. Magnetic defects are typically considered to be responsible for inhibiting long-range magnetism due to the atomic-scale disorder. Here we demonstrate novel behavior in which magnetic defects tune the magnetic ground states in the transition metal dichalcogenide (TMDC) bulk crystal Fe$_{1/3+\delta}$NbS$_2$.

\begin{figure}
\includegraphics[width=1.\columnwidth,clip,angle =0]{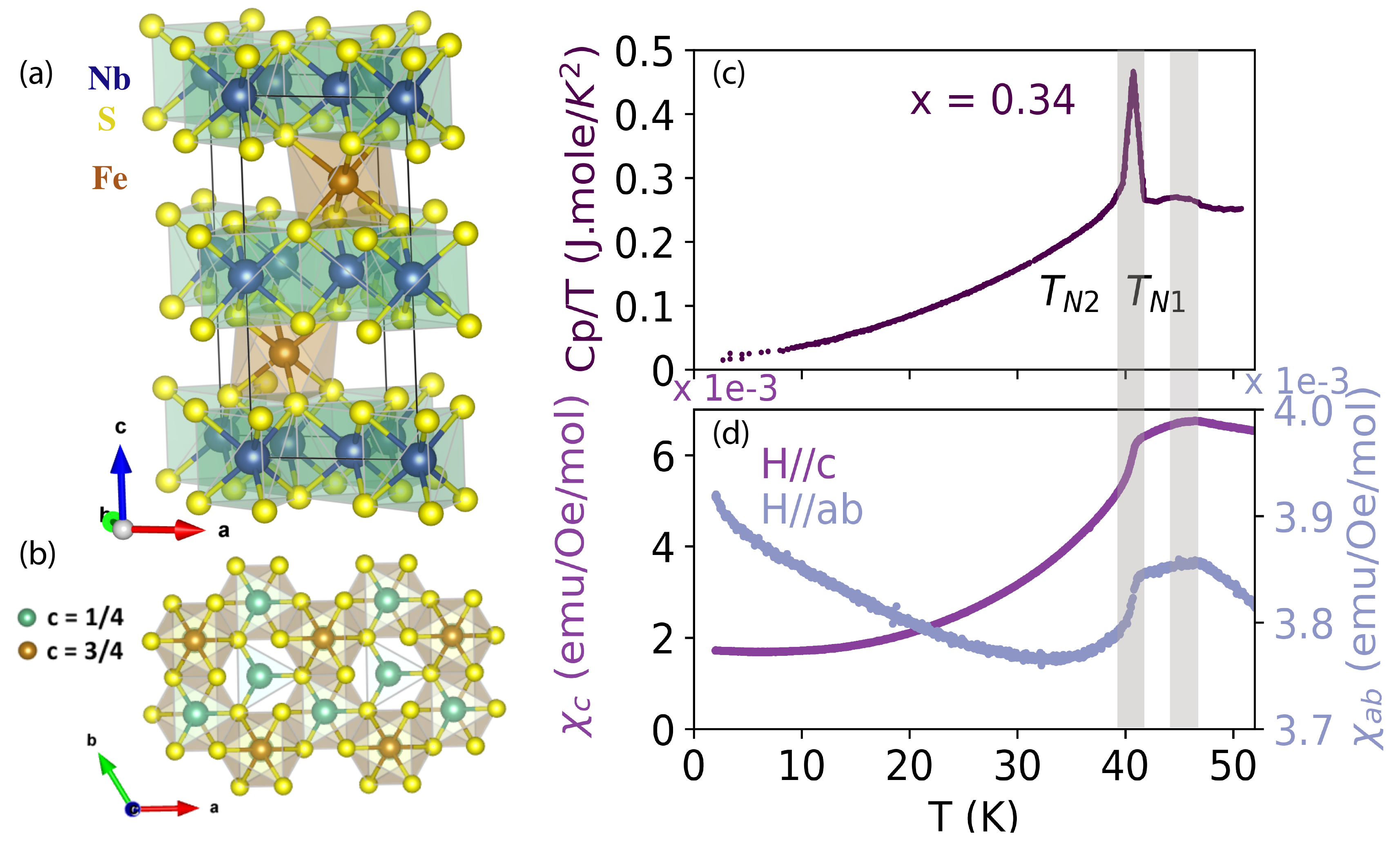}
\caption{\label{structure} (a) Crystallographic structure of Fe$_{1/3}$NbS$_2$ and (b) corresponding  view in the $ab$ plane. Upper ($c$ = 3/4) and lower ($c$ = 1/4) Fe atoms form triangular lattice layers.  (c) Specific heat and (d) magnetization measurement  in the low temperature region for $x$ = 0.34 sample. The shaded regions mark two anomalies, identified as antiferromagnetic transitions by neutron scattering measurements. }
\label{fig1}
\end{figure}

Fe$_{1/3+\delta}$NbS$_2$ is a member of a large class of intercalated TMDCs, M$_x$TA$_2$ family (M = 3d transition metal; T = Nb, Ta; A = S, Se)  \cite{parkin1980,friend1977}. The host material is a prototypical example of a charge density wave system; recently these systems have been attracting major attention because of other exotic properties, such as possible quantum spin liquid phases and 2D superconductivity \cite{wilson1974,naito1981,castro2001,guillam2008,law2017,ad2020}. The vdW bonding between chalcogen atoms of adjacent 2H-TA$_2$ layers allows the ready intercalation of magnetic atoms. The intercalated atoms order into a stacked $\sqrt3 \times \sqrt3$ superlattice when x = $1/3$ \cite{boswell1978}. This family of compounds shares the same crystal structure with a non-centro-symmetric space group $P6_3 22$ and a bi-layer triangular arrangement of the intercalated atoms (Fig. \ref{fig1} (a-b)). The broken inversion symmetry results in an in-plane Dzyaloshinskii-Moriya (DM) interaction between-plane interaction in addition to the competing bilinear exchange interactions with their concomitant geometric frustration. In addition, as a metallic system, there is a strong interaction between the conduction electrons and the local moments via the Ruderman-Kittel-Kasuya-Yosida (RKKY) mechanism. Depending on the host 2H-TA$_2$ layer and the intercalated species, the family exhibits a fascinating variety of magnetic and electronic properties \cite{HULLIGER1970,ANZENHOFER1970,togawa2012,braam2015,Kousaka_2016,kousaka2009,karna2019,Parkin_1983,lu2020}  in bulk samples.

In the intercalated variant M$_{1/3}$NbS$_2$ subgroup, chiral helimagnetism was observed for the Cr and Mn species \cite{togawa2012,braam2015,Kousaka_2016,kousaka2009,karna2019,adam2018}; the V and Co compounds exhibit a spin structure characterized by ferromagnetic planes stacked antiferromagnetically with canted in-plane moments \cite{Parkin_1983,hall2021,lu2020}. Novel physical properties were reported in these species, including the anomalous Hall effect, an electrical magnetochiral effect and magnetic soliton confinement \cite{togawa2015,ghimire2018,aoki2019}.  Most materials in this family are characterized by an easy-plane anisotropy and mostly dominant ferromagnetic interactions.  In contrast, the intercalated Fe version displays predominantly antiferromagnetic correlations and a strong easy-axis anisotropy \cite{gorochov1979,parkin1980,yamamura2004}.

Recently, a resurgence of interest in the Fe version has been sparked by the demonstration of intriguing spintronic properties in bulk Fe$_{1/3+\delta}$NbS$_2$ crystals \cite{nair2020,maniv2021_1}. Both current-induced resistance switching and magnetic memory effects were reported below the N\'{e}el transition temperature $T_N \sim$ 45 K. Moreover, the relevant spintronic properties were found to depend sensitively on the intercalation ratio $x$ (= $1/3+\delta$) \cite{maniv2021_2}. By decreasing the ratio slightly below 1/3, the system exhibited a much more prominent spintronic response concomitant with dramatic spin-glass-like behavior below the AFM N\'{e}el temperature.   There are so far only a few known examples of current-induced switching behavior in AFM single crystal compounds \cite{wadley2016,bodnar2018}. The mechanism is believed to entail an applied current inducing a spin-polarization due to the combination of the breaking of inversion symmetry and Rashba spin-orbit coupling \cite{prb2008_manchon}. It has been argued that the reported resistance switching in the off-stoichiometric sample of Fe$_{1/3+\delta}$NbS$_2$ somehow relates to the observed spin-glass behavior \cite{maniv2021_1,maniv2021_2}, thence providing a possible new way to explore AFM spintronic devices. Therefore, a complete understanding of the magnetic ground states and magnetic correlations as a function of the intercalation ratio are essential to uncover the mechanism of the observed interesting spintronic properties.  Further, the only relevant information about the magnetic structures which currently exists derives from neutron powder diffraction measurements carried out decades ago at low temperatures \cite{laar1970}. In addition to the spintronic motivation, this system is of intrinsic interest as a vdW material with interesting and, as we shall see, novel magnetic properties.

In this paper, we report detailed neutron scattering measurements on high quality single crystals of Fe-intercalated TMDC Fe$_{1/3+\delta}$NbS$_2$  with $x$ spanning 1/3.  Surprisingly, we found highly tunable magnetic phases in the bulk crystal that are more versatile than the single phase reported in previous work. By a comprehensive experimental investigation together with  modeling of the magnetic structures, we determined that there are two long-ranged anti-ferromagnetically ordered states and that one can tune from one state to the other by varying $x$ subtly from less than to greater than 1/3, that is, by varying from Fe vacancies to Fe interstitials.  ($\delta \sim \pm 0.01$). The stoichiometric sample with $x$=1/3, on the other hand, exhibits both magnetic structures characterized by two successive magnetic phase transitions upon cooling. In the Discussion section of this paper, we discuss this tunability and its implications to the fascinating spintronic behavior exhibited by these materials. This finding is the first example of such unusual switching and exchange bias behaviors in the intercalated TMDCs M$_x$TA$_2$ family; it can provide an archetypal case for magnetic defect-induced switching of the magnetic state in bulk magnetic vdW systems.

\section{Methods}
High quality single crystals were synthesized using a chemical vapor transport method with a polycrystalline precursor made from Fe, Nb and S elements in the ratio of $x$:1:2 \cite{boswell1978}. The values of the actual intercalation ratio $x$ of individual single crystals were determined by energy dispersive X-ray spectroscopy (EDX).  Room temperature single crystal X-ray diffraction patterns were measured at the ChemXray facility, UC Berkeley. Magnetization measurements were performed using a Quantum Design MPMS-3 system. The heat capacity was measured in a Quantum Design PPMS system. Neutron scattering experiments were carried out at several instrumental stations. Single-crystal diffraction mapping at temperatures $T$ = 38 and 5 K with co-aligned crystals (mosaicity $\sim 5 ^{\circ}$) in the range of $x=0.32\sim0.34$ employed the MACS spectrometer at NCNR \cite{etde_22152337}.  The data were collected with $E_f$ = 5 meV with a double focusing monochromator and a Be filter placed before and after the sample. To investigate accurately the tunable magnetic state, single crystal neutron diffraction measurements with one crystal were carried out on SPINS, BT-7 at NCNR and HB1A at HFIR for different intercalation ratios: $x$ = 0.31,  0.32, 0.33, 0.34 and 0.35 with masses of 12, 23, 9, 3, and 27 mg respectively. Measurements were conducted with a PG (002) monochromator and analyzer using $E_f$ = 5 meV, 14.7 meV and 14.48 meV neutrons on SPINS, BT-7 and HB1A respectively.
We discuss the density functional theory (DFT) calculation strongly related to our experimental results. The DFT calculations utilized the Perdew-Burke-Ernzerhof (PBE) functional  and added a Hubbard U correction accounting for the Fe $d$ electrons. For details of the DFT calculations we refer the reader to Reference \cite{weber2021}.

\begin{figure}
\includegraphics[width=0.8\columnwidth,clip,angle =0]{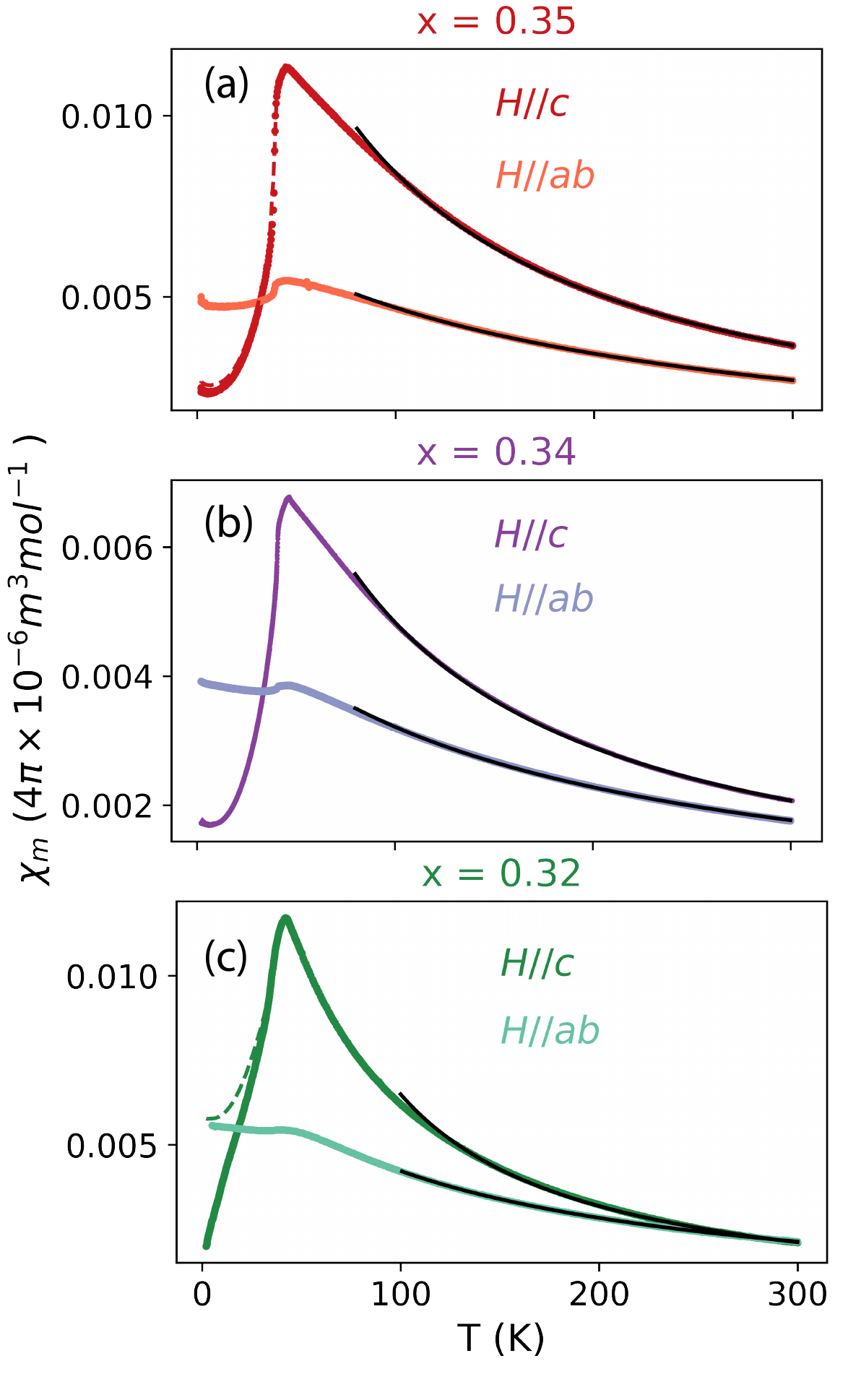}
\caption{\label{fig2} Magnetization measurements with an applied field $H = 1000$Oe along the $c$ axis and in the $ab$ plane for (a) $x >$ 1/3 (b) $x$ close to 1/3 and (c) $x <$ 1/3 sample. The dashed and solid lines correspond to the measurements with field-cooled and zero field cooled processes, respectively. The solid black lines are the results of the Curie Weiss fits. }
\end{figure}

\begin{figure*}
\includegraphics[width=2.\columnwidth,clip,angle =0]{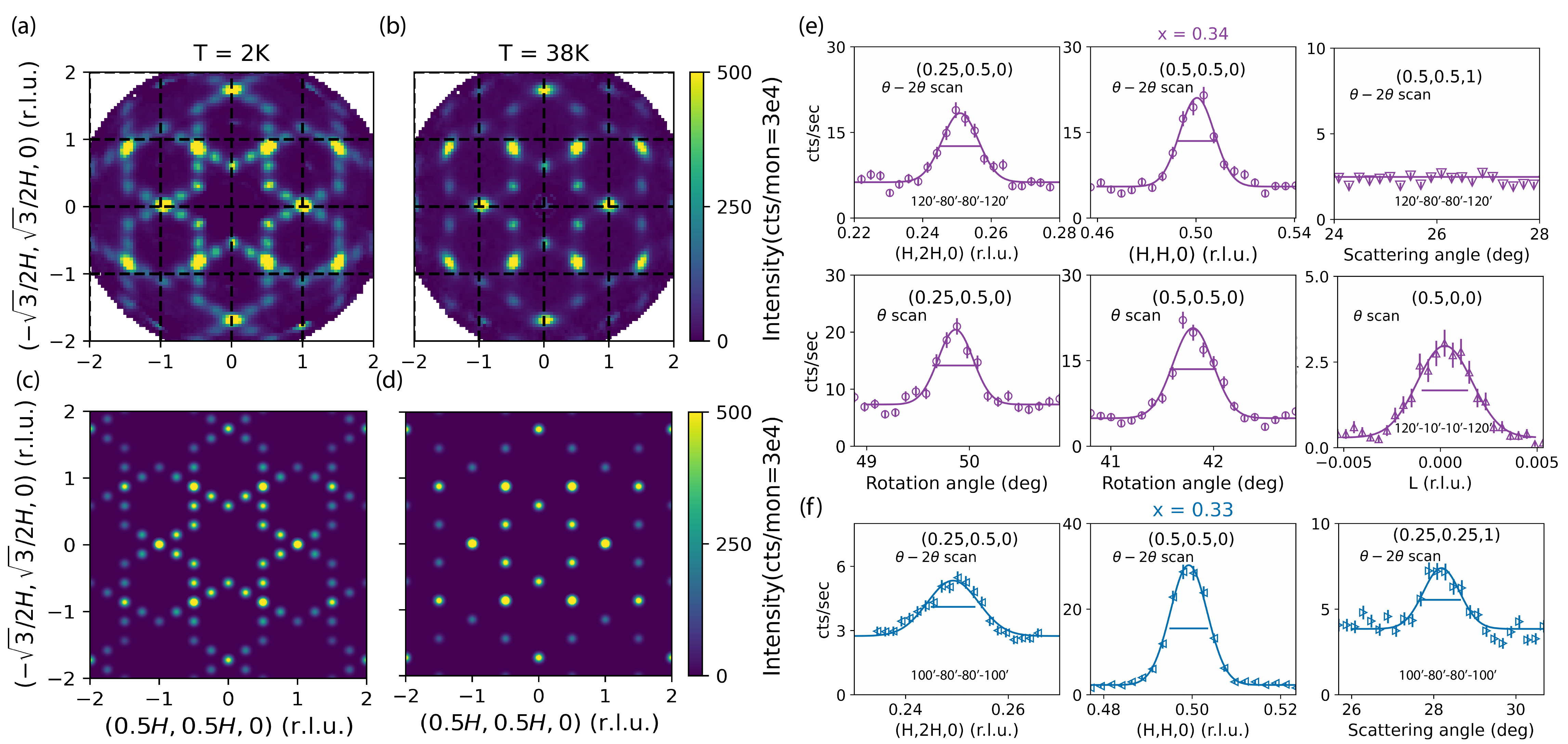}
\caption{\label{fig3}  (a-b) Symmetrized single crystal neutron diffraction patterns collected at MACS  by co-aligned crystals of Fe$_x$NbS$_2$ ($x=0.32 \sim0.34$) mounted in the $(HK0)$ scattering plane at T = 2 K and 38 K. A data set acquired at $T$ = 60 K was subtracted as the background.  (c) \& (d) Calculated diffraction patterns for given spin configurations. The details are described in the main text. Representative transverse ($\theta$) and longitudinal ($\theta$-$2\theta$) scans (dots) measured on one single crystal close to stoichiometric ratio with (e) $x=0.34$ at BT7 and (f) $x=0.33$ at SPINS with $T$ = 5 K. The markers of $\bigcirc$, $\bigtriangledown$, $\bigtriangleup$ denote the data collected in the $(HK0)$, $(HHL)$, $(H0L)$ scattering planes at BT7; and $\triangleleft$, $\triangleright$ for the $(HK0)$ and $(HHL)$ scattering planes at SPINS, compatible with the markers in Fig. \ref{cal_obs}. The horizontal bars denote the instrument $Q$-resolutions. The solid lines are the results of the fits to a Gaussian line-shape. The corresponding configurations of the collimations are written in the panel. Error bars in all figures represent one standard deviation. }
\label{fig3}
\end{figure*}

\section{Experimental results}
The crystallographic structure of Fe$_{1/3}$NbS$_2$ is identical to that of other species in this family, described by the space group $P 6_322$,  with a triangular sub-lattice of iron ions intercalated in the honeycomb 2H-NbS$_2$ (Fig. \ref{fig1} (a)).  One crystallographic unit cell contains two iron sites at coordinates (1/3, 2/3,1/4) and (2/3, 1/3, 3/4) respectively. They occupy the vacant octahedral sites stacking between the prismatic NbS$_2$ layers and form two triangular superlattice planes (Fig. \ref{fig1} (b)). The single crystal X-ray diffraction pattern has been refined in the space group $P 6_322$ with $R_1$ value of 4.62\%, consistent with a previous report \cite{boswell1978}.

The un-intercalated host is a $d$-band metal with one electron on the Nb ion. Charge transfer from the Fe ions to the Nb band results in divalent oxidation states of the Fe with localized $d$ electrons on the intercalated Fe ions \cite{parkin1980}. We present magnetic susceptibility and specific heat measurements for the $x=0.34$ sample in Fig. \ref{fig1} (c-d). Two successive anomalies occur at $T_{N1} \sim$ 45 K and   $T_{N2} \sim$ 41 K; these features are also observed in the specific heat data. Curie Weiss fits to the magnetic susceptibility in the paramagnetic region (Fig. \ref{fig2} (b)) yield values for the paramagnetic effective moment   $\mu_{eff}$ = 4.3(2)$\mu_B$  and Curie Weiss temperature $\theta_{cw}$ = -49 K along the $c$ axis; $\mu_{eff}$ = 4.0(2)$\mu_B$ and $\theta_{cw}$ = -143 K in the $ab$ plane. These values are consistent within the range of previous reports \cite{HULLIGER1970,ANZENHOFER1970,laar1970,gorochov1979,doi1991} with effective spin $S$ = 2.  The negative Curie Weiss temperature suggests that antiferromagnetic exchange interactions are dominant.  The derived single-ion anisotropy $D$ is approximately 2 meV \cite{dj2017}. In the off-stoichiometric sample with $x < 1/3$, one transition was identified\cite{maniv2021_2}; and a bifurcation between zero-field-cooled (ZFC) and field-cooled (FC) susceptibility data was observed, indicating spin-glass-like behavior.  In the $x>1/3$ sample, a small bifurcation between ZFC and FC data was observed below $T_{f} \sim 10$K \cite{maniv2021_2}. The characterizations of other single crystals used for neutron diffraction experiments in this paper are shown in the appendix (Fig. \ref{suscp_all}). Highly anisotropic magnetization was observed in all magnetically ordered samples (Fig. \ref{fig2}).  The sensitivity to the intercalation ratio $x$ of the bulk magnetic and thermodynamic properties, as well as the associated intriguing spintronic properties, clearly call for a detailed experimental study of the $x$-dependence of the magnetic ground states in this bilayer triangular lattice system.

\begin{figure*}
\includegraphics[width=1.5\columnwidth,clip,angle =0]{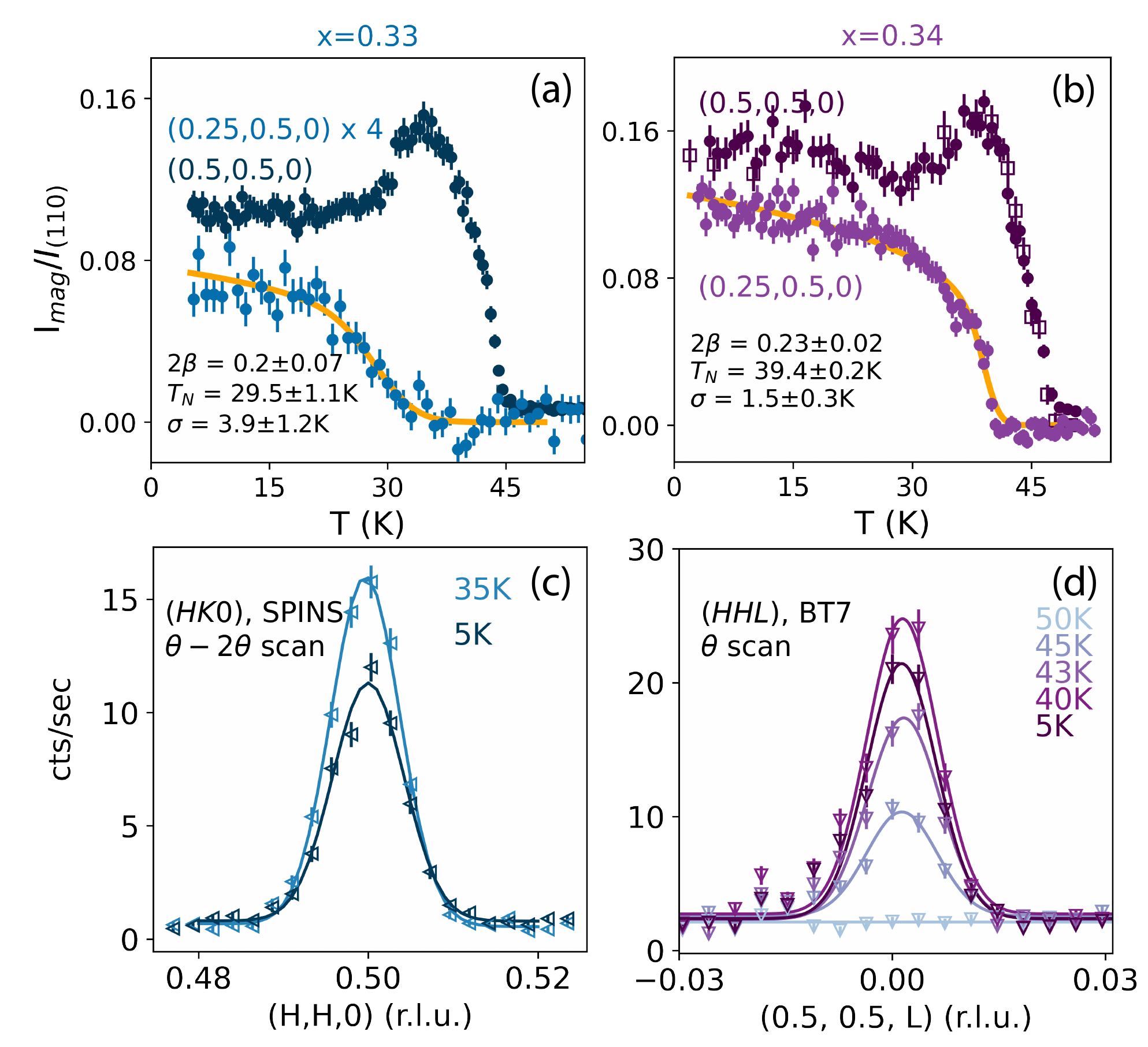}
\caption{\label{fig4} Temperature dependence of the magnetic peak intensities  at $\bf{Q_2}$ = (0.25, 0.5, 0)  and $\bf{Q_1}$ = (0.5, 0.5, 0) for two nearly stoichiometric samples (a) $x$= 0.33 on SPINS and (b) $x$= 0.34 on BT7.  The peak intensities are scaled to match the integrated intensities (empty squares) and both are normalized by the integrated intensity of the nuclear peak (110). Orange lines are the results of fits to the power law function with a thermal Gaussian distribution of $T_N$ ( $I \propto \int_{0}^{\infty} (1-\frac{T}{t_N})^{2\beta} \frac{1}{\sqrt{2\pi}\sigma} e^{-(t_N-T_N)^2/2\sigma^2} dt_N$) \cite{birgeneau1973,birgeneau1977}. 
Representative motor scans at $\bf{Q_1}$  at elevated temperatures: (c) $\theta$-$2\theta$ scan in the $(HK0)$  plane for $x$= 0.33 and (d) $\theta$ scan in $(HHL)$  plane for $x$= 0.34. These  correspond to in-plane and out-of-plane  $Q$-scans along the $HH$ and $L$ directions, respectively.  The solid lines are the results of fits to a Gaussian function.}
\label{fig4}
\end{figure*}

\subsection{Neutron scattering measurement }
We first employed neutron diffraction scattering measurements in the $(HK0)$ scattering plane to study magnetic transitions for an assembly of co-aligned crystals.  These experiments were carried out at MACS, which is well suited for a broad momentum survey. The diffraction pattern, with the data at $T =$ 60 K ($> T_N$) subtracted clearly shows two antiferromagnetic phases (Fig. \ref{fig3} (a \& b)). At $T = 38 $ K, superlattice peaks are observed at wave vector transfer $Q$s associated with the propagation wave vector $\bf{k_1}$ = (0.5, 0, 0). At $T = $2 K, another phase associated with the second propagation wave vector $\bf {k_2}$ = (0.5, 0.25, 0) appears leading to additional magnetic Bragg peaks. The pattern displays a six-fold symmetry; this is the result of three magnetic domains with $Z_3$ symmetry. From the measurements, the most intense peaks associated with $\bf{k_1}$ and $\bf{k_2}$ have wave vector transfers of $\bf{Q_1}$ = (0.5, 0.5, 0) (or 6 equivalent positions) and $\bf{Q_2}$ = (0.25, 0.5, 0) (or 12 equivalent positions), respectively. 

The measurement on MACS were carried out on a set of co-aligned single crystals with $x$ in the range of 0.32 to 0.34. To obtain more information and specifically to elucidate the x-dependence of the magnetic structures, we measured individual high quality single crystals close to stoichiometry ($x$ = 0.33, 0.34), under-intercalated ($x$ = 0.31, 0.32) and over-intercalated ($x$ = 0.35).

\subsubsection{Nearly stoichiometric $x$ = 1/3 sample}
We measured two crystals with $x$ = 0.33 and 0.34 separately at SPINS and BT7. Representative transverse and longitudinal, namely $\theta$ and $\theta - 2\theta$, scans at 5 K are shown in Fig. \ref{fig3} (e \& f). For the  $x$ = 0.34 sample, the magnetic peak at $\bf{Q_2}$ = (0.25, 0.5, 0) has comparable intensity with the peak at $\bf{Q_1}$ = (0.5, 0.5, 0) (Fig. \ref{fig3} (e)). Both peaks have their full width at half maximum (FWHM) determined by the instrumental Q-resolution, marked by horizontal bars in the plots. The $\theta$ scan with the crystal in the $(H0L)$ plane, equivalent to an $L$ scan, displays also a resolution-limited peak (Fig. \ref{fig3} (e-VI)), indicating three dimensional long-range order even though the structures are lamellar. For the  $x=$0.33 sample the relative intensity of peaks between $\bf{Q_1}$ and $\bf{Q_2}$ (Fig. \ref{fig3} (f)) are dramatically different from that with $x$ = 0.34, having more intensity related to $k_1$ = (0.5, 0, 0). We also collected superlattice peaks at a series of  reciprocal lattice positions, $(0.5, 0.5, L)$ by varying $L$.  The intensity decreases gradually with increasing $L$ value; the intensity following roughly the square of the magnetic form factor, manifesting the magnetic nature of the superlattice peaks.

To study the temperature evolution of the two magnetic phases, the intensities at peak position $\bf{Q_1}$ and $\bf{Q_2}$ were measured as a function of temperature for the two samples, as shown in Fig. \ref{fig4} (a \& b). The magnetic peak intensities are scaled to comply with the integrated areas of the peaks measured from the motor scans, and normalized by the integrated area of the nuclear Bragg peak (110). The samples were measured in the $(HK0)$ scattering plane for these two plots. Both samples display the onsets of two magnetic transitions, consistent with the transition temperature anomalies observed in the bulk characterization measurements. The first transition is identified at $T_{N1} \sim$ 45 K based on a guide to the eye. To extract the power law exponent $2\beta$ and $T_{N2}$, we assume a Gaussian distribution of transition temperatures within the bulk crystal in the power law function \cite{birgeneau1973,birgeneau1977}:
\begin{equation}
\int_{0}^{\infty} (1-\frac{T}{t_N})^{2\beta} \frac{1}{\sqrt{2\pi}\sigma} e^{-(t_N-T_N)^2/2\sigma^2} dt_N
\label{eq0}
\end{equation}
The fits provide the results $T_{N2}$ = 30(1) K and 39.4(2) K with the thermal width of $\sigma = 4(1)$ K and $1.5(3)$ K, and the power law exponent  $2\beta$ = 0.20(7) and 0.23(2) for the $x$= 0.33 and 0.34 crystals, respectively. The values for $2\beta$ are close to that for the ideal 2D Ising model, $2\beta$ = 0.25 although, because of the large spread in $T_{N}$, one should not over-interpret this result.  Specifically, we cannot rule out a weakly first order transition.

Interestingly, both nearly stoichiometric samples display an increase of the magnetic peak intensity at $\bf{Q_1}$ below T$_{N1}$, followed by a partial drop of the intensity below $T_{N2}$. This rules out the scenario that the stoichiometric sample is simply composed of partial under- and over-intercalated regions; otherwise we should simply see two separated order parameter curves. This unusual feature is also confirmed in the $\theta - 2\theta$ and $\theta$ motor scans at elevated temperatures in Fig. \ref{fig4} (c \& d). These measurements were carried out in the spectrometer configuration with the crystal mounted in the $(HK0)$ and $(HHL)$ planes. Correspondingly, motor scans traversing across $\bf{Q_1}$ are equivalent to scans along the $HH$ and $L$ directions respectively. The magnetic peaks at intermediate temperatures ($T$ = 35 K in $x$ = 0.33 and 40 K in $x$ = 0.34) show higher intensities than the data at 5 K and a constant resolution-limited width from the Gaussian peak fits. These results preclude explanations due to the change of the magnetic correlations from 3D to 2D with decreasing temperature, which can lead to the broadening of the peak in the out of plane direction thereby reducing the peak intensity simultaneously within the plane.

\subsubsection{Off-stoichiometric samples}

\begin{figure}
\includegraphics[width=1\columnwidth,clip,angle =0]{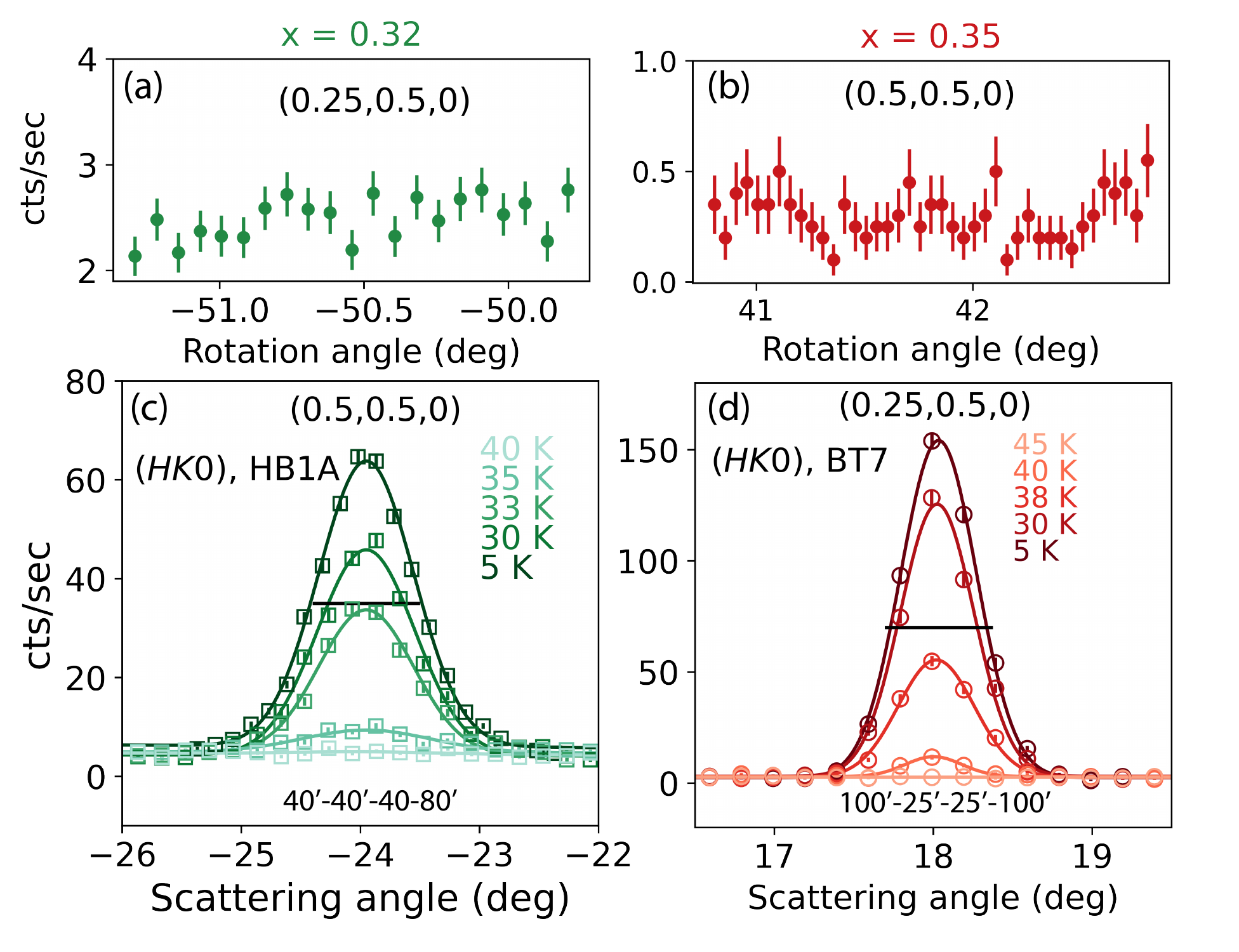}
\caption{\label{fig4_motor}  
Sample rotation $\theta$ scans at the given positions at $T$ = 5 K for (a) $x$= 0.32 and (b) $x$= 0.35, showing no detectable signals. Representative temperature dependent $\theta$-$2\theta$ scans for (c) $x$= 0.32 at $\bf{Q_1}$ = (0.5, 0.5, 0) and (d) $x$ = 0.35 at $\bf{Q_2}$ = (0.25, 0.5, 0). The solid lines are results of fits to the Gaussian function with the resolution shown in the horizontal black line at 5 K. 
}
\label{fig4_motor}
\end{figure}

\begin{figure}
\includegraphics[width=0.8\columnwidth,clip,angle =0]{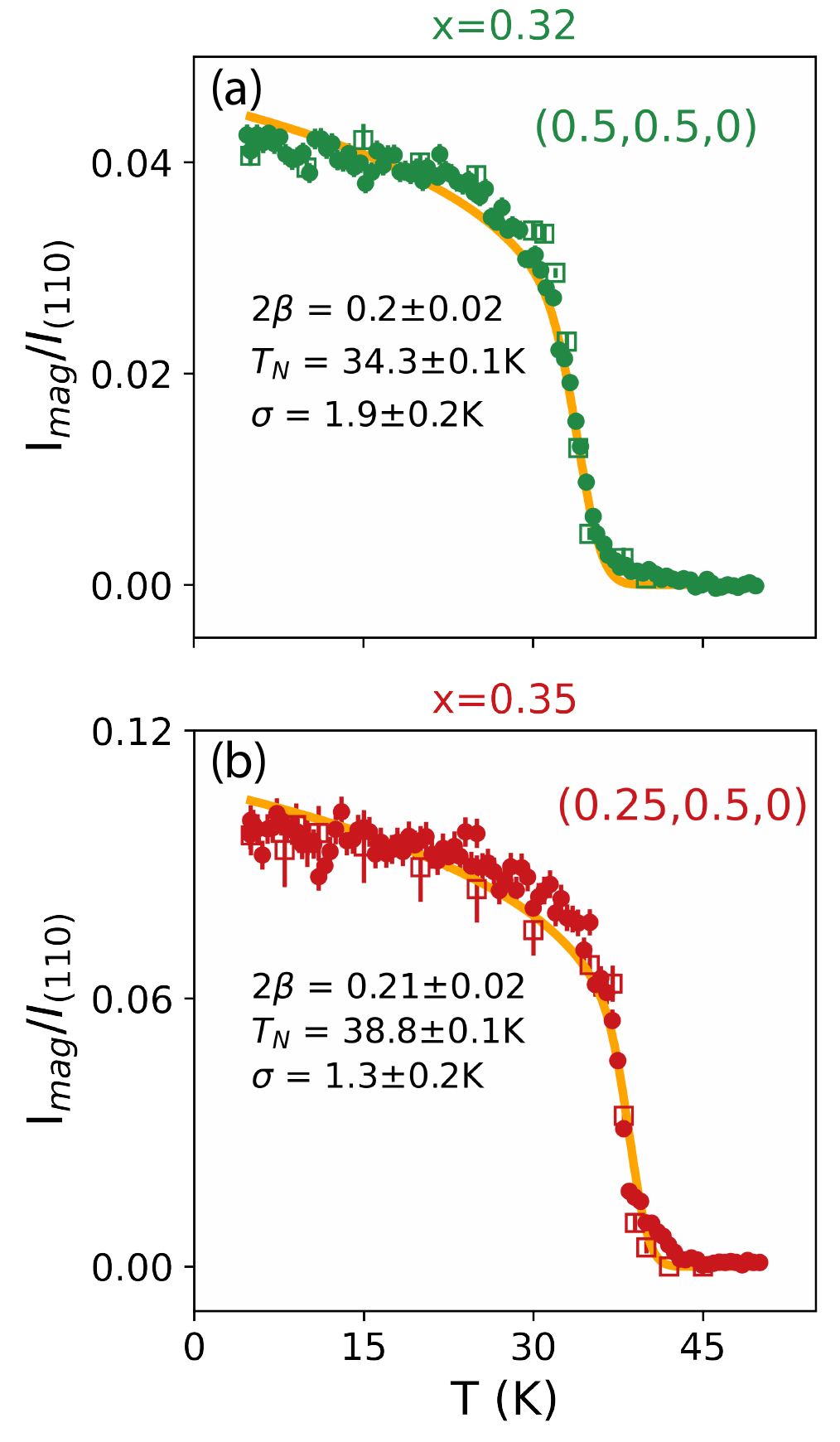}
\caption{\label{fig6} Temperature dependence of the magnetic peak intensity (filled dots) for (a) under-intercalated sample  $x$= 0.32 at $\bf{Q_1}$ = (0.5, 0.5, 0)  on HB1A and (b) over-intercalated sample $x$= 0.35 at $\bf{Q_2}$ = (0.25, 0.5, 0) on BT7.  Empty squares are integrated intensities extracted from the Gaussian fits to the $\theta$-$2\theta$ scans in Fig. \ref{fig4_motor}. Orange lines are the results of fits to the power law function with a Gaussian distribution of $T_N$ \cite{birgeneau1973}.
}
\label{fig6}
\end{figure}

To investigate the magnetic states and spin-glass-like physics in the off-stoichiometric samples, we measured two under-intercalated samples ($x$ = 0.31, 0.32) and one over-intercalated sample ($x$ = 0.35) in Fe$_x$NbS$_2$. The results turn out to be quite striking.  In the $x$ = 0.32 sample, we observed only magnetic peaks associated with wave vector $\bf{k_1}$, and no detectable peaks at the positions related to $\bf{k_2}$ (Fig. \ref{fig4_motor} (a \& c)). In contrast, we observed only peaks associated with $\bf{k_2}$, not with $\bf{k_1}$, in the $x$ = 0.35 sample (Fig. \ref{fig4_motor} (b \& d)).  The strongest intensity is observed at $\bf{Q_1}$ = (0.5, 0.5, 0) and $\bf{Q_2}$ = (0.25, 0.5, 0)  respectively for each sample. These peak positions were used to study the temperature-dependent behavior for each crystal. 

The onset of the peaks at the two positions upon cooling clearly manifests magnetic transitions. (Fig. \ref{fig6}). From fits to the Gaussian-broadened power law function (Eq. \ref{eq0}), we obtain  $T_N$ of 34.2(1) K and 38.8(1) K with widths $\sigma$ of 1.9(2) K and 1.3(2) K, and power law exponents $2\beta$ of 0.20(2) and 0.21(2) for the $x$ = 0.32 and 0.35 samples, respectively.  The transition temperature in the $x$ = 0.32 crystal is consistent with the second kink of $\chi_{ab}$ (appendix Fig. \ref{suscp_all} (c)). While in $x$ = 0.35, the transition temperature coincides with the peak anomaly in the susceptibility measurement  \cite{maniv2021_2}. The extracted values of the power law exponents, as well as for the stoichiometric sample, are consistent with the value for the 2D Ising system (2$\beta$ = 0.25) \cite{pelissetto2002,onsager} as we noted previously.

The width of the magnetic Bragg peaks in both samples agrees within the measurement uncertainties with the instrumental resolution, thence implying long range AFM order. Naively, this might be seen to be unexpected since the magnetization measurements manifest a bifurcation between the ZFC and FC processes and a slow relaxation of the magnetization. Specifically, we might have expected to observe a short-ranged magnetically ordered state in light of the apparent spin glass behavior.

In the $x$ = 0.31 crystal, we examined scans along the high symmetry directions and also carried out a 2D mapping in $(HK0)$ plane at 5 K on SPINS. To our surprise, we found no short- or long-ranged magnetic signal above the background level below $T_f$ or $T_N$. This could be due to the in-plane disorder that destroys magnetic order, or that the magnetic signals were sufficiently broad that they could not be distinguished from the background.

\subsubsection{Field cooled neutron scattering measurements}
We also employed neutron diffraction measurements in the presence of an applied magnetic field at MACS to investigate any relevant spin-glass behavior. Interestingly, we observed a broadening of the magnetic superlattice peak at wave vector $\bf{Q_1}$ by cooling the crystal across $T_N$ under an 8 Tesla ($T$) magnetic field. This broadening is evident by viewing the diffraction pattern in a ZFC measurement with the pattern obtained after subtracting an equivalent FC measurement as shown in Fig. \ref{fig7}. The $Q$-cut across the position of $(-0.5,-0.5,0)$ in the difference pattern exhibits more intensity at the peak center and two symmetric wings with negative net counts after the subtraction (Fig. \ref{fig7} (b)). That implies a different line shape of the magnetic peak in the FC process compared with the ZFC process. Such peak broadening on field cooling was also observed in other dilute two-dimensional Ising antiferromagnets \cite{birgeneau1983,birgeneau1977}, in which the broadening is attributed to the random staggered magnetic field generated by the applied magnetic field.  

To sum up our single crystal neutron scattering measurements, we have obtained the following principal magnetic properties in Fe$_x$NbS$_2$ with varying intercalation ratio $x$ but with identical crystallographic structures. 
1) Strong magnetic intensities at in-plane positions suggesting that the spins are oriented along the $c$ axis, consistent with the highly anisotropic magnetization data. 2) Two types of magnetic phases associated with wave vector $\bf{k_1}$ = (0.5, 0, 0) and $\bf{k_2}$ = (0.25, 0.5, 0) were observed. We observed magnetic peaks related to only $\bf{k_1}$ in samples with $x < $ 1/3, both $\bf{k_1}$ and $\bf{k_2}$ in stoichiometric samples, $x \sim$ 1/3, and $\bf{k_2}$ alone in over-intercalated crystals, that is, $x >$ 1/3. 3) In crystals with $x \sim$ 1/3, there are two successive magnetic transitions, showing a rise-and-fall feature in the peak intensity curve. 4) All samples, except for the heavily under-intercalated sample ($x$ = 0.31), exhibit resolution-limited peaks implying long-range order within the given resolution. The fitted power law exponent $\beta$ is consistent with 2D Ising behavior ($\beta$ = 0.125)  \cite{onsager}, although the uncertainties are large and we cannot rule out a weakly first order transition due to the spread of the transition temperature $T_N$ . 5) No clear features related to spin-glass physics are evident from neutron diffraction measurements under zero field; however,  magnetic peak broadening presumably due to induced staggered random field effects was observed in the field-cooled process.

\begin{figure}
\includegraphics[width=1.\columnwidth,clip,angle =0]{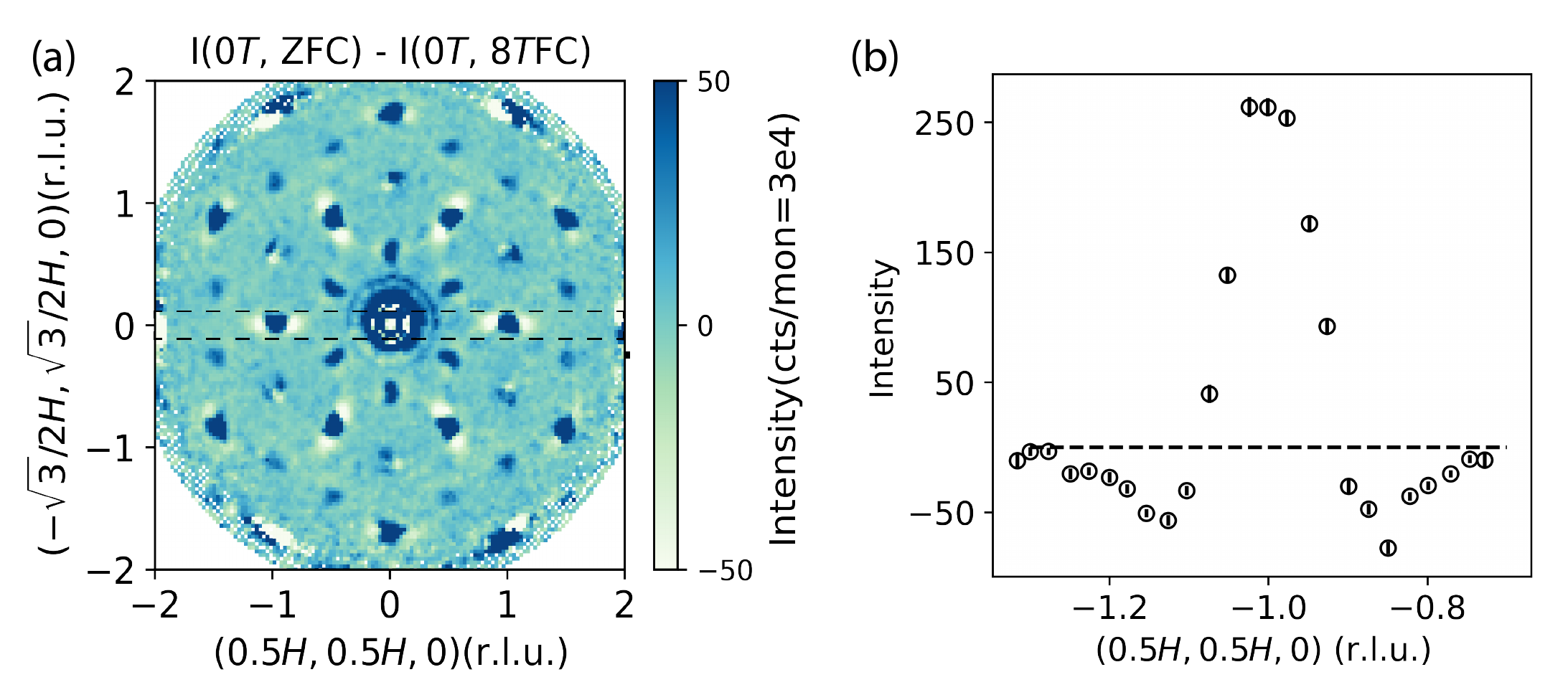}
\caption{\label{fig7} (a) Single crystal neutron diffraction pattern in $(HK0)$ plane at 2 K with zero-field-cooled process, with data collected with cooling in an 8 Tesla ($T$) magnetic field subtracted. The data were folded with six-fold rotational symmetry and expanded to the full rotation angle for  presentation purposes. (b) The cut along ($0.5H,0.5H,0$) with integrating the range of $H$ = [-0.07, 0.07]  r.l.u. in ($-\sqrt3/2H, \sqrt3/2H,0$) direction, as marked by the dashed rectangle in (a). These are the data obtained by subtracting the 8 $T$ field cooled measurement.  The dashed line denotes the base line with zero intensity in (b).
}
\label{fig7}
\end{figure}

\begin{figure*}
\includegraphics[width=1.8\columnwidth,clip,angle =0]{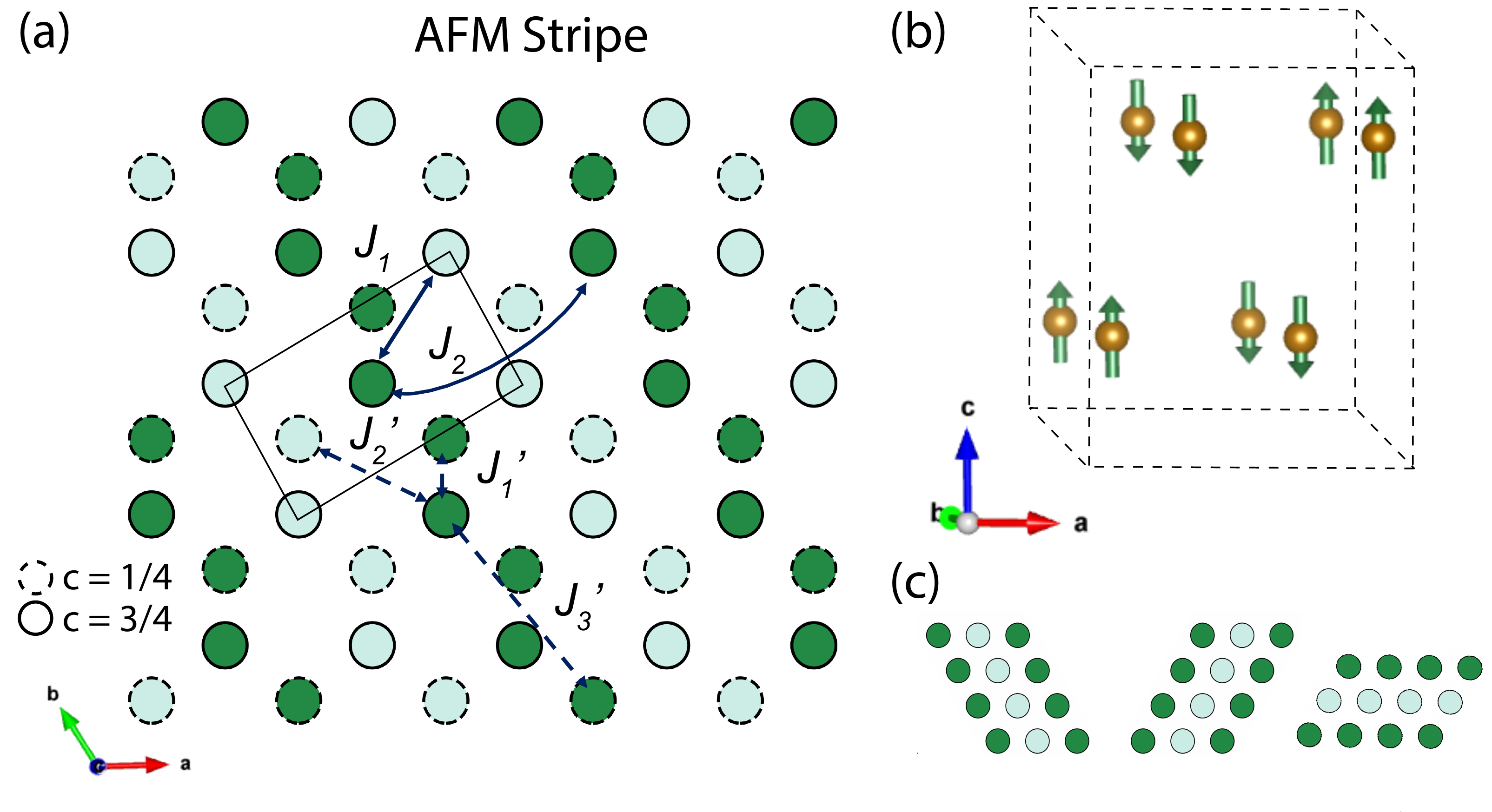}
\caption{\label{spin1}  AFM stripe magnetic structure associated with the $\bf{k_1}$ = (0.5, 0, 0) domain in Fe$_{x}$NbS$_2$ ($x < $ 1/3): view in (a) $ab$ plane and (b) three dimensions.  `AFM-' is defined when two Fe atoms in one unit cell have anti-parallel spins. Circles with solid and dashed outlines in (a) represent two Fe layers at $c$ = 3/4 and $c$ = 1/4.  Dark and light colors denote spins up and down. Solid rectangle depicts the smallest magnetic unit cell. (c) Plots of three equivalent domain directions within one Fe triangular lattice layer. }
\label{spin1}
\end{figure*}

\begin{figure*}
\includegraphics[width=1.8\columnwidth,clip,angle =0]{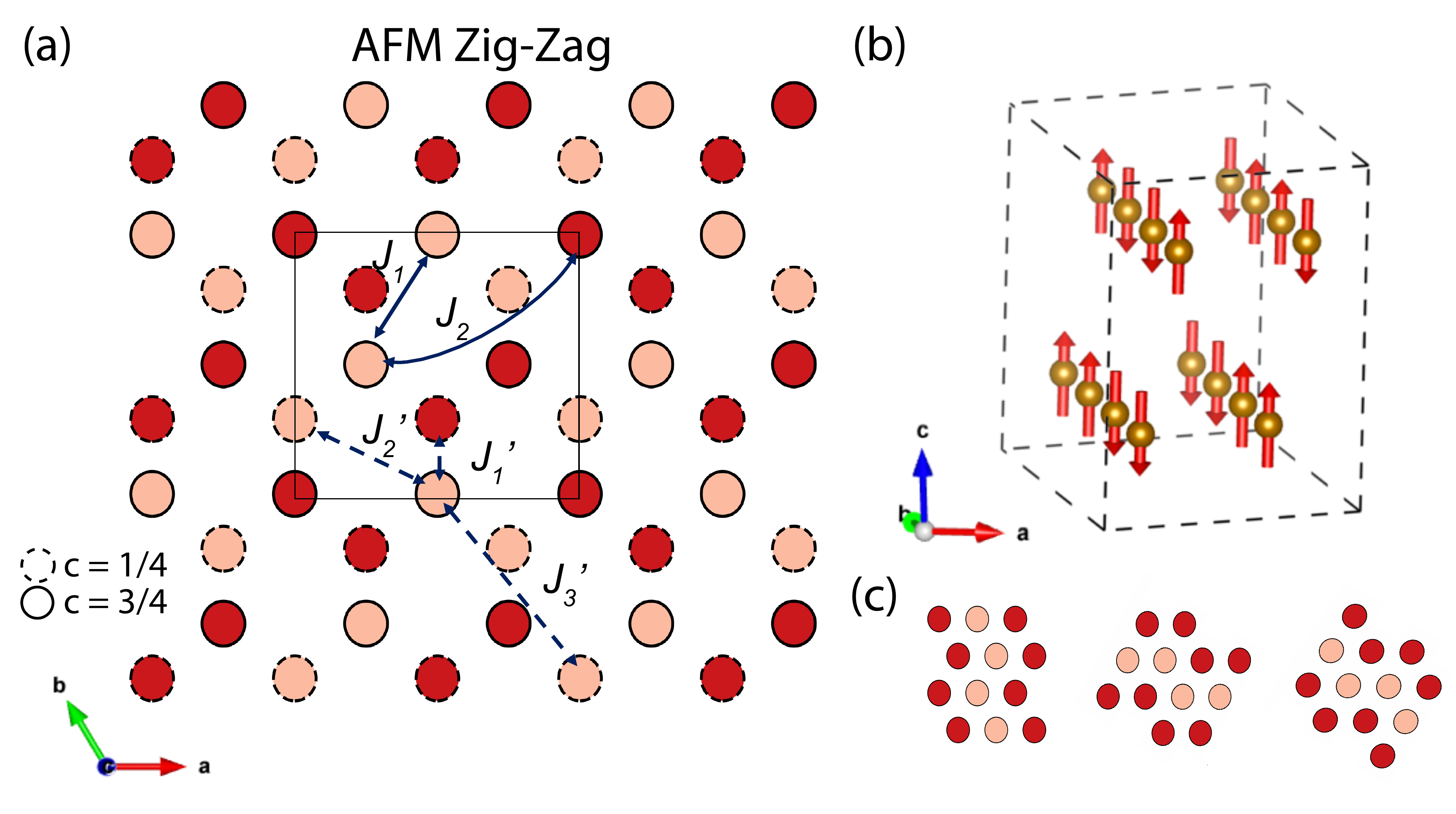}
\caption{\label{spin2}  AFM zigzag magnetically ordered phase associated with the $\bf{k_2}$ = (0.5, 0.25, 0) domain in Fe$_{x}$NbS$_2$ ($x > $ 1/3): view in (a) $ab$ plane and (b) three dimensions. `AFM-' is defined when two Fe atoms in one unit cell have anti-parallel spins. Circles with solid and dashed outlines in (a) represent two Fe layers at $c$ = 3/4 and $c$ = 1/4.  Dark and light colors denote spins up and down. Solid rectangle depicts the smallest magnetic unit cell. (c) Plots of three equivalent domain directions within one Fe triangular lattice layer. }
\label{spin2}
\end{figure*}

\subsection{Magnetic structure determination}
For a systematic analysis of possible magnetic structures associated with $\bf{k_1}$ and $\bf{k_2}$, we use representation analysis in SARAH \cite{sarah} and Fullprof \cite{fullprof}, and calculate magnetic scattering intensities.  Given the crystallographic symmetry  $P 6_3 22$, the $2c$ Wyckoff position for the Fe atoms and propagation wave vector $\bf{k}$, group theory analysis describes that  the magnetic representation $\Gamma_{mag}$ can be decomposed into irreducible representations (IRs) and their corresponding Basis vectors (BVs). According to Landau theory, the magnetic symmetry can be described by one IR for each transition. This information is then implemented to perform model calculations for the determination of the magnetic structure. 
For single-crystal diffraction, the measured magnetic coherent cross section follows the expression \cite{squires_2012}: 
\begin{equation}
\frac{d\sigma}{d\Omega} = N_M \frac{(2\pi)^3}{V_M} p^2\sum_{\bf{G}_\textit{M}} \delta(\bf{Q} - \bf{G}_\textit{M}) |\bf{F}_\perp (Q)|^2 
\label{eq1}
\end{equation}
Here $|\bf{F}_\perp (Q)|^2 = |\bf{F}_\textit{M}(\bf{Q})|^2 -  |\hat{e} \cdot \bf{F}_\textit{M}(\bf{Q})|^2$ contains the static magnetic structure factor and magnetic form factor and represents the component of the spin axis perpendicular to \textbf{Q}.  $\bf{G_M}$ is the wave vector transfer associated with the reciprocal lattice vector $\bf{\tau}$ as $\bf{G_M} = \bf{\tau}  \pm \bf{k}$ and a single propagation vector $\bf {k}$. $N_M$ and $V_M$ are the number and volume of the magnetic unit cell respectively and $p = 2.695 fm$. The magnetic structure factor, $\bf{F}_\textit{M}$, is related to the spin configuration as:
\begin{equation}
\bf{F}_\textit{M}(\textbf{Q})= \sum_\textit{j} \textit{f}(\bf{Q}) \bf{S}_{k\textit{,j}} e^{i \bf{Q \cdot r}_\textit{j}},
\label{eq2}
\end{equation}
where $\bf{S}_{k\textit{,j}}$ is the spin moment for atom  $j$ at the position $r_j$ within a magnetic unit cell, and can be written by the BVs in irreducible representation analysis. By this formalism we can calculate magnetic scattering intensities for different spin structures, and determine the configuration most accordant with the data.

\begin{table}[t]
\setlength{\tabcolsep}{8pt}
\caption{\label{tab1}Basis vectors (BVs) $\psi_i$ of IRs for two Fe atoms in the unit cell (Fe1: (0.333, 0.667, 0.25), Fe2: (0.667, 0.333, 0.75)) associated with propagation vector $\bf{k_1}$ = (0.5, 0, 0). BVs are defined by the crystallographic axes.}
\begin{tabular}{c|c|c|c}
\hline
IR 					      & BV 		& Fe1  & Fe2  \\ \hline
\hline
$\Gamma_1$			       & $\psi_1$ & (2 1 0) & (2 1 0)  \\ \hline
\multirow{2}{*}{$\Gamma_2$} & $\psi_2$ & (0 -1 0) & (0 1 0) \\ \cline{2-4}
					     & $\psi_3$ & (0 0 1) & (0 0 1) \\ \hline
\multirow{2}{*}{$\Gamma_3$} & $\psi_4$ & (0 -1 0) & (0 -1 0) \\ \cline{2-4}
					     & $\psi_5$ & (0 0 1) & (0 0 -1) \\ \hline
$\Gamma_4$			       & $\psi_6$ & (2 1 0) & (-2 -1 0)  \\ \hline
\hline
\end{tabular}
\vspace{-1.em}
\end{table}

First, we describe the representation analysis for the two types of propagation vectors $\bf{k_1}$ and $\bf{k_2}$ and discuss the choice of BVs supported by the observed data. 

\subsubsection{Phase $\bf{k_1}$ = (0.5, 0, 0)}
For the propagation vector $\bf{k_1}$ = (0.5, 0, 0), the magnetic representation $\Gamma_{mag}$ can decomposed into IRs $\Gamma_{mag} = \Gamma_1 + 2\Gamma_2 + 2 \Gamma_3 + \Gamma_4$ with corresponding BVs listed in  Table. \ref{tab1}. Since the moment direction has been determined to be predominantly along the $c$ axis by both the magnetic susceptibility and neutron data, only $\Gamma_2$ ($\psi_2,\psi_3$) and $\Gamma_3$ ($\psi_4,\psi_5$) are relevant. For the same reason, we concentrate on the BV $\psi_3$ and $\psi_5$. 
The difference between them is two Fe atoms in one unit cell (Fig. \ref{structure} (a)) oriented parallel or anti-parallel respectively.  The calculated magnetic scattering patterns (Fig. \ref{fig3} (d) and Fig. \ref{FigS2} in appendix) with $\psi_5$   agree with the data, showing an antiparallel stacking between two Fe spins. This is consistent with the strongest intensity being observed at $\bf{Q_1}$ = (0.5, 0.5, 0). Though $\psi_4$  is also allowed by group theory analysis, however, no peak feature is observed at the position \textbf{Q} = (0.5, 0.5, 1) (Fig. \ref{fig3} (e)) disfavoring the spin component related to that peak position, suggesting an absence of any measurable in-plane moment associated with $\bf{k_1}$.

The spin configuration corresponding to $\psi_5$ is shown in Fig. \ref{spin1} (a-c). 
It consists of spins oriented in the same direction along one crystal axis and alternating along the other one, forming a stripe pattern elongated along an in-plane high symmetric crystal axis. Two Fe atoms with different $c$ coordinates stack antiferromagnetically. We named this configuration  `AFM stripe' for simplicity. The magnetic unit cell is $2$ times the size of the structural unit cell. Note that there are three equivalent $k$ vectors ((0.5, 0, 0), (0, 0.5, 0) and (0.5, -0.5, 0)), corresponding to three magnetic domains along three directions (Fig. \ref{spin1} (c)).

\subsubsection{Phase $\bf{k_2}$ = (0.25, 0.5, 0)}

\begin{table}[t]
\setlength{\tabcolsep}{8pt}
\caption{\label{tab2}Basis vectors (BVs) $\psi_i$ of IRs for two Fe atoms  in unit cell associated with propagation vector $\bf{k_2}$ = (0.25, 0.5, 0). BVs are defined by the crystallographic axes. }
\begin{tabular}{c|c|c|c}
\hline
IR 					      & BV 		& Fe1  & Fe2  \\ \hline
\hline
\multirow{3}{*}{$\Gamma_1$} & $\psi_1$ & (1 0 0) & (-1 -1 0) \\ \cline{2-4}
					     & $\psi_2$ & (0 1 0) & (0 1 0) \\ \cline{2-4}
					     & $\psi_3$ & (0 0 1) & (0 0 -1) \\ \hline
\multirow{3}{*}{$\Gamma_2$} & $\psi_4$ & (1 0 0) & (1 1 0) \\ \cline{2-4}
					      & $\psi_5$ & (0 1 0) & (0 -1 0) \\ \cline{2-4}
					     & $\psi_6$ & (0 0 1) & (0 0 1) \\ \hline
\hline
\end{tabular}
\vspace{-1.em}
\end{table}

\begin{figure*}
\includegraphics[width=1.8\columnwidth,clip,angle =0]{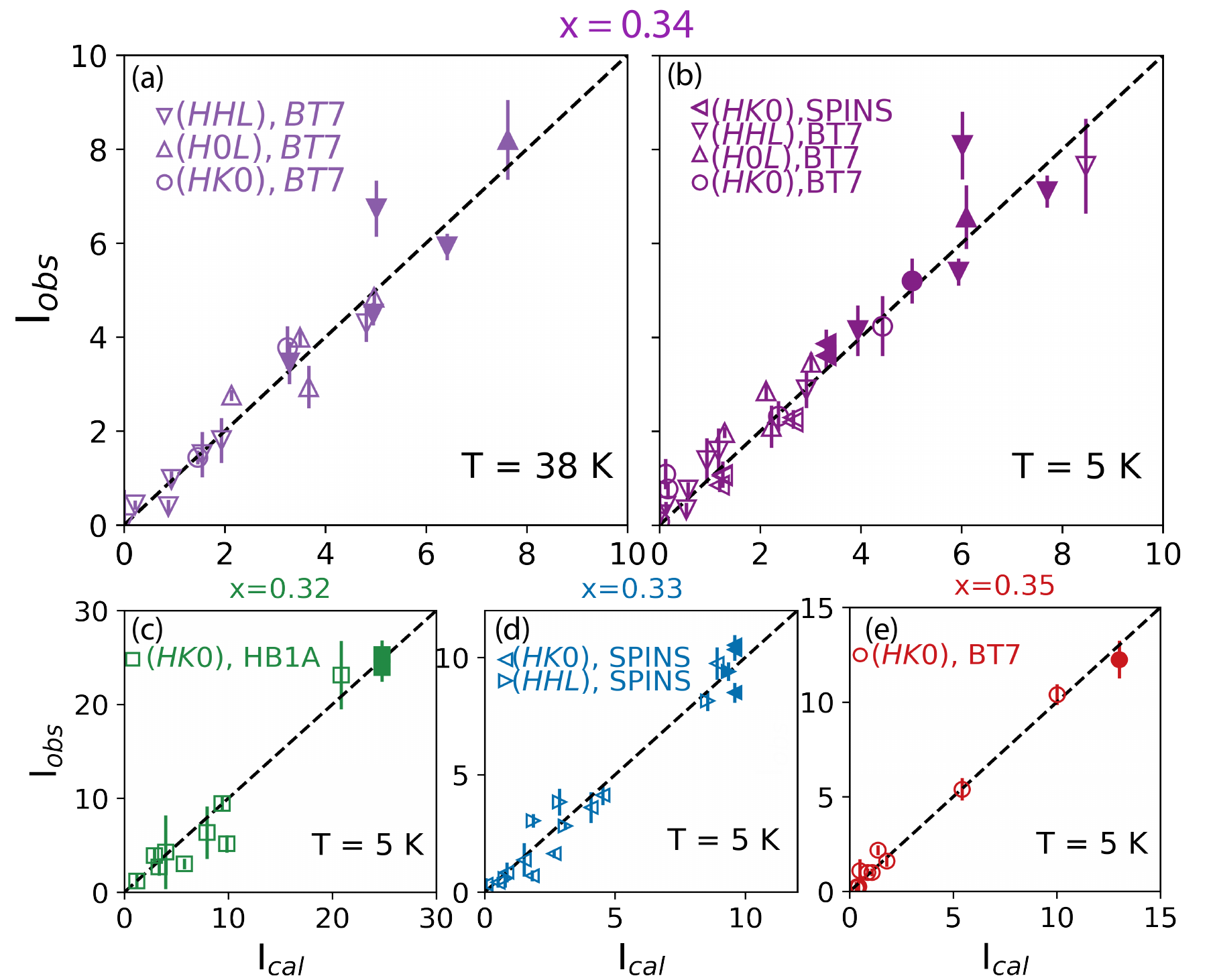}
\caption{\label{cal_obs} Observed versus calculated intensities of nuclear (filled symbols) and magnetic (empty symbols) peaks at $T$ = (a) 38 K and (b) 5 K for Fe$_x$NbS$_2$ crystal with  $x=0.34$, (c) $x=0.32$, (d) $x=0.33$ and $x=0.35$. Symbols in different types are data collected in different scattering planes and instruments according to the legends in each panel.
The calculated and observed intensities of peaks under different scattering geometries have been scaled simultaneously in order to be presented within the same frame. }
\end{figure*}

For the propagation vector $\bf{k_2}$ = (0.25, 0.5, 0), the magnetic representation $\Gamma_{mag}$ decomposes into IRs $\Gamma_{mag} = 3\Gamma_1 + 3\Gamma_2$  with corresponding BVs listed in the Table \ref{tab2}. Six BVs describe a collinear ($\psi_2,\psi_3, \psi_5,\psi_6$) and non-collinear ($\psi_1,\psi_4$) spin configuration.  $\psi_2$ and $\psi_6$ depict two parallel Fe spins, while $\psi_3$ and $\psi_5$  represent anti-parallel spins in one unit cell. By qualitatively comparing these to the diffraction pattern associated with $\bf{k_2}$ (Fig. \ref{fig3} (a)), the calculated patterns (Fig. \ref{fig3} (c) and Fig. \ref{FigS2} in the appendix) that are described by  $\psi_3$ and $\psi_5$ clearly follow selection rules for the magnetic peaks that are consistent with the observation. The other BVs result in unwanted reflections, for example $\bf{Q}=$ (0.25, 0.25, 0).

The corresponding spin configuration is displayed in Fig. \ref{spin2} (a-c). The difference between $\psi_3$ and $\psi_5$ is spin moments directed out-of-plane and in-plane, respectively.  Within the layer, spins point in the sequence of $++--$ along one crystal axis (+ and - denotes spins up and down for $\psi_3$).  Two Fe atoms in one-unit cell have spins pointing in opposite directions. Since connecting the same direction of the Fe spins within one-layer institutes a zig-zag route, we named this configuration as `AFM zigzag' for simplicity. The minimum magnetic unit cell is $4$ times the structural unit cell, and orthohexagonal. Note that there are six equivalent $k$ vectors ((0.5, 0.25, 0), (0.5, -0.75, 0), (0.75, -0.25, 0), (0.25, 0.5, 0), (-0.75, 0.5, 0) and (-0.25, 0.75, 0)), leading to three magnetic domains along three directions (Fig. \ref{spin2} (c)). 

Next, we quantitatively determine the magnetic structures for off and nearly stoichiometric Fe$_x$NbS$_2$ samples.

\subsubsection{Spin structure}
The neutron scattering data for the off-stoichiometric samples display a single magnetic transition with wave vector $\bf{k_1}$ and $\bf{k_2}$ in the $x$ = 0.32 and $x$ = 0.35 crystals, respectively. In the $x$ = 0.32 sample, since the strongest peak in the $(HK0)$ plane is at $\bf{Q}$ = (0.5, 0.5, 0), the spin structure with spins  along the $c$ axis is described by the basis vector of $\psi_5$ in the irreducible representation (IR) $\Gamma_3$ (Table \ref{tab1}). The ordered moment was obtained as $m = 2.6(3) \mu_B$ from comparison between the observed and calculated intensities  (Fig. \ref{cal_obs} (c)) by using Eq. (2-3) and a normalization factor from the nuclear peaks. The spin configuration can be described as AFM stripe with the moments oriented along the $c$ axis (Fig. \ref{spin1} (b)).

In the over-intercalated $x$ = 0.35 sample, all of the magnetic reflections are related to the wave vector $\bf{k_2}$. Since only one IR is allowed for a second order phase transition, $\psi_3$ in $\Gamma_1$ (Table. \ref{tab2}) was assigned to provide consistent results with the observed magnetic intensities (Fig. \ref{cal_obs} (e)). The ordered moment was obtained as $m = 3.0(3) \mu_B$; and the spin configuration can be described as  AFM zig-zag with moments along $c$ axis (Fig. \ref{spin2} (b)).

In the nearly stoichiometric samples with $x \sim$ 1/3, there are two magnetic transitions. Below $T_{N1}$, the spin structure can be ascribed to the AFM stripe configuration (Fig. \ref{spin1} (b)) depicted by $\psi_5$ in $\Gamma_3$ with ordered moment of 2.9(3) $\mu_B$ (Fig. \ref{cal_obs} (a)). Below $T_{N2}$,  to elaborate the rise-and-fall feature of magnetic peak at $\bf{Q_1}$ = (0.5, 0.5, 0) and the second phase transition, one possible scenario is to assign in-plane component associated with zig-zag configuration (Table \ref{tab2}: $\psi_5$ in $\Gamma_2$ ), which is allowed by the group theory and IR analysis. However, the calculated tilting angle (see the appendix) contradicts the large $c$-axis magnetic anisotropy found in our  susceptibility measurements and, furthermore, would require a DM interaction orders of magnitude larger than that allowed for by theory.

Alternatively, the rise-and-fall feature can be viewed as simply the zig-zag phase developing at the expense of the stripe phase. This can readily occur with decreasing temperature when the energy of two magnetic phases are nearly degenerate and the relative energies of the two phases changes subtly as a function of temperature. That is, the delicate energy balance between the two phases changes around $T_{N2}$ so that increasing regions of the sample favor the zig-zag phase as the temperature is decreased. This can also happen if, as the zig-zag phase grows, the domain boundaries of the stripe phase are converted to the zig-zag configuration.   Real space imaging of the domains would help elucidate this growth process. The redistribution of two magnetic phases is consistent with the rounding of the $T_{N}$ (Fig. \ref{fig4} (a-b)), indicating a small spread in the Fe ratio across the sample. In this scenario, the calculated intensities with ratio of $\sim$ 75 \%  and $\sim$ 35 \% stripe phase for $x$ =  0.33 and $x$ = 0.34 samples, respectively, are consistent with the observed patterns (Fig. \ref{cal_obs} (b) \& (d))  at 5 K.  The ordered moment is extracted as 3.2(3) $\mu_B$ and 3.5(3) $\mu_B$ correspondingly. The smaller value of these moments from the saturated moment under high field ($\sim$ 4 $\mu_B$ per Fe) is likely due to errors in the normalization factor because of the limited number of nuclear Bragg peaks.

\section{Discussions}
\subsection{High degenerate two magnetic phases}
In general, one finds that magnetic defects typically suppress transition temperatures and reduce  magnetic correlations. Here, both the ordered moments and transition temperatures are slightly reduced for off-stoichiometric samples. The remarkable observation here is the dramatic difference in the spin structures tuned by a small change in the concentration and nature of the magnetic defects, namely from vacancies to interstitials. Our single crystal neutron diffraction measurements reveal that the spin structure changes from purely stripe to purely zig-zag by varying $\delta$ from $\sim -0.01$ to 0.01 with both phases coexisting in near-stoichiometric samples.   These results demonstrate the first example of flexible tuning of the magnetic ground state by a subtle change of magnetic defects in the intercalation complexes of the Nb and Ta dichalcogenides and, more generally, a rare example in magnetic vdW compounds.

In the non-centrosymmetric intercalation species, several characteristic magnetic interactions are relevant. Two anisotropic exchange interactions are considered in Fe$_{1/3+\delta}$NbS$_2$.  First, single-ion anisotropy and, possibly, anisotropic exchange ($\sim$ 2 meV) result in highly anisotropic uniaxial Ising behavior, distinct from the easy-plane anisotropy observed in other T$_x$NbS$_2$ species studied so far \cite{friend1977}. Second, the Dzyaloshinskii-Moriya (DM) anti-symmetric interaction originates from the loss of inversion symmetry.  Specifically, the interlayer DM interactions with an in-plane component \cite{mankovsky2016} could theoretically produce a small in-plane moment. Unfortunately,  the extreme sensitivity of the magnetic ground state to the Fe concentration makes the determination of any small tilt angle of the spins indicated by zero field anisotropic magnetoresistance  (ZF-AMR) measurement \cite{maniv2021_2} extremely difficult.

Both the Ruderman-Kittel-Kasuya-Yosida (RKKY) interaction and the super-exchange interaction were considered as the relevant mechanisms for the magnetic ordering in this system \cite{parkin1980,friend1977}.   The former is long-ranged and variable in both sign and magnitude; it relies on the separation of localized moments and the Fermi wave vector \cite{ruderman1954,Yosida1957,kasuya1956}.  The latter is relatively short-ranged and the sign and magnitude are often determined by application of the Goodenough-Kanamori rules \cite{goodenough1955,kanamori1959}. Since two dramatically different ordered phases are facilitated by a small concentration of magnetic defects, the super-exchange interaction would be barely affected without a change of the local structure. Alternatively, the change of magnetic defects from vacancies to interstitials, presumably, could influence the RKKY interaction, especially for the interlayer exchange coupling with its larger Fe-Fe distance. This scenario is embedded in the oscillatory character of RKKY interaction \cite{ruderman1954,Yosida1957,kasuya1956,aristov1997,Ko2011}; it is analogous to the alternating exchange couplings in transition-metal layers separated by a non-magnetic metal spacer \cite{parkin1991_prl,parkin1991}.

Recent density functional theory (DFT) studies \cite{shannon2020, weber2021} of the AFM stripe and AFM zigzag magnetic ground states strongly support our experimental results. 
To partially account for enhanced localizations of Fe $d$ electrons, a Hubbard U corrections was added in our DFT calculations. Both $\mathrm{U}=0.3$ $\mathrm{eV}$  and $\mathrm{U}=0.9$ $\mathrm{eV}$ predict an easy-axis anisotropy along $[001]$, consistent with experiment. PBE+U energies for magnetic orderings corresponding to AFM stripe ($\bf{k}_1$= (0.5, 0, 0))  and AFM zigzag ($\bf{k}_2$= (0.25, 0.5, 0)) are reported differ in energy by at most a few (1-3) $\mathrm{meV}$ per $\mathrm{Fe}$ atom.  For context, this energy scale is significantly smaller (2 $\mathrm{meV}/k_B=23.2$ $\mathrm{K}$) than the onset temperature of either magnetic phase for near-stoichiometry samples, rendering the stripe and zigzag phases effectively degenerate. Moreover, intriguingly, the relative energy ordering of AFM stripe and AFM zigzag phases switches in going from PBE+U with $\mathrm{U}=0.3$ $\mathrm{eV}$ to $\mathrm{U}=0.9$ $\mathrm{eV}$ \cite{weber2021}. The AFM stripe is lower in energy by 0.9 meV/Fe by using $\mathrm{U}$ = 0.3 eV, whereas the AFM zigzag is lower by 2.5 meV/Fe by using $\mathrm{U}$ = 0.9 eV.

 \begin{table}
\begin{ruledtabular}
\begin{tabular}{| c | c | c | c | c | c |}
 & $J_1$ & $J_{1}'$ & $J_{2}'$ & $J_2$ & $J_{3}'$ \\ \hline
$\mathrm{U}=0.3$ $\mathrm{eV}$ & $+0.76$ & $+0.49$ & $-0.20$ & $-0.006$ & $-0.07$\\ \hline
$\mathrm{U}=0.9$ $\mathrm{eV}$ & $+0.57$ & $+0.28$ & $-0.16$ & $-0.14$ & $-0.09$\\
\end{tabular}
\end{ruledtabular}
\caption{Heisenberg spin exchange constants, in $\mathrm{meV}$/$\mathrm{Fe}$ atom, calculated with PBE+U for $\mathrm{U}=0.3$ $\mathrm{eV}$ and $\mathrm{U}=0.9$ $\mathrm{eV}$. Positive values ($J>0$) are AFM coupling constants in our notation, and negative ($J<0$) couplings are FM. The prime refers to interplanar couplings. 
}\label{tab:Jcouplings} 
\end{table}

The near-degeneracy and competition between AFM stripe and zigzag phases near stoichiometry can be further understood by a minimal Heisenberg model \cite{shannon2020, weber2021}, neglecting the single-ion anisotropy since this contributions cancels when calculating differences in energy between [001] oriented collinear magnetic orders. We  highlight the results of the prior work related to our experiments in what follows. The mean-field energy with classical spin $S$ can be written as \cite{weber2021}:  
\begin{multline}
H=E_0+\sum_{\langle ij \rangle}J_1S^2+\sum_{\langle \langle ij \rangle \rangle}J_2S^2+\sum_{{\langle ij \rangle}'}J_{1}'S^2 \\
+\sum_{{\langle \langle ij \rangle \rangle}'}J_{2}'S^2+\sum_{{\langle \langle \langle ij \rangle \rangle \rangle}'}J_{3}'S^2,
 \label{eq:HeisHam}
 \end{multline}
Where one, two and three pairs of brackets denote Heisenberg exchange constants between equivalent nearest, next-nearest  and third-nearest neighbor interactions respectively; and the prime refers to inter-layer interactions (Fig. \ref{spin1} (a) \& Fig. \ref{spin2} (a)).  PBE+U-derived Heisenberg exchange constants for both $\mathrm{U}$ values examined from \cite{weber2021} are given in $\mathrm{meV}$ per $\mathrm{Fe}$ atom in Table \ref{tab:Jcouplings}. Based on the AFM nearest neighbor interactions $J_1$ and $J_{1}'$ alone, the mean-field energies for AFM stripe and zigzag are degenerate; and are primarily responsible for the antiferromagnetism within and between the layers in both structures. The degeneracy is broken by the relative small values of the next nearest neighbor interactions $J_2$ and $J_{2}'$ as well as third-nearest neighbor inter-layer interaction $J_{3}'$. The energy difference between the two phases is $E_{stripe}-E_{zigzag}=4J_{2}'S^2-4J_2S^2-8J_{3}'S^2$ \cite{weber2021}. The AFM stripe  phase is then favored when $|J_2'| >|J_2|+2|J_{3}'|$, whereas the AFM zigzag phase is energetically favored when $|J_{2}'|<|J_2|+2|J_{3}'|$.

The relative change in magnitude and even signs of three exchange constants can be attributed to the high degeneracy of the two magnetic phases. As a possible microscopic mechanism, we note that on the one hand,  the interlayer exchange interactions originated via RKKY mechanism are weak due to the long separation distance ($\sim 9-10$ \AA) and further have an oscillatory nature. On the other hand, magnetic Fe defects that reside within or between layers can give rise to changes in the Fermi surface. Our preliminary photoemission work reveals a rapid change of the Fermi surface size from $x < 1/3$  to $x > 1/3$. This provides evidence that magnetic defects would affect the couplings between localized moments via the conduction electrons. Accordingly, the values or even sign of three exchange constants would be quite sensitive to $x$, leading to the tuning between the two AFM phases by magnetic defects from  $x < 1/3$  to $x > 1/3$. 
As a result of the nearly degenerate states, the delicate balance of the two magnetic phases, which are spatially separated in the $x= 1/3$ sample, can be changed causing one phase to win over the other leading to the rise-and-fall feature in the order parameter curve. The knob could be subtle changes in the RKKY interactions with decreasing the temperature, or magneto-elastic interactions that would turn on when magnetic ordering sets in for the two different phases. Further calculations could elucidate the possible mechanisms.

\subsection{Relation to the spintronic features}
Magnetic defects not only tune the magnetic ground states, but they also influence the intriguing spintronic features in Fe$_{1/3+\delta}$NbS$_2$ \cite{maniv2021_1,maniv2021_2}. 
By injecting a current pulse along the $[100]$ direction,  for $x < 1/3$ and $x>1/3$ samples,  the change in transverse resistance is positive and negative, respectively. In addition, the devices display more active responses in off-stoichiometric samples, either below or above the $x$ = 1/3 sample; while for $x = 1/3$ the amplitude of the resistance switching is dramatically diminished. Our neutron work provides fundamental information on the magnetic ground states in samples with different Fe ratios that display rapid changes in the spintronic behaviors as a function of Fe concentrations.   

 First, in the stoichiometric $x = 1/3$ sample, the evolution of the two magnetic phases is reflected in the response to electrical-current and magnetic field where both the switching resistance and the ZF-AMR  reveal a sign change upon lowering the temperature\cite{maniv2021_2}. This suggests that differing crystal symmetries in the two magnetic phases may play important roles in such effects. Moreover, based on PBE+U calculations of the electronic structure and transport properties \cite{weber2021}, the AFM stripe and AFM zigzag phases have opposite in-plane transport anisotropy, that is, $R_{xx}<R_{yy}$ for stripe and $R_{xx}>R_{yy}$ for zigzag (Here the $xx$ subscript defines the resistance measured parallel to the stripe/zigzag single domain). If we hypothesize that the resistance switching is caused by a current-induced repopulation of magnetic domains via the Rashba coupling \cite{little2020,weber2021}, the current pulse could lead to the sign change in the resistance between the two magnetic phases according to the calculations \cite{weber2021}.  The suppression of resistance switching with decreasing the temperature and compared to the off-stochiometric samples could then be simply ascribed to the partial cancellation of the resistance changes.
 
 Second, in $x < 1/3$ and $x>1/3$  sample our neutron experiments clearly demonstrate single long-ranged stripe order and zigzag order respectively, with both revealing three magnetic domains.  Further, as discussed above, the PBE+U calculations predict opposite signs for the in-plane transport anisotropy in the two phases.  These provide important bases for understanding the rich observed switching behavior.  However, the latter also requires a detailed understanding of the actual switching mechanism,  which has not yet been definitively identified.  We leave that as a subject of future research, both experimental and theoretical. Finally, the apparent absence of AFM order in our heavily under-intercalated sample ($x$ = 0.31) is surprising. Further studies of samples in this regime are clearly required.

\section{Conclusions}
To conclude, we have performed single crystal neutron diffraction experiments in the Fe intercalated transition metal dichalcogenide material Fe$_{1/3+\delta}$NbS$_2$, which recently has been shown to exhibit intriguing resistance switching and magnetic memory effects.   Two long-range ordered magnetic phases, specifically AFM stripe order with wave vector $k_1$ = (0.5, 0, 0) and AFM zig-zag order with $k_2$ = (0.25, 0.5, 0), have been found and they can be sensitively tuned by the Fe concentration as one goes from the under-intercatated to over-intercaleted region of the phase diagram. This  arises from the nearly degenerate energies for the two spin structures, supported by our DFT calculation. Two phases can be tuned from one to the other due to the oscillating nature of RKKY interaction $J$ and the competition between secondary intra- and inter-layer interactions. Two successive magnetic transitions are observed in stoichiometric samples;  the emergence of the second magnetic phase is consistent with the remarkable near-degeneracy in energy of the two states. We provide crucial information on magnetic ground states that form the basis for understanding the interesting spintronic behaviors. Our discovery of the highly tunable magnetic phases in this bulk sample open up new, intriguing opportunities to manipulate magnetic states and, concomitantly, the spintronic properties by magnetic defects.

\section{Acknowledgments}
The authors would like to thanks Edith Bourret-Courchesne, Didier Perrodin and Drew Onken for initial help of precursors synthesis, Yu He for help of some MPMS characterizations, Xiang Chen for some EDX measurements, Zhentao Wang and Zhenglu Li  for fruitful discussions, and Nick Settineri for single crystal X-ray measurement at ChemXray facility. This work is funded by the U.S. Department of Energy, Office of Science, Office of Basic Energy Sciences, Materials Sciences and Engineering Division under Contract No. DE-AC02-05-CH11231 within the Quantum Materials Program (KC2202).  The work from Weber and Neaton were supported by the Center for Novel Pathways to Quantum Coherence in Materials, an Energy Frontier Research Center funded by the US Department of Energy, Director, Office of Science, Office of Basic Energy Sciences under Contract No. DE-AC02-05CH11231. Access to MACS was provided by the Center for High Resolution Neutron Scattering, a partnership between the National Institute of Standards and Technology and the National Science Foundation under Agreement No. DMR-1508249.
The identification of any commercial product or trade name does not imply endorsement or recommendation by the National Institute of Standards and Technology, nor does it imply that the materials or equipment identified are necessarily the best available for the purpose.
A portion of this research used resources at the High Flux Isotope Reactor, a DOE Office of Science User Facility operated by the Oak Ridge National Laboratory.

\section{Appendix}
The appendix includes the following information: 1) characterization of the other single crystals that were used in the neutron diffraction experiments; 2) remarks on the possibility of a small in-plane moment developed below T$_{N2}$ in the x = 1/3 sample; 3) the calculated pattern for each basis vector associated with $k_1$ = (0.5, 0, 0) and $k_2$ = (0.25, 0.5, 0).

\begin{figure}
\includegraphics[width=1\columnwidth,clip,angle =0]{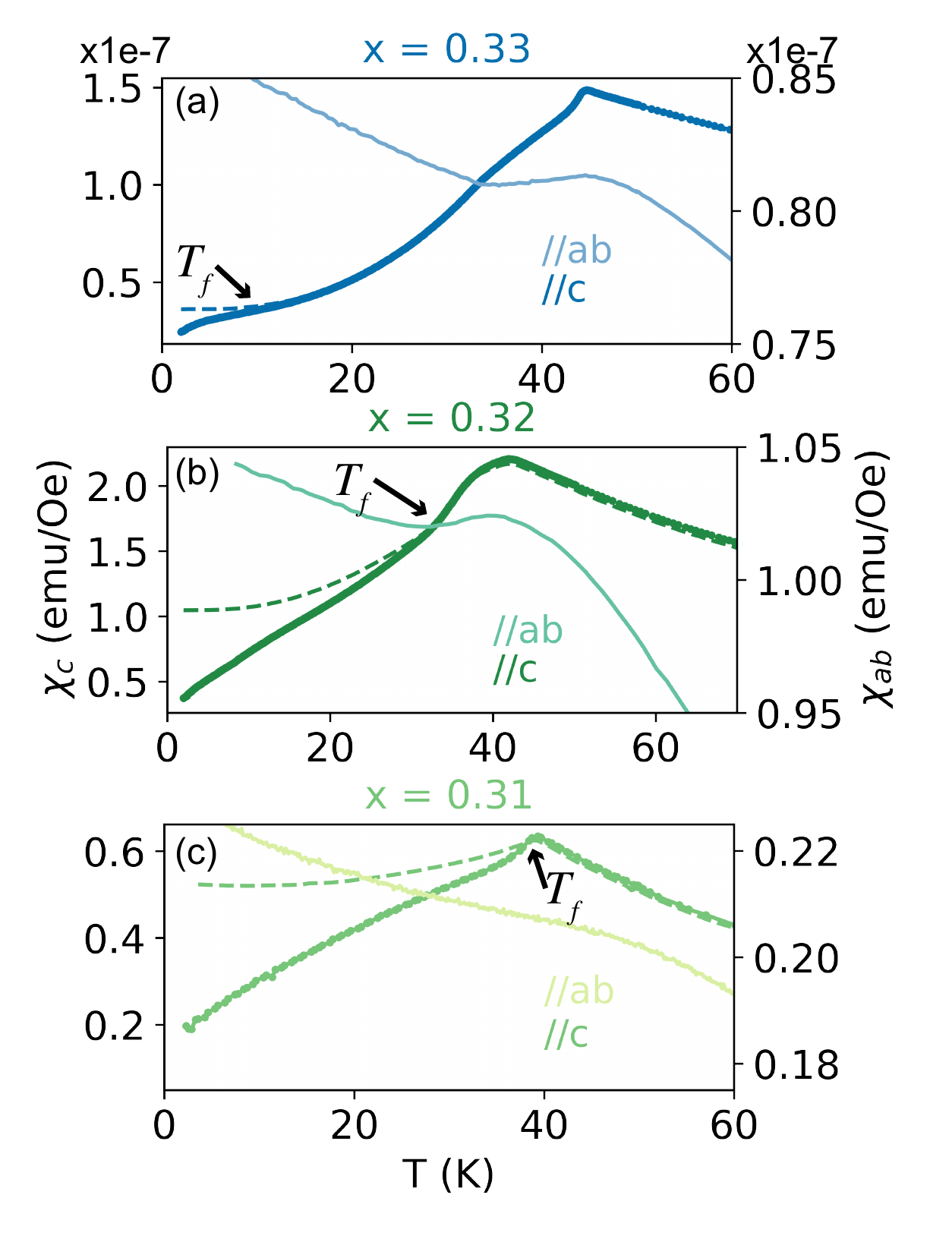}
\caption{\label{suscp_all} Magnetization measurements for $x$ = (a) 0.33, (b) 0.32 and (c) 0.31 with applied field along $c$ and in $ab$-plane. The dashed and solid lines corresponding to the measurements with field-cooled and zero field cooled process. The measurements are used from the same sample in the neutron diffraction experiment. 
}
\end{figure}

 The magnetization measurements for the other measured neutron samples are shown in Fig. \ref{suscp_all}.  The separation between the zero field cooled (ZFC) and field cooled (FC) susceptibility is dependent upon the deviation from the stoichiometric ratio of 1/3.  ZFC and FC curves separate at a characteristic temperature $T_f$; such a separation does not occur in samples with x very near 1/3. In the crystal with x = 0.33, the susceptibility along the $c$-axis exhibits one anomalous peak followed by a broad hump with decreasing temperature; correspondingly two kinks in the in-plane susceptibility $\chi_{ab}$, are shown at $T_{N1} \sim$ 32 K and $T_{N2} \sim$ 45 K. In the single crystal with x = 0.32, the second kink in $\chi_{ab}$ occurs around $T_f \sim 32$ K. In the crystal with x = 0.31, $T_f$ is close to the peak anomaly in the $c$ axis susceptibility around 40 K. In the x = 0.35 sample, the characterization data show the transition T$_N \sim 40$ K. Both off-stochiometric  and stoichiometric samples exhibit strong uniaxial anisotropy in their susceptibilities (Fig. \ref{fig2}). For x $>$ 1/3,  Curie Weiss fits to the susceptibility  in the paramagnetic region yield values of the paramagnetic effective moment   $\mu_{eff}$ = 5.0(3)$\mu_B$  and Curie Weiss temperature $\theta_{cw}$ = -50(2) K along the $c$ axis; $\mu_{eff}$ = 5.0(3)$\mu_B$ and $\theta_{cw}$ = -165(5) K in $ab$ plane. For x $<$ 1/3, the Curie Weiss fits in the paramagnetic region yield the values of paramagnetic effective moment   $\mu_{eff}$ = 5.0(3)$\mu_B$  and Curie Weiss temperature $\theta_{cw}$ = -24(1) K along the $c$ axis; $\mu_{eff}$ = 4.7(2)$\mu_B$ and $\theta_{cw}$ = -104(2) K in the $ab$ plane.

In the samples with x $\sim$ 1/3 because of the complications presented by the coexistence of two different magnetic structures, it is not possible to say anything meaningful about any in-plane moment.  However, any such in-plane moment would be caused by the interlayer DM interaction which is small compared to both the primary exchange and the c-axis anisotropy.

The single crystal neutron diffraction intensities are calculated according to the formula Eq. 2 and Eq. 3 in the main text. We utilize the BVs vector that describe the spin configuration and calculate the magnetic intensities. We derived the selection rule for each propagation wave vector as:  $\delta_{2h} \delta_k$ for the wave vector $\bf{k}$ = (0.5,0,0) domain, and  $\delta_{2h} \delta_{h+2k}$ for the wave vector $\bf{k}$= (0.5,0.25,0) domain. Here $h$ and $k$ are Miller indices for the wave vector transfer $Q$. The calculation includes three domains with equal weights and the square of the magnetic form factor. The normalization factor for the magnetic peaks $NC_{m}$ is obtained from the ratio between the calculated square of the structure factor and the integrated area of the nuclear peaks $NC_{n}$. The relation between them is: $NC_{m} = \frac{N_{n}V_{m}}{N_{m}*V_{n}}*NC_{n}$, where $V$ and $N$ stand for the volume and number of magnetic/nuclear unit cell respectively. By this normalization, we can obtain the ordered moment size by comparing the calculated and measured intensities as shown in Fig. \ref{cal_obs}.

\begin{figure}
\includegraphics[width=1\columnwidth,clip,angle =0]{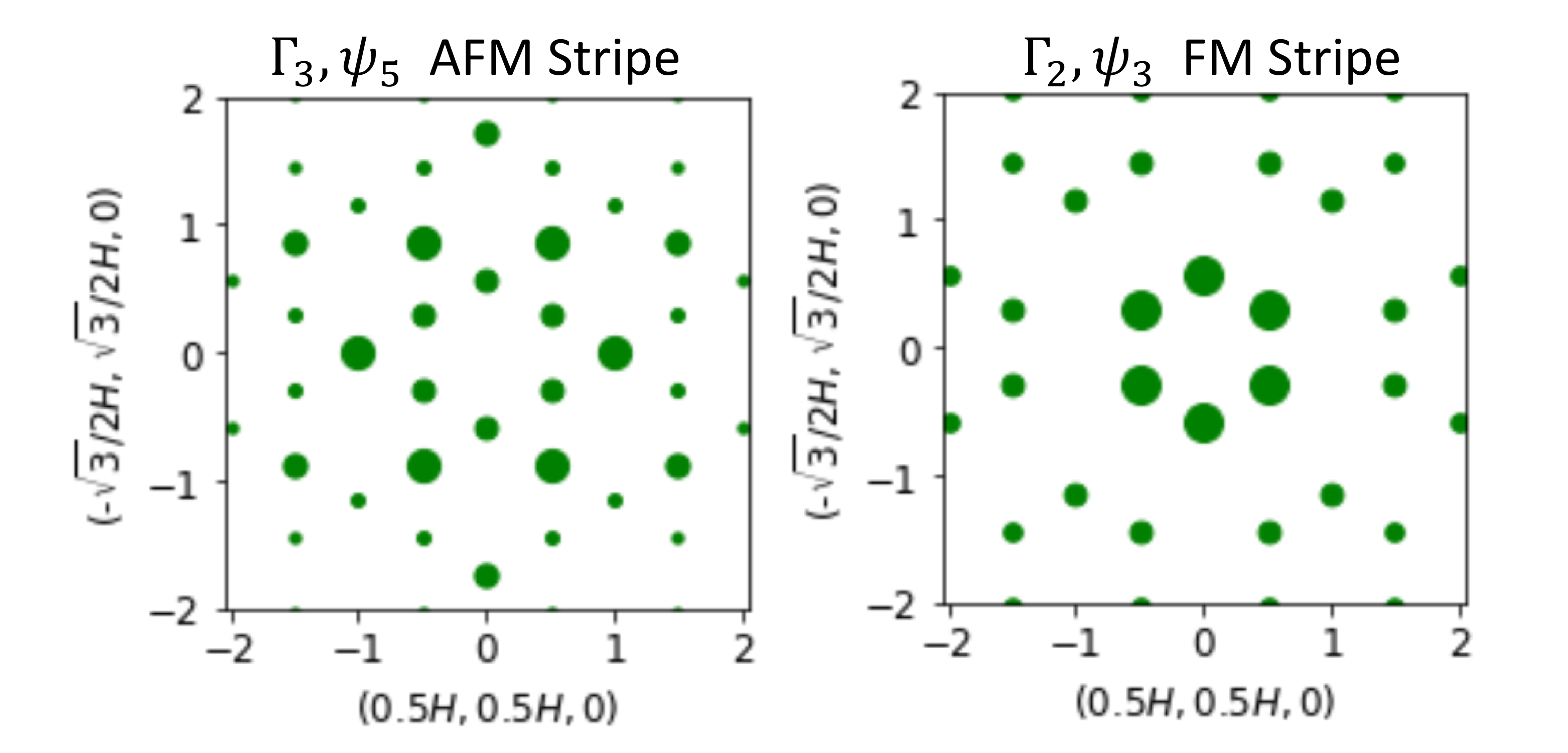}
\caption{\label{FigS1} Calculated intensities for given irreducible representation and basis vector associated with $\bf{k_1}$ = (0.5, 0, 0) and other 2 equivalent $k$s , describing a spin configuration of AFM stripe (left) and FM stripe (right) with moment direction along $c$ axis. The size of dots represent the intensities of peaks, including contributions of all equivalent domains.}
\end{figure}

\begin{figure}
\includegraphics[width=1.\columnwidth,clip,angle =0]{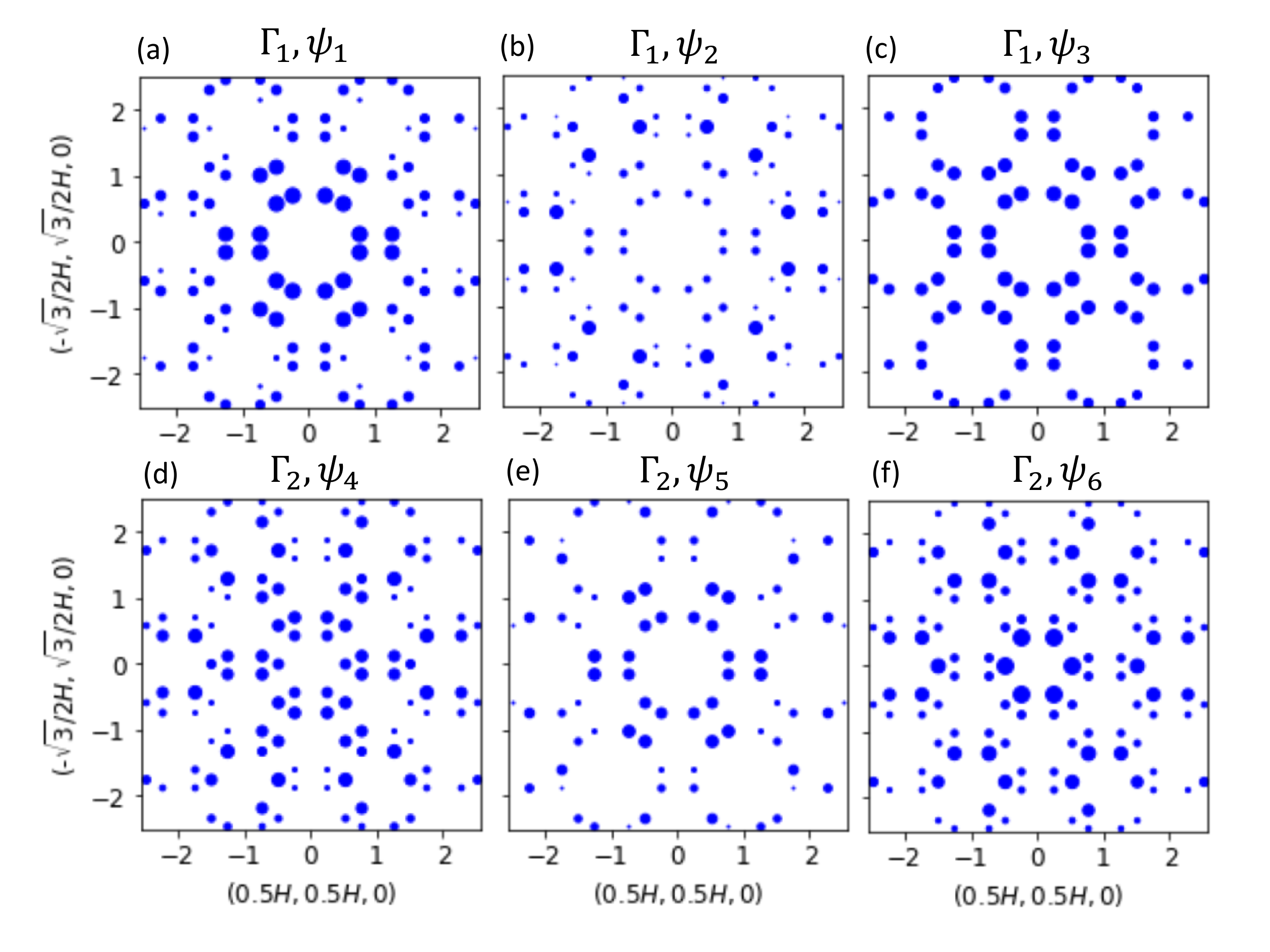}
\caption{\label{FigS2} Calculated intensities for given irreducible representation (IR) and basis vector associated with $\bf{k_2}$ = (0.5, 0.25, 0) and other 5 equivalent $k$s. Plots of (a-c) present the simulation for $\Gamma_1$ and (d-f) for $\Gamma_2$ IR, corresponding to the magnetic space group symmetry $P_c 2_12_12$ and  $P_c 2_12_12_1$ respectively. The size of dots represent the intensities, including contributions of all equivalent domains. 
}
\end{figure}

\bibliography{bibfile.bib}

\begin{thebibliography}{68}%
\makeatletter
\providecommand \@ifxundefined [1]{%
 \@ifx{#1\undefined}
}%
\providecommand \@ifnum [1]{%
 \ifnum #1\expandafter \@firstoftwo
 \else \expandafter \@secondoftwo
 \fi
}%
\providecommand \@ifx [1]{%
 \ifx #1\expandafter \@firstoftwo
 \else \expandafter \@secondoftwo
 \fi
}%
\providecommand \natexlab [1]{#1}%
\providecommand \enquote  [1]{``#1''}%
\providecommand \bibnamefont  [1]{#1}%
\providecommand \bibfnamefont [1]{#1}%
\providecommand \citenamefont [1]{#1}%
\providecommand \href@noop [0]{\@secondoftwo}%
\providecommand \href [0]{\begingroup \@sanitize@url \@href}%
\providecommand \@href[1]{\@@startlink{#1}\@@href}%
\providecommand \@@href[1]{\endgroup#1\@@endlink}%
\providecommand \@sanitize@url [0]{\catcode `\\12\catcode `\$12\catcode
  `\&12\catcode `\#12\catcode `\^12\catcode `\_12\catcode `\%12\relax}%
\providecommand \@@startlink[1]{}%
\providecommand \@@endlink[0]{}%
\providecommand \url  [0]{\begingroup\@sanitize@url \@url }%
\providecommand \@url [1]{\endgroup\@href {#1}{\urlprefix }}%
\providecommand \urlprefix  [0]{URL }%
\providecommand \Eprint [0]{\href }%
\providecommand \doibase [0]{https://doi.org/}%
\providecommand \selectlanguage [0]{\@gobble}%
\providecommand \bibinfo  [0]{\@secondoftwo}%
\providecommand \bibfield  [0]{\@secondoftwo}%
\providecommand \translation [1]{[#1]}%
\providecommand \BibitemOpen [0]{}%
\providecommand \bibitemStop [0]{}%
\providecommand \bibitemNoStop [0]{.\EOS\space}%
\providecommand \EOS [0]{\spacefactor3000\relax}%
\providecommand \BibitemShut  [1]{\csname bibitem#1\endcsname}%
\let\auto@bib@innerbib\@empty
\bibitem [{\citenamefont {Maniv}\ \emph
  {et~al.}(2021{\natexlab{a}})\citenamefont {Maniv}, \citenamefont {Murphy},
  \citenamefont {Haley}, \citenamefont {Doyle}, \citenamefont {John},
  \citenamefont {Maniv}, \citenamefont {Ramakrishna}, \citenamefont {Tang},
  \citenamefont {Ercius}, \citenamefont {Ramesh}, \citenamefont {Reyes},
  \citenamefont {Long},\ and\ \citenamefont {Analytis}}]{maniv2021_1}%
  \BibitemOpen
  \bibfield  {author} {\bibinfo {author} {\bibfnamefont {E.}~\bibnamefont
  {Maniv}}, \bibinfo {author} {\bibfnamefont {R.~A.}\ \bibnamefont {Murphy}},
  \bibinfo {author} {\bibfnamefont {S.~C.}\ \bibnamefont {Haley}}, \bibinfo
  {author} {\bibfnamefont {S.}~\bibnamefont {Doyle}}, \bibinfo {author}
  {\bibfnamefont {C.}~\bibnamefont {John}}, \bibinfo {author} {\bibfnamefont
  {A.}~\bibnamefont {Maniv}}, \bibinfo {author} {\bibfnamefont {S.~K.}\
  \bibnamefont {Ramakrishna}}, \bibinfo {author} {\bibfnamefont {Y.-L.}\
  \bibnamefont {Tang}}, \bibinfo {author} {\bibfnamefont {P.}~\bibnamefont
  {Ercius}}, \bibinfo {author} {\bibfnamefont {R.}~\bibnamefont {Ramesh}},
  \bibinfo {author} {\bibfnamefont {A.~P.}\ \bibnamefont {Reyes}}, \bibinfo
  {author} {\bibfnamefont {J.~R.}\ \bibnamefont {Long}},\ and\ \bibinfo
  {author} {\bibfnamefont {J.~G.}\ \bibnamefont {Analytis}},\ }\bibfield
  {title} {\bibinfo {title} {Exchange bias due to coupling between coexisting
  antiferromagnetic and spin-glass orders},\ }\bibfield  {journal} {\bibinfo
  {journal} {Nature Physics}\ }\href
  {https://doi.org/10.1038/s41567-020-01123-w} {10.1038/s41567-020-01123-w}
  (\bibinfo {year} {2021}{\natexlab{a}})\BibitemShut {NoStop}%
\bibitem [{\citenamefont {Maniv}\ \emph
  {et~al.}(2021{\natexlab{b}})\citenamefont {Maniv}, \citenamefont {Nair},
  \citenamefont {Haley}, \citenamefont {Doyle}, \citenamefont {John},
  \citenamefont {Cabrini}, \citenamefont {Maniv}, \citenamefont {Ramakrishna},
  \citenamefont {Tang}, \citenamefont {Ercius}, \citenamefont {Ramesh},
  \citenamefont {Tserkovnyak}, \citenamefont {Reyes},\ and\ \citenamefont
  {Analytis}}]{maniv2021_2}%
  \BibitemOpen
  \bibfield  {author} {\bibinfo {author} {\bibfnamefont {E.}~\bibnamefont
  {Maniv}}, \bibinfo {author} {\bibfnamefont {N.~L.}\ \bibnamefont {Nair}},
  \bibinfo {author} {\bibfnamefont {S.~C.}\ \bibnamefont {Haley}}, \bibinfo
  {author} {\bibfnamefont {S.}~\bibnamefont {Doyle}}, \bibinfo {author}
  {\bibfnamefont {C.}~\bibnamefont {John}}, \bibinfo {author} {\bibfnamefont
  {S.}~\bibnamefont {Cabrini}}, \bibinfo {author} {\bibfnamefont
  {A.}~\bibnamefont {Maniv}}, \bibinfo {author} {\bibfnamefont {S.~K.}\
  \bibnamefont {Ramakrishna}}, \bibinfo {author} {\bibfnamefont {Y.-L.}\
  \bibnamefont {Tang}}, \bibinfo {author} {\bibfnamefont {P.}~\bibnamefont
  {Ercius}}, \bibinfo {author} {\bibfnamefont {R.}~\bibnamefont {Ramesh}},
  \bibinfo {author} {\bibfnamefont {Y.}~\bibnamefont {Tserkovnyak}}, \bibinfo
  {author} {\bibfnamefont {A.~P.}\ \bibnamefont {Reyes}},\ and\ \bibinfo
  {author} {\bibfnamefont {J.~G.}\ \bibnamefont {Analytis}},\ }\bibfield
  {title} {\bibinfo {title} {Antiferromagnetic switching driven by the
  collective dynamics of a coexisting spin glass},\ }\bibfield  {journal}
  {\bibinfo  {journal} {Science Advances}\ }\textbf {\bibinfo {volume} {7}},\
  \href {https://doi.org/10.1126/sciadv.abd8452} {10.1126/sciadv.abd8452}
  (\bibinfo {year} {2021}{\natexlab{b}})\BibitemShut {NoStop}%
\bibitem [{\citenamefont {Burch}\ \emph {et~al.}(2018)\citenamefont {Burch},
  \citenamefont {Mandrus},\ and\ \citenamefont {Park}}]{kenneth2018}%
  \BibitemOpen
  \bibfield  {author} {\bibinfo {author} {\bibfnamefont {K.~S.}\ \bibnamefont
  {Burch}}, \bibinfo {author} {\bibfnamefont {D.}~\bibnamefont {Mandrus}},\
  and\ \bibinfo {author} {\bibfnamefont {J.-G.}\ \bibnamefont {Park}},\
  }\bibfield  {title} {\bibinfo {title} {Magnetism in two-dimensional van der
  waals materials},\ }\href {https://doi.org/10.1038/s41586-018-0631-z}
  {\bibfield  {journal} {\bibinfo  {journal} {Nature}\ }\textbf {\bibinfo
  {volume} {563}},\ \bibinfo {pages} {47} (\bibinfo {year} {2018})}\BibitemShut
  {NoStop}%
\bibitem [{\citenamefont {Park}(2016)}]{jgp2016}%
  \BibitemOpen
  \bibfield  {author} {\bibinfo {author} {\bibfnamefont {J.-G.}\ \bibnamefont
  {Park}},\ }\bibfield  {title} {\bibinfo {title} {Opportunities and challenges
  of 2d magnetic van der waals materials: magnetic graphene?},\ }\href
  {https://doi.org/10.1088/0953-8984/28/30/301001} {\bibfield  {journal}
  {\bibinfo  {journal} {Journal of Physics: Condensed Matter}\ }\textbf
  {\bibinfo {volume} {28}},\ \bibinfo {pages} {301001} (\bibinfo {year}
  {2016})}\BibitemShut {NoStop}%
\bibitem [{\citenamefont {Ajayan}\ \emph {et~al.}(2016)\citenamefont {Ajayan},
  \citenamefont {Kim},\ and\ \citenamefont {Banerjee}}]{pulickel2016}%
  \BibitemOpen
  \bibfield  {author} {\bibinfo {author} {\bibfnamefont {P.}~\bibnamefont
  {Ajayan}}, \bibinfo {author} {\bibfnamefont {P.}~\bibnamefont {Kim}},\ and\
  \bibinfo {author} {\bibfnamefont {K.}~\bibnamefont {Banerjee}},\ }\bibfield
  {title} {\bibinfo {title} {Two-dimensional van der waals materials},\
  }\href@noop {} {\bibfield  {journal} {\bibinfo  {journal} {Phys. Today}\
  }\textbf {\bibinfo {volume} {69}} (\bibinfo {year} {2016})}\BibitemShut
  {NoStop}%
\bibitem [{\citenamefont {Gong}\ and\ \citenamefont {Zhang}(2019)}]{cheng2019}%
  \BibitemOpen
  \bibfield  {author} {\bibinfo {author} {\bibfnamefont {C.}~\bibnamefont
  {Gong}}\ and\ \bibinfo {author} {\bibfnamefont {X.}~\bibnamefont {Zhang}},\
  }\bibfield  {title} {\bibinfo {title} {Two-dimensional magnetic crystals and
  emergent heterostructure devices},\ }\bibfield  {journal} {\bibinfo
  {journal} {Science}\ }\textbf {\bibinfo {volume} {363}},\ \href
  {https://doi.org/10.1126/science.aav4450} {10.1126/science.aav4450} (\bibinfo
  {year} {2019})\BibitemShut {NoStop}%
\bibitem [{\citenamefont {Mak}\ \emph {et~al.}(2019)\citenamefont {Mak},
  \citenamefont {Shan},\ and\ \citenamefont {Ralph}}]{kin2019}%
  \BibitemOpen
  \bibfield  {author} {\bibinfo {author} {\bibfnamefont {K.~F.}\ \bibnamefont
  {Mak}}, \bibinfo {author} {\bibfnamefont {J.}~\bibnamefont {Shan}},\ and\
  \bibinfo {author} {\bibfnamefont {D.~C.}\ \bibnamefont {Ralph}},\ }\bibfield
  {title} {\bibinfo {title} {Probing and controlling magnetic states in 2d
  layered magnetic materials},\ }\href
  {https://doi.org/10.1038/s42254-019-0110-y} {\bibfield  {journal} {\bibinfo
  {journal} {Nature Reviews Physics}\ }\textbf {\bibinfo {volume} {1}},\
  \bibinfo {pages} {646} (\bibinfo {year} {2019})}\BibitemShut {NoStop}%
\bibitem [{\citenamefont {Gibertini}\ \emph {et~al.}(2019)\citenamefont
  {Gibertini}, \citenamefont {Koperski}, \citenamefont {Morpurgo},\ and\
  \citenamefont {Novoselov}}]{gibertini2019}%
  \BibitemOpen
  \bibfield  {author} {\bibinfo {author} {\bibfnamefont {M.}~\bibnamefont
  {Gibertini}}, \bibinfo {author} {\bibfnamefont {M.}~\bibnamefont {Koperski}},
  \bibinfo {author} {\bibfnamefont {A.~F.}\ \bibnamefont {Morpurgo}},\ and\
  \bibinfo {author} {\bibfnamefont {K.~S.}\ \bibnamefont {Novoselov}},\
  }\bibfield  {title} {\bibinfo {title} {Magnetic 2d materials and
  heterostructures},\ }\href {https://doi.org/10.1038/s41565-019-0438-6}
  {\bibfield  {journal} {\bibinfo  {journal} {Nature Nanotechnology}\ }\textbf
  {\bibinfo {volume} {14}},\ \bibinfo {pages} {408} (\bibinfo {year}
  {2019})}\BibitemShut {NoStop}%
\bibitem [{\citenamefont {Li}\ \emph {et~al.}(2019)\citenamefont {Li},
  \citenamefont {Jiang}, \citenamefont {Sivadas}, \citenamefont {Wang},
  \citenamefont {Xu}, \citenamefont {Weber}, \citenamefont {Goldberger},
  \citenamefont {Watanabe}, \citenamefont {Taniguchi}, \citenamefont {Fennie},
  \citenamefont {Fai~Mak},\ and\ \citenamefont {Shan}}]{ting2019}%
  \BibitemOpen
  \bibfield  {author} {\bibinfo {author} {\bibfnamefont {T.}~\bibnamefont
  {Li}}, \bibinfo {author} {\bibfnamefont {S.}~\bibnamefont {Jiang}}, \bibinfo
  {author} {\bibfnamefont {N.}~\bibnamefont {Sivadas}}, \bibinfo {author}
  {\bibfnamefont {Z.}~\bibnamefont {Wang}}, \bibinfo {author} {\bibfnamefont
  {Y.}~\bibnamefont {Xu}}, \bibinfo {author} {\bibfnamefont {D.}~\bibnamefont
  {Weber}}, \bibinfo {author} {\bibfnamefont {J.~E.}\ \bibnamefont
  {Goldberger}}, \bibinfo {author} {\bibfnamefont {K.}~\bibnamefont
  {Watanabe}}, \bibinfo {author} {\bibfnamefont {T.}~\bibnamefont {Taniguchi}},
  \bibinfo {author} {\bibfnamefont {C.~J.}\ \bibnamefont {Fennie}}, \bibinfo
  {author} {\bibfnamefont {K.}~\bibnamefont {Fai~Mak}},\ and\ \bibinfo {author}
  {\bibfnamefont {J.}~\bibnamefont {Shan}},\ }\bibfield  {title} {\bibinfo
  {title} {Pressure-controlled interlayer magnetism in atomically thin cri3},\
  }\href {https://doi.org/10.1038/s41563-019-0506-1} {\bibfield  {journal}
  {\bibinfo  {journal} {Nature Materials}\ }\textbf {\bibinfo {volume} {18}},\
  \bibinfo {pages} {1303} (\bibinfo {year} {2019})}\BibitemShut {NoStop}%
\bibitem [{\citenamefont {Song}\ \emph {et~al.}(2019)\citenamefont {Song},
  \citenamefont {Fei}, \citenamefont {Yankowitz}, \citenamefont {Lin},
  \citenamefont {Jiang}, \citenamefont {Hwangbo}, \citenamefont {Zhang},
  \citenamefont {Sun}, \citenamefont {Taniguchi}, \citenamefont {Watanabe},
  \citenamefont {McGuire}, \citenamefont {Graf}, \citenamefont {Cao},
  \citenamefont {Chu}, \citenamefont {Cobden}, \citenamefont {Dean},
  \citenamefont {Xiao},\ and\ \citenamefont {Xu}}]{tiancheng2019}%
  \BibitemOpen
  \bibfield  {author} {\bibinfo {author} {\bibfnamefont {T.}~\bibnamefont
  {Song}}, \bibinfo {author} {\bibfnamefont {Z.}~\bibnamefont {Fei}}, \bibinfo
  {author} {\bibfnamefont {M.}~\bibnamefont {Yankowitz}}, \bibinfo {author}
  {\bibfnamefont {Z.}~\bibnamefont {Lin}}, \bibinfo {author} {\bibfnamefont
  {Q.}~\bibnamefont {Jiang}}, \bibinfo {author} {\bibfnamefont
  {K.}~\bibnamefont {Hwangbo}}, \bibinfo {author} {\bibfnamefont
  {Q.}~\bibnamefont {Zhang}}, \bibinfo {author} {\bibfnamefont
  {B.}~\bibnamefont {Sun}}, \bibinfo {author} {\bibfnamefont {T.}~\bibnamefont
  {Taniguchi}}, \bibinfo {author} {\bibfnamefont {K.}~\bibnamefont {Watanabe}},
  \bibinfo {author} {\bibfnamefont {M.~A.}\ \bibnamefont {McGuire}}, \bibinfo
  {author} {\bibfnamefont {D.}~\bibnamefont {Graf}}, \bibinfo {author}
  {\bibfnamefont {T.}~\bibnamefont {Cao}}, \bibinfo {author} {\bibfnamefont
  {J.-H.}\ \bibnamefont {Chu}}, \bibinfo {author} {\bibfnamefont {D.~H.}\
  \bibnamefont {Cobden}}, \bibinfo {author} {\bibfnamefont {C.~R.}\
  \bibnamefont {Dean}}, \bibinfo {author} {\bibfnamefont {D.}~\bibnamefont
  {Xiao}},\ and\ \bibinfo {author} {\bibfnamefont {X.}~\bibnamefont {Xu}},\
  }\bibfield  {title} {\bibinfo {title} {Switching 2d magnetic states via
  pressure tuning of layer stacking},\ }\href
  {https://doi.org/10.1038/s41563-019-0505-2} {\bibfield  {journal} {\bibinfo
  {journal} {Nature Materials}\ }\textbf {\bibinfo {volume} {18}},\ \bibinfo
  {pages} {1298} (\bibinfo {year} {2019})}\BibitemShut {NoStop}%
\bibitem [{\citenamefont {Coak}\ \emph {et~al.}(2021)\citenamefont {Coak},
  \citenamefont {Jarvis}, \citenamefont {Hamidov}, \citenamefont {Wildes},
  \citenamefont {Paddison}, \citenamefont {Liu}, \citenamefont {Haines},
  \citenamefont {Dang}, \citenamefont {Kichanov}, \citenamefont {Savenko},
  \citenamefont {Lee}, \citenamefont {Kratochv\'{\i}lov\'a}, \citenamefont
  {Klotz}, \citenamefont {Hansen}, \citenamefont {Kozlenko}, \citenamefont
  {Park},\ and\ \citenamefont {Saxena}}]{matthew2021}%
  \BibitemOpen
  \bibfield  {author} {\bibinfo {author} {\bibfnamefont {M.~J.}\ \bibnamefont
  {Coak}}, \bibinfo {author} {\bibfnamefont {D.~M.}\ \bibnamefont {Jarvis}},
  \bibinfo {author} {\bibfnamefont {H.}~\bibnamefont {Hamidov}}, \bibinfo
  {author} {\bibfnamefont {A.~R.}\ \bibnamefont {Wildes}}, \bibinfo {author}
  {\bibfnamefont {J.~A.~M.}\ \bibnamefont {Paddison}}, \bibinfo {author}
  {\bibfnamefont {C.}~\bibnamefont {Liu}}, \bibinfo {author} {\bibfnamefont
  {C.~R.~S.}\ \bibnamefont {Haines}}, \bibinfo {author} {\bibfnamefont {N.~T.}\
  \bibnamefont {Dang}}, \bibinfo {author} {\bibfnamefont {S.~E.}\ \bibnamefont
  {Kichanov}}, \bibinfo {author} {\bibfnamefont {B.~N.}\ \bibnamefont
  {Savenko}}, \bibinfo {author} {\bibfnamefont {S.}~\bibnamefont {Lee}},
  \bibinfo {author} {\bibfnamefont {M.}~\bibnamefont {Kratochv\'{\i}lov\'a}},
  \bibinfo {author} {\bibfnamefont {S.}~\bibnamefont {Klotz}}, \bibinfo
  {author} {\bibfnamefont {T.~C.}\ \bibnamefont {Hansen}}, \bibinfo {author}
  {\bibfnamefont {D.~P.}\ \bibnamefont {Kozlenko}}, \bibinfo {author}
  {\bibfnamefont {J.-G.}\ \bibnamefont {Park}},\ and\ \bibinfo {author}
  {\bibfnamefont {S.~S.}\ \bibnamefont {Saxena}},\ }\bibfield  {title}
  {\bibinfo {title} {Emergent magnetic phases in pressure-tuned van der waals
  antiferromagnet ${\mathrm{feps}}_{3}$},\ }\href
  {https://doi.org/10.1103/PhysRevX.11.011024} {\bibfield  {journal} {\bibinfo
  {journal} {Phys. Rev. X}\ }\textbf {\bibinfo {volume} {11}},\ \bibinfo
  {pages} {011024} (\bibinfo {year} {2021})}\BibitemShut {NoStop}%
\bibitem [{\citenamefont {Haines}\ \emph {et~al.}(2018)\citenamefont {Haines},
  \citenamefont {Coak}, \citenamefont {Wildes}, \citenamefont {Lampronti},
  \citenamefont {Liu}, \citenamefont {Nahai-Williamson}, \citenamefont
  {Hamidov}, \citenamefont {Daisenberger},\ and\ \citenamefont
  {Saxena}}]{hanies2018}%
  \BibitemOpen
  \bibfield  {author} {\bibinfo {author} {\bibfnamefont {C.~R.~S.}\
  \bibnamefont {Haines}}, \bibinfo {author} {\bibfnamefont {M.~J.}\
  \bibnamefont {Coak}}, \bibinfo {author} {\bibfnamefont {A.~R.}\ \bibnamefont
  {Wildes}}, \bibinfo {author} {\bibfnamefont {G.~I.}\ \bibnamefont
  {Lampronti}}, \bibinfo {author} {\bibfnamefont {C.}~\bibnamefont {Liu}},
  \bibinfo {author} {\bibfnamefont {P.}~\bibnamefont {Nahai-Williamson}},
  \bibinfo {author} {\bibfnamefont {H.}~\bibnamefont {Hamidov}}, \bibinfo
  {author} {\bibfnamefont {D.}~\bibnamefont {Daisenberger}},\ and\ \bibinfo
  {author} {\bibfnamefont {S.~S.}\ \bibnamefont {Saxena}},\ }\bibfield  {title}
  {\bibinfo {title} {Pressure-induced electronic and structural phase evolution
  in the van der waals compound ${\mathrm{feps}}_{3}$},\ }\href
  {https://doi.org/10.1103/PhysRevLett.121.266801} {\bibfield  {journal}
  {\bibinfo  {journal} {Phys. Rev. Lett.}\ }\textbf {\bibinfo {volume} {121}},\
  \bibinfo {pages} {266801} (\bibinfo {year} {2018})}\BibitemShut {NoStop}%
\bibitem [{\citenamefont {May}\ \emph {et~al.}(2020)\citenamefont {May},
  \citenamefont {Du}, \citenamefont {Cooper},\ and\ \citenamefont
  {McGuire}}]{andrew2020}%
  \BibitemOpen
  \bibfield  {author} {\bibinfo {author} {\bibfnamefont {A.~F.}\ \bibnamefont
  {May}}, \bibinfo {author} {\bibfnamefont {M.-H.}\ \bibnamefont {Du}},
  \bibinfo {author} {\bibfnamefont {V.~R.}\ \bibnamefont {Cooper}},\ and\
  \bibinfo {author} {\bibfnamefont {M.~A.}\ \bibnamefont {McGuire}},\
  }\bibfield  {title} {\bibinfo {title} {Tuning magnetic order in the van der
  waals metal ${\mathrm{fe}}_{5}{\mathrm{gete}}_{2}$ by cobalt substitution},\
  }\href {https://doi.org/10.1103/PhysRevMaterials.4.074008} {\bibfield
  {journal} {\bibinfo  {journal} {Phys. Rev. Materials}\ }\textbf {\bibinfo
  {volume} {4}},\ \bibinfo {pages} {074008} (\bibinfo {year}
  {2020})}\BibitemShut {NoStop}%
\bibitem [{\citenamefont {Drachuck}\ \emph {et~al.}(2018)\citenamefont
  {Drachuck}, \citenamefont {Salman}, \citenamefont {Masters}, \citenamefont
  {Taufour}, \citenamefont {Lamichhane}, \citenamefont {Lin}, \citenamefont
  {Straszheim}, \citenamefont {Bud'ko},\ and\ \citenamefont
  {Canfield}}]{gil2018}%
  \BibitemOpen
  \bibfield  {author} {\bibinfo {author} {\bibfnamefont {G.}~\bibnamefont
  {Drachuck}}, \bibinfo {author} {\bibfnamefont {Z.}~\bibnamefont {Salman}},
  \bibinfo {author} {\bibfnamefont {M.~W.}\ \bibnamefont {Masters}}, \bibinfo
  {author} {\bibfnamefont {V.}~\bibnamefont {Taufour}}, \bibinfo {author}
  {\bibfnamefont {T.~N.}\ \bibnamefont {Lamichhane}}, \bibinfo {author}
  {\bibfnamefont {Q.}~\bibnamefont {Lin}}, \bibinfo {author} {\bibfnamefont
  {W.~E.}\ \bibnamefont {Straszheim}}, \bibinfo {author} {\bibfnamefont
  {S.~L.}\ \bibnamefont {Bud'ko}},\ and\ \bibinfo {author} {\bibfnamefont
  {P.~C.}\ \bibnamefont {Canfield}},\ }\bibfield  {title} {\bibinfo {title}
  {Effect of nickel substitution on magnetism in the layered van der waals
  ferromagnet ${\mathrm{fe}}_{3}{\mathrm{gete}}_{2}$},\ }\href
  {https://doi.org/10.1103/PhysRevB.98.144434} {\bibfield  {journal} {\bibinfo
  {journal} {Phys. Rev. B}\ }\textbf {\bibinfo {volume} {98}},\ \bibinfo
  {pages} {144434} (\bibinfo {year} {2018})}\BibitemShut {NoStop}%
\bibitem [{\citenamefont {Parkin}\ and\ \citenamefont
  {Friend}(1980)}]{parkin1980}%
  \BibitemOpen
  \bibfield  {author} {\bibinfo {author} {\bibfnamefont {S.~S.~P.}\
  \bibnamefont {Parkin}}\ and\ \bibinfo {author} {\bibfnamefont {R.~H.}\
  \bibnamefont {Friend}},\ }\bibfield  {title} {\bibinfo {title} {3d
  transition-metal intercalates of the niobium and tantalum dichalcogenides. i.
  magnetic properties},\ }\href {https://doi.org/10.1080/13642818008245370}
  {\bibfield  {journal} {\bibinfo  {journal} {Philosophical Magazine B}\
  }\textbf {\bibinfo {volume} {41}},\ \bibinfo {pages} {65} (\bibinfo {year}
  {1980})}\BibitemShut {NoStop}%
\bibitem [{\citenamefont {Friend}\ \emph {et~al.}(1977)\citenamefont {Friend},
  \citenamefont {Beal},\ and\ \citenamefont {Yoffe}}]{friend1977}%
  \BibitemOpen
  \bibfield  {author} {\bibinfo {author} {\bibfnamefont {R.~H.}\ \bibnamefont
  {Friend}}, \bibinfo {author} {\bibfnamefont {A.~R.}\ \bibnamefont {Beal}},\
  and\ \bibinfo {author} {\bibfnamefont {A.~D.}\ \bibnamefont {Yoffe}},\
  }\bibfield  {title} {\bibinfo {title} {Electrical and magnetic properties of
  some first row transition metal intercalates of niobium disulphide},\ }\href
  {https://doi.org/10.1080/14786437708232952} {\bibfield  {journal} {\bibinfo
  {journal} {The Philosophical Magazine: A Journal of Theoretical Experimental
  and Applied Physics}\ }\textbf {\bibinfo {volume} {35}},\ \bibinfo {pages}
  {1269} (\bibinfo {year} {1977})}\BibitemShut {NoStop}%
\bibitem [{\citenamefont {Wilson}\ \emph {et~al.}(1975)\citenamefont {Wilson},
  \citenamefont {Salvo},\ and\ \citenamefont {Mahajan}}]{wilson1974}%
  \BibitemOpen
  \bibfield  {author} {\bibinfo {author} {\bibfnamefont {J.}~\bibnamefont
  {Wilson}}, \bibinfo {author} {\bibfnamefont {F.~D.}\ \bibnamefont {Salvo}},\
  and\ \bibinfo {author} {\bibfnamefont {S.}~\bibnamefont {Mahajan}},\
  }\bibfield  {title} {\bibinfo {title} {Charge-density waves and superlattices
  in the metallic layered transition metal dichalcogenides},\ }\href
  {https://doi.org/10.1080/00018737500101391} {\bibfield  {journal} {\bibinfo
  {journal} {Advances in Physics}\ }\textbf {\bibinfo {volume} {24}},\ \bibinfo
  {pages} {117} (\bibinfo {year} {1975})}\BibitemShut {NoStop}%
\bibitem [{\citenamefont {Naito}\ and\ \citenamefont
  {Tanaka}(1982)}]{naito1981}%
  \BibitemOpen
  \bibfield  {author} {\bibinfo {author} {\bibfnamefont {M.}~\bibnamefont
  {Naito}}\ and\ \bibinfo {author} {\bibfnamefont {S.}~\bibnamefont {Tanaka}},\
  }\bibfield  {title} {\bibinfo {title} {Electrical transport properties in
  2h-nbs2, -nbse2, -tas2 and -tase2},\ }\href
  {https://doi.org/10.1143/JPSJ.51.219} {\bibfield  {journal} {\bibinfo
  {journal} {Journal of the Physical Society of Japan}\ }\textbf {\bibinfo
  {volume} {51}},\ \bibinfo {pages} {219} (\bibinfo {year} {1982})}\BibitemShut
  {NoStop}%
\bibitem [{\citenamefont {Castro~Neto}(2001)}]{castro2001}%
  \BibitemOpen
  \bibfield  {author} {\bibinfo {author} {\bibfnamefont {A.~H.}\ \bibnamefont
  {Castro~Neto}},\ }\bibfield  {title} {\bibinfo {title} {Charge density wave,
  superconductivity, and anomalous metallic behavior in 2d transition metal
  dichalcogenides},\ }\href {https://doi.org/10.1103/PhysRevLett.86.4382}
  {\bibfield  {journal} {\bibinfo  {journal} {Phys. Rev. Lett.}\ }\textbf
  {\bibinfo {volume} {86}},\ \bibinfo {pages} {4382} (\bibinfo {year}
  {2001})}\BibitemShut {NoStop}%
\bibitem [{\citenamefont {Guillam\'on}\ \emph {et~al.}(2008)\citenamefont
  {Guillam\'on}, \citenamefont {Suderow}, \citenamefont {Vieira}, \citenamefont
  {Cario}, \citenamefont {Diener},\ and\ \citenamefont
  {Rodi\`ere}}]{guillam2008}%
  \BibitemOpen
  \bibfield  {author} {\bibinfo {author} {\bibfnamefont {I.}~\bibnamefont
  {Guillam\'on}}, \bibinfo {author} {\bibfnamefont {H.}~\bibnamefont
  {Suderow}}, \bibinfo {author} {\bibfnamefont {S.}~\bibnamefont {Vieira}},
  \bibinfo {author} {\bibfnamefont {L.}~\bibnamefont {Cario}}, \bibinfo
  {author} {\bibfnamefont {P.}~\bibnamefont {Diener}},\ and\ \bibinfo {author}
  {\bibfnamefont {P.}~\bibnamefont {Rodi\`ere}},\ }\bibfield  {title} {\bibinfo
  {title} {Superconducting density of states and vortex cores of
  2h-${\mathrm{nbs}}_{2}$},\ }\href
  {https://doi.org/10.1103/PhysRevLett.101.166407} {\bibfield  {journal}
  {\bibinfo  {journal} {Phys. Rev. Lett.}\ }\textbf {\bibinfo {volume} {101}},\
  \bibinfo {pages} {166407} (\bibinfo {year} {2008})}\BibitemShut {NoStop}%
\bibitem [{\citenamefont {Law}\ and\ \citenamefont {Lee}(2017)}]{law2017}%
  \BibitemOpen
  \bibfield  {author} {\bibinfo {author} {\bibfnamefont {K.~T.}\ \bibnamefont
  {Law}}\ and\ \bibinfo {author} {\bibfnamefont {P.~A.}\ \bibnamefont {Lee}},\
  }\bibfield  {title} {\bibinfo {title} {1t-tas2 as a quantum spin liquid},\
  }\href {https://doi.org/10.1073/pnas.1706769114} {\bibfield  {journal}
  {\bibinfo  {journal} {Proceedings of the National Academy of Sciences}\
  }\textbf {\bibinfo {volume} {114}},\ \bibinfo {pages} {6996} (\bibinfo {year}
  {2017})}\BibitemShut {NoStop}%
\bibitem [{\citenamefont {Devarakonda}\ \emph {et~al.}(2020)\citenamefont
  {Devarakonda}, \citenamefont {Inoue}, \citenamefont {Fang}, \citenamefont
  {Ozsoy-Keskinbora}, \citenamefont {Suzuki}, \citenamefont {Kriener},
  \citenamefont {Fu}, \citenamefont {Kaxiras}, \citenamefont {Bell},\ and\
  \citenamefont {Checkelsky}}]{ad2020}%
  \BibitemOpen
  \bibfield  {author} {\bibinfo {author} {\bibfnamefont {A.}~\bibnamefont
  {Devarakonda}}, \bibinfo {author} {\bibfnamefont {H.}~\bibnamefont {Inoue}},
  \bibinfo {author} {\bibfnamefont {S.}~\bibnamefont {Fang}}, \bibinfo {author}
  {\bibfnamefont {C.}~\bibnamefont {Ozsoy-Keskinbora}}, \bibinfo {author}
  {\bibfnamefont {T.}~\bibnamefont {Suzuki}}, \bibinfo {author} {\bibfnamefont
  {M.}~\bibnamefont {Kriener}}, \bibinfo {author} {\bibfnamefont
  {L.}~\bibnamefont {Fu}}, \bibinfo {author} {\bibfnamefont {E.}~\bibnamefont
  {Kaxiras}}, \bibinfo {author} {\bibfnamefont {D.~C.}\ \bibnamefont {Bell}},\
  and\ \bibinfo {author} {\bibfnamefont {J.~G.}\ \bibnamefont {Checkelsky}},\
  }\bibfield  {title} {\bibinfo {title} {Clean 2d superconductivity in a bulk
  van der waals superlattice},\ }\href
  {https://doi.org/10.1126/science.aaz6643} {\bibfield  {journal} {\bibinfo
  {journal} {Science}\ }\textbf {\bibinfo {volume} {370}},\ \bibinfo {pages}
  {231} (\bibinfo {year} {2020})}\BibitemShut {NoStop}%
\bibitem [{\citenamefont {Boswell}\ \emph {et~al.}(1978)\citenamefont
  {Boswell}, \citenamefont {Prodan}, \citenamefont {Vaughan},\ and\
  \citenamefont {Corbett}}]{boswell1978}%
  \BibitemOpen
  \bibfield  {author} {\bibinfo {author} {\bibfnamefont {F.}~\bibnamefont
  {Boswell}}, \bibinfo {author} {\bibfnamefont {A.}~\bibnamefont {Prodan}},
  \bibinfo {author} {\bibfnamefont {W.~R.}\ \bibnamefont {Vaughan}},\ and\
  \bibinfo {author} {\bibfnamefont {J.}~\bibnamefont {Corbett}},\ }\bibfield
  {title} {\bibinfo {title} {{On the ordering of Fe atoms in FexNbS2}},\
  }\href@noop {} {\bibfield  {journal} {\bibinfo  {journal} {physics status
  solidi (a)}\ }\textbf {\bibinfo {volume} {45}} (\bibinfo {year}
  {1978})}\BibitemShut {NoStop}%
\bibitem [{\citenamefont {Hulliger}\ and\ \citenamefont
  {Pobitschka}(1970)}]{HULLIGER1970}%
  \BibitemOpen
  \bibfield  {author} {\bibinfo {author} {\bibfnamefont {F.}~\bibnamefont
  {Hulliger}}\ and\ \bibinfo {author} {\bibfnamefont {E.}~\bibnamefont
  {Pobitschka}},\ }\bibfield  {title} {\bibinfo {title} {On the magnetic
  behavior of new 2hîžnbs2-type derivatives},\ }\href
  {https://doi.org/https://doi.org/10.1016/0022-4596(70)90001-0} {\bibfield
  {journal} {\bibinfo  {journal} {Journal of Solid State Chemistry}\ }\textbf
  {\bibinfo {volume} {1}},\ \bibinfo {pages} {117 } (\bibinfo {year}
  {1970})}\BibitemShut {NoStop}%
\bibitem [{\citenamefont {Anzenhofer}\ \emph {et~al.}(1970)\citenamefont
  {Anzenhofer}, \citenamefont {{Van Den Berg}}, \citenamefont {Cossee},\ and\
  \citenamefont {Helle}}]{ANZENHOFER1970}%
  \BibitemOpen
  \bibfield  {author} {\bibinfo {author} {\bibfnamefont {K.}~\bibnamefont
  {Anzenhofer}}, \bibinfo {author} {\bibfnamefont {J.}~\bibnamefont {{Van Den
  Berg}}}, \bibinfo {author} {\bibfnamefont {P.}~\bibnamefont {Cossee}},\ and\
  \bibinfo {author} {\bibfnamefont {J.}~\bibnamefont {Helle}},\ }\bibfield
  {title} {\bibinfo {title} {The crystal structure and magnetic
  susceptibilities of mnnb3s6, fenb3s6, conb3s6 and ninb3s6},\ }\href
  {https://doi.org/https://doi.org/10.1016/0022-3697(70)90315-X} {\bibfield
  {journal} {\bibinfo  {journal} {Journal of Physics and Chemistry of Solids}\
  }\textbf {\bibinfo {volume} {31}},\ \bibinfo {pages} {1057 } (\bibinfo {year}
  {1970})}\BibitemShut {NoStop}%
\bibitem [{\citenamefont {Togawa}\ \emph {et~al.}(2012)\citenamefont {Togawa},
  \citenamefont {Koyama}, \citenamefont {Takayanagi}, \citenamefont {Mori},
  \citenamefont {Kousaka}, \citenamefont {Akimitsu}, \citenamefont {Nishihara},
  \citenamefont {Inoue}, \citenamefont {Ovchinnikov},\ and\ \citenamefont
  {Kishine}}]{togawa2012}%
  \BibitemOpen
  \bibfield  {author} {\bibinfo {author} {\bibfnamefont {Y.}~\bibnamefont
  {Togawa}}, \bibinfo {author} {\bibfnamefont {T.}~\bibnamefont {Koyama}},
  \bibinfo {author} {\bibfnamefont {K.}~\bibnamefont {Takayanagi}}, \bibinfo
  {author} {\bibfnamefont {S.}~\bibnamefont {Mori}}, \bibinfo {author}
  {\bibfnamefont {Y.}~\bibnamefont {Kousaka}}, \bibinfo {author} {\bibfnamefont
  {J.}~\bibnamefont {Akimitsu}}, \bibinfo {author} {\bibfnamefont
  {S.}~\bibnamefont {Nishihara}}, \bibinfo {author} {\bibfnamefont
  {K.}~\bibnamefont {Inoue}}, \bibinfo {author} {\bibfnamefont {A.~S.}\
  \bibnamefont {Ovchinnikov}},\ and\ \bibinfo {author} {\bibfnamefont
  {J.}~\bibnamefont {Kishine}},\ }\bibfield  {title} {\bibinfo {title} {Chiral
  magnetic soliton lattice on a chiral helimagnet},\ }\href
  {https://doi.org/10.1103/PhysRevLett.108.107202} {\bibfield  {journal}
  {\bibinfo  {journal} {Phys. Rev. Lett.}\ }\textbf {\bibinfo {volume} {108}},\
  \bibinfo {pages} {107202} (\bibinfo {year} {2012})}\BibitemShut {NoStop}%
\bibitem [{\citenamefont {Braam}\ \emph {et~al.}(2015)\citenamefont {Braam},
  \citenamefont {Gomez}, \citenamefont {Tezok}, \citenamefont {de~Mello},
  \citenamefont {Li}, \citenamefont {Mandrus}, \citenamefont {Kee},\ and\
  \citenamefont {Sonier}}]{braam2015}%
  \BibitemOpen
  \bibfield  {author} {\bibinfo {author} {\bibfnamefont {D.}~\bibnamefont
  {Braam}}, \bibinfo {author} {\bibfnamefont {C.}~\bibnamefont {Gomez}},
  \bibinfo {author} {\bibfnamefont {S.}~\bibnamefont {Tezok}}, \bibinfo
  {author} {\bibfnamefont {E.~V.~L.}\ \bibnamefont {de~Mello}}, \bibinfo
  {author} {\bibfnamefont {L.}~\bibnamefont {Li}}, \bibinfo {author}
  {\bibfnamefont {D.}~\bibnamefont {Mandrus}}, \bibinfo {author} {\bibfnamefont
  {H.-Y.}\ \bibnamefont {Kee}},\ and\ \bibinfo {author} {\bibfnamefont {J.~E.}\
  \bibnamefont {Sonier}},\ }\bibfield  {title} {\bibinfo {title} {Magnetic
  properties of the helimagnet ${\mathrm{cr}}_{1/3}{\mathrm{nbs}}_{2}$ observed
  by $\ensuremath{\mu}\mathrm{SR}$},\ }\href
  {https://doi.org/10.1103/PhysRevB.91.144407} {\bibfield  {journal} {\bibinfo
  {journal} {Phys. Rev. B}\ }\textbf {\bibinfo {volume} {91}},\ \bibinfo
  {pages} {144407} (\bibinfo {year} {2015})}\BibitemShut {NoStop}%
\bibitem [{\citenamefont {Kousaka}\ \emph {et~al.}(2016)\citenamefont
  {Kousaka}, \citenamefont {Ogura}, \citenamefont {Zhang}, \citenamefont
  {Miao}, \citenamefont {Lee}, \citenamefont {Torii}, \citenamefont {Kamiyama},
  \citenamefont {Campo}, \citenamefont {Inoue},\ and\ \citenamefont
  {Akimitsu}}]{Kousaka_2016}%
  \BibitemOpen
  \bibfield  {author} {\bibinfo {author} {\bibfnamefont {Y.}~\bibnamefont
  {Kousaka}}, \bibinfo {author} {\bibfnamefont {T.}~\bibnamefont {Ogura}},
  \bibinfo {author} {\bibfnamefont {J.}~\bibnamefont {Zhang}}, \bibinfo
  {author} {\bibfnamefont {P.}~\bibnamefont {Miao}}, \bibinfo {author}
  {\bibfnamefont {S.}~\bibnamefont {Lee}}, \bibinfo {author} {\bibfnamefont
  {S.}~\bibnamefont {Torii}}, \bibinfo {author} {\bibfnamefont
  {T.}~\bibnamefont {Kamiyama}}, \bibinfo {author} {\bibfnamefont
  {J.}~\bibnamefont {Campo}}, \bibinfo {author} {\bibfnamefont
  {K.}~\bibnamefont {Inoue}},\ and\ \bibinfo {author} {\bibfnamefont
  {J.}~\bibnamefont {Akimitsu}},\ }\bibfield  {title} {\bibinfo {title} {Long
  periodic helimagnetic ordering in {CrM}3s6(m = nb and ta)},\ }\href
  {https://doi.org/10.1088/1742-6596/746/1/012061} {\bibfield  {journal}
  {\bibinfo  {journal} {Journal of Physics: Conference Series}\ }\textbf
  {\bibinfo {volume} {746}},\ \bibinfo {pages} {012061} (\bibinfo {year}
  {2016})}\BibitemShut {NoStop}%
\bibitem [{\citenamefont {Kousaka}\ \emph {et~al.}(2009)\citenamefont
  {Kousaka}, \citenamefont {Nakao}, \citenamefont {Kishine}, \citenamefont
  {Akita}, \citenamefont {Inoue},\ and\ \citenamefont
  {Akimitsu}}]{kousaka2009}%
  \BibitemOpen
  \bibfield  {author} {\bibinfo {author} {\bibfnamefont {Y.}~\bibnamefont
  {Kousaka}}, \bibinfo {author} {\bibfnamefont {Y.}~\bibnamefont {Nakao}},
  \bibinfo {author} {\bibfnamefont {J.}~\bibnamefont {Kishine}}, \bibinfo
  {author} {\bibfnamefont {M.}~\bibnamefont {Akita}}, \bibinfo {author}
  {\bibfnamefont {K.}~\bibnamefont {Inoue}},\ and\ \bibinfo {author}
  {\bibfnamefont {J.}~\bibnamefont {Akimitsu}},\ }\bibfield  {title} {\bibinfo
  {title} {Chiral helimagnetism in t1/3nbs2 (t=cr and mn)},\ }\href
  {https://doi.org/https://doi.org/10.1016/j.nima.2008.11.040} {\bibfield
  {journal} {\bibinfo  {journal} {Nuclear Instruments and Methods in Physics
  Research Section A: Accelerators, Spectrometers, Detectors and Associated
  Equipment}\ }\textbf {\bibinfo {volume} {600}},\ \bibinfo {pages} {250}
  (\bibinfo {year} {2009})}\BibitemShut {NoStop}%
\bibitem [{\citenamefont {Karna}\ \emph {et~al.}(2019)\citenamefont {Karna},
  \citenamefont {Womack}, \citenamefont {Chapai}, \citenamefont {Young},
  \citenamefont {Marshall}, \citenamefont {Xie}, \citenamefont {Graf},
  \citenamefont {Wu}, \citenamefont {Cao}, \citenamefont {DeBeer-Schmitt},
  \citenamefont {Adams}, \citenamefont {Jin},\ and\ \citenamefont
  {DiTusa}}]{karna2019}%
  \BibitemOpen
  \bibfield  {author} {\bibinfo {author} {\bibfnamefont {S.~K.}\ \bibnamefont
  {Karna}}, \bibinfo {author} {\bibfnamefont {F.~N.}\ \bibnamefont {Womack}},
  \bibinfo {author} {\bibfnamefont {R.}~\bibnamefont {Chapai}}, \bibinfo
  {author} {\bibfnamefont {D.~P.}\ \bibnamefont {Young}}, \bibinfo {author}
  {\bibfnamefont {M.}~\bibnamefont {Marshall}}, \bibinfo {author}
  {\bibfnamefont {W.}~\bibnamefont {Xie}}, \bibinfo {author} {\bibfnamefont
  {D.}~\bibnamefont {Graf}}, \bibinfo {author} {\bibfnamefont {Y.}~\bibnamefont
  {Wu}}, \bibinfo {author} {\bibfnamefont {H.}~\bibnamefont {Cao}}, \bibinfo
  {author} {\bibfnamefont {L.}~\bibnamefont {DeBeer-Schmitt}}, \bibinfo
  {author} {\bibfnamefont {P.~W.}\ \bibnamefont {Adams}}, \bibinfo {author}
  {\bibfnamefont {R.}~\bibnamefont {Jin}},\ and\ \bibinfo {author}
  {\bibfnamefont {J.~F.}\ \bibnamefont {DiTusa}},\ }\bibfield  {title}
  {\bibinfo {title} {Consequences of magnetic ordering in chiral
  $\mathrm{M}{\mathrm{n}}_{1/3}\mathrm{Nb}{\mathrm{s}}_{2}$},\ }\href
  {https://doi.org/10.1103/PhysRevB.100.184413} {\bibfield  {journal} {\bibinfo
   {journal} {Phys. Rev. B}\ }\textbf {\bibinfo {volume} {100}},\ \bibinfo
  {pages} {184413} (\bibinfo {year} {2019})}\BibitemShut {NoStop}%
\bibitem [{\citenamefont {Parkin}\ \emph {et~al.}(1983)\citenamefont {Parkin},
  \citenamefont {Marseglia},\ and\ \citenamefont {Brown}}]{Parkin_1983}%
  \BibitemOpen
  \bibfield  {author} {\bibinfo {author} {\bibfnamefont {S.~S.~P.}\
  \bibnamefont {Parkin}}, \bibinfo {author} {\bibfnamefont {E.~A.}\
  \bibnamefont {Marseglia}},\ and\ \bibinfo {author} {\bibfnamefont {P.~J.}\
  \bibnamefont {Brown}},\ }\bibfield  {title} {\bibinfo {title} {Magnetic
  structure of co1/3nbs2and co1/3tas2},\ }\href
  {https://doi.org/10.1088/0022-3719/16/14/016} {\bibfield  {journal} {\bibinfo
   {journal} {Journal of Physics C: Solid State Physics}\ }\textbf {\bibinfo
  {volume} {16}},\ \bibinfo {pages} {2765} (\bibinfo {year}
  {1983})}\BibitemShut {NoStop}%
\bibitem [{\citenamefont {Lu}\ \emph {et~al.}(2020)\citenamefont {Lu},
  \citenamefont {Sapkota}, \citenamefont {DeBeer-Schmitt}, \citenamefont {Wu},
  \citenamefont {Cao}, \citenamefont {Mannella}, \citenamefont {Mandrus},
  \citenamefont {Aczel},\ and\ \citenamefont {MacDougall}}]{lu2020}%
  \BibitemOpen
  \bibfield  {author} {\bibinfo {author} {\bibfnamefont {K.}~\bibnamefont
  {Lu}}, \bibinfo {author} {\bibfnamefont {D.}~\bibnamefont {Sapkota}},
  \bibinfo {author} {\bibfnamefont {L.}~\bibnamefont {DeBeer-Schmitt}},
  \bibinfo {author} {\bibfnamefont {Y.}~\bibnamefont {Wu}}, \bibinfo {author}
  {\bibfnamefont {H.~B.}\ \bibnamefont {Cao}}, \bibinfo {author} {\bibfnamefont
  {N.}~\bibnamefont {Mannella}}, \bibinfo {author} {\bibfnamefont
  {D.}~\bibnamefont {Mandrus}}, \bibinfo {author} {\bibfnamefont {A.~A.}\
  \bibnamefont {Aczel}},\ and\ \bibinfo {author} {\bibfnamefont {G.~J.}\
  \bibnamefont {MacDougall}},\ }\bibfield  {title} {\bibinfo {title} {Canted
  antiferromagnetic order in the monoaxial chiral magnets
  ${\mathrm{v}}_{1/3}{\mathrm{tas}}_{2}$ and
  ${\mathrm{v}}_{1/3}{\mathrm{nbs}}_{2}$},\ }\href
  {https://doi.org/10.1103/PhysRevMaterials.4.054416} {\bibfield  {journal}
  {\bibinfo  {journal} {Phys. Rev. Materials}\ }\textbf {\bibinfo {volume}
  {4}},\ \bibinfo {pages} {054416} (\bibinfo {year} {2020})}\BibitemShut
  {NoStop}%
\bibitem [{\citenamefont {Aczel}\ \emph {et~al.}(2018)\citenamefont {Aczel},
  \citenamefont {Debeer-Schmitt}, \citenamefont {Williams}, \citenamefont
  {McGuire}, \citenamefont {Ghimire}, \citenamefont {Li},\ and\ \citenamefont
  {Mandrus}}]{adam2018}%
  \BibitemOpen
  \bibfield  {author} {\bibinfo {author} {\bibfnamefont {A.~A.}\ \bibnamefont
  {Aczel}}, \bibinfo {author} {\bibfnamefont {L.~M.}\ \bibnamefont
  {Debeer-Schmitt}}, \bibinfo {author} {\bibfnamefont {T.~J.}\ \bibnamefont
  {Williams}}, \bibinfo {author} {\bibfnamefont {M.}~\bibnamefont {McGuire}},
  \bibinfo {author} {\bibfnamefont {N.}~\bibnamefont {Ghimire}}, \bibinfo
  {author} {\bibfnamefont {L.}~\bibnamefont {Li}},\ and\ \bibinfo {author}
  {\bibfnamefont {D.}~\bibnamefont {Mandrus}},\ }\bibfield  {title} {\bibinfo
  {title} {Extended exchange interactions stabilize long-period magnetic
  structures in cr1/3nbs2},\ }\href@noop {} {\bibfield  {journal} {\bibinfo
  {journal} {Applied Physics Letters}\ }\textbf {\bibinfo {volume} {113}},\
  \bibinfo {pages} {032404} (\bibinfo {year} {2018})}\BibitemShut {NoStop}%
\bibitem [{\citenamefont {Hall}\ \emph {et~al.}(2021)\citenamefont {Hall},
  \citenamefont {Khalyavin}, \citenamefont {Manuel}, \citenamefont {Mayoh},
  \citenamefont {Orlandi}, \citenamefont {Petrenko}, \citenamefont {Lees},\
  and\ \citenamefont {Balakrishnan}}]{hall2021}%
  \BibitemOpen
  \bibfield  {author} {\bibinfo {author} {\bibfnamefont {A.~E.}\ \bibnamefont
  {Hall}}, \bibinfo {author} {\bibfnamefont {D.~D.}\ \bibnamefont {Khalyavin}},
  \bibinfo {author} {\bibfnamefont {P.}~\bibnamefont {Manuel}}, \bibinfo
  {author} {\bibfnamefont {D.~A.}\ \bibnamefont {Mayoh}}, \bibinfo {author}
  {\bibfnamefont {F.}~\bibnamefont {Orlandi}}, \bibinfo {author} {\bibfnamefont
  {O.~A.}\ \bibnamefont {Petrenko}}, \bibinfo {author} {\bibfnamefont {M.~R.}\
  \bibnamefont {Lees}},\ and\ \bibinfo {author} {\bibfnamefont
  {G.}~\bibnamefont {Balakrishnan}},\ }\bibfield  {title} {\bibinfo {title}
  {Magnetic structure investigation of the intercalated transition metal
  dichalcogenide ${\mathrm{v}}_{1/3}{\mathrm{nbs}}_{2}$},\ }\href
  {https://doi.org/10.1103/PhysRevB.103.174431} {\bibfield  {journal} {\bibinfo
   {journal} {Phys. Rev. B}\ }\textbf {\bibinfo {volume} {103}},\ \bibinfo
  {pages} {174431} (\bibinfo {year} {2021})}\BibitemShut {NoStop}%
\bibitem [{\citenamefont {Togawa}\ \emph {et~al.}(2015)\citenamefont {Togawa},
  \citenamefont {Koyama}, \citenamefont {Nishimori}, \citenamefont {Matsumoto},
  \citenamefont {McVitie}, \citenamefont {McGrouther}, \citenamefont {Stamps},
  \citenamefont {Kousaka}, \citenamefont {Akimitsu}, \citenamefont {Nishihara},
  \citenamefont {Inoue}, \citenamefont {Bostrem}, \citenamefont {Sinitsyn},
  \citenamefont {Ovchinnikov},\ and\ \citenamefont {Kishine}}]{togawa2015}%
  \BibitemOpen
  \bibfield  {author} {\bibinfo {author} {\bibfnamefont {Y.}~\bibnamefont
  {Togawa}}, \bibinfo {author} {\bibfnamefont {T.}~\bibnamefont {Koyama}},
  \bibinfo {author} {\bibfnamefont {Y.}~\bibnamefont {Nishimori}}, \bibinfo
  {author} {\bibfnamefont {Y.}~\bibnamefont {Matsumoto}}, \bibinfo {author}
  {\bibfnamefont {S.}~\bibnamefont {McVitie}}, \bibinfo {author} {\bibfnamefont
  {D.}~\bibnamefont {McGrouther}}, \bibinfo {author} {\bibfnamefont {R.~L.}\
  \bibnamefont {Stamps}}, \bibinfo {author} {\bibfnamefont {Y.}~\bibnamefont
  {Kousaka}}, \bibinfo {author} {\bibfnamefont {J.}~\bibnamefont {Akimitsu}},
  \bibinfo {author} {\bibfnamefont {S.}~\bibnamefont {Nishihara}}, \bibinfo
  {author} {\bibfnamefont {K.}~\bibnamefont {Inoue}}, \bibinfo {author}
  {\bibfnamefont {I.~G.}\ \bibnamefont {Bostrem}}, \bibinfo {author}
  {\bibfnamefont {V.~E.}\ \bibnamefont {Sinitsyn}}, \bibinfo {author}
  {\bibfnamefont {A.~S.}\ \bibnamefont {Ovchinnikov}},\ and\ \bibinfo {author}
  {\bibfnamefont {J.}~\bibnamefont {Kishine}},\ }\bibfield  {title} {\bibinfo
  {title} {Magnetic soliton confinement and discretization effects arising from
  macroscopic coherence in a chiral spin soliton lattice},\ }\href
  {https://doi.org/10.1103/PhysRevB.92.220412} {\bibfield  {journal} {\bibinfo
  {journal} {Phys. Rev. B}\ }\textbf {\bibinfo {volume} {92}},\ \bibinfo
  {pages} {220412} (\bibinfo {year} {2015})}\BibitemShut {NoStop}%
\bibitem [{\citenamefont {Ghimire}\ \emph {et~al.}(2018)\citenamefont
  {Ghimire}, \citenamefont {Botana}, \citenamefont {Jiang}, \citenamefont
  {Zhang}, \citenamefont {Chen},\ and\ \citenamefont {Mitchell}}]{ghimire2018}%
  \BibitemOpen
  \bibfield  {author} {\bibinfo {author} {\bibfnamefont {N.~J.}\ \bibnamefont
  {Ghimire}}, \bibinfo {author} {\bibfnamefont {A.~S.}\ \bibnamefont {Botana}},
  \bibinfo {author} {\bibfnamefont {J.~S.}\ \bibnamefont {Jiang}}, \bibinfo
  {author} {\bibfnamefont {J.}~\bibnamefont {Zhang}}, \bibinfo {author}
  {\bibfnamefont {Y.~S.}\ \bibnamefont {Chen}},\ and\ \bibinfo {author}
  {\bibfnamefont {J.~F.}\ \bibnamefont {Mitchell}},\ }\bibfield  {title}
  {\bibinfo {title} {Large anomalous hall effect in the chiral-lattice
  antiferromagnet conb3s6},\ }\href
  {https://doi.org/10.1038/s41467-018-05756-7} {\bibfield  {journal} {\bibinfo
  {journal} {Nature Communications}\ }\textbf {\bibinfo {volume} {9}},\
  \bibinfo {pages} {3280} (\bibinfo {year} {2018})}\BibitemShut {NoStop}%
\bibitem [{\citenamefont {Aoki}\ \emph {et~al.}(2019)\citenamefont {Aoki},
  \citenamefont {Kousaka},\ and\ \citenamefont {Togawa}}]{aoki2019}%
  \BibitemOpen
  \bibfield  {author} {\bibinfo {author} {\bibfnamefont {R.}~\bibnamefont
  {Aoki}}, \bibinfo {author} {\bibfnamefont {Y.}~\bibnamefont {Kousaka}},\ and\
  \bibinfo {author} {\bibfnamefont {Y.}~\bibnamefont {Togawa}},\ }\bibfield
  {title} {\bibinfo {title} {Anomalous nonreciprocal electrical transport on
  chiral magnetic order},\ }\href
  {https://doi.org/10.1103/PhysRevLett.122.057206} {\bibfield  {journal}
  {\bibinfo  {journal} {Phys. Rev. Lett.}\ }\textbf {\bibinfo {volume} {122}},\
  \bibinfo {pages} {057206} (\bibinfo {year} {2019})}\BibitemShut {NoStop}%
\bibitem [{\citenamefont {Gorochov}\ \emph {et~al.}(1981)\citenamefont
  {Gorochov}, \citenamefont {Blanc-soreau}, \citenamefont {Rouxel},
  \citenamefont {Imbert},\ and\ \citenamefont {Jehanno}}]{gorochov1979}%
  \BibitemOpen
  \bibfield  {author} {\bibinfo {author} {\bibfnamefont {O.}~\bibnamefont
  {Gorochov}}, \bibinfo {author} {\bibfnamefont {A.~L.}\ \bibnamefont
  {Blanc-soreau}}, \bibinfo {author} {\bibfnamefont {J.}~\bibnamefont
  {Rouxel}}, \bibinfo {author} {\bibfnamefont {P.}~\bibnamefont {Imbert}},\
  and\ \bibinfo {author} {\bibfnamefont {G.}~\bibnamefont {Jehanno}},\
  }\bibfield  {title} {\bibinfo {title} {Transport properties, magnetic
  susceptibility and mÃ¶ssbauer spectroscopy of fe0.25nbs2 and fe0.33nbs2},\
  }\href {https://doi.org/10.1080/01418638108222164} {\bibfield  {journal}
  {\bibinfo  {journal} {Philosophical Magazine B}\ }\textbf {\bibinfo {volume}
  {43}},\ \bibinfo {pages} {621} (\bibinfo {year} {1981})}\BibitemShut
  {NoStop}%
\bibitem [{\citenamefont {Yamamura}\ \emph {et~al.}(2004)\citenamefont
  {Yamamura}, \citenamefont {Moriyama}, \citenamefont {Tsuji}, \citenamefont
  {Iwasa}, \citenamefont {Koyano}, \citenamefont {Katayama},\ and\
  \citenamefont {Ito}}]{yamamura2004}%
  \BibitemOpen
  \bibfield  {author} {\bibinfo {author} {\bibfnamefont {Y.}~\bibnamefont
  {Yamamura}}, \bibinfo {author} {\bibfnamefont {S.}~\bibnamefont {Moriyama}},
  \bibinfo {author} {\bibfnamefont {T.}~\bibnamefont {Tsuji}}, \bibinfo
  {author} {\bibfnamefont {Y.}~\bibnamefont {Iwasa}}, \bibinfo {author}
  {\bibfnamefont {M.}~\bibnamefont {Koyano}}, \bibinfo {author} {\bibfnamefont
  {S.}~\bibnamefont {Katayama}},\ and\ \bibinfo {author} {\bibfnamefont
  {M.}~\bibnamefont {Ito}},\ }\bibfield  {title} {\bibinfo {title} {Heat
  capacity and phase transition of fexnbs2 at low temperature},\ }\href
  {https://doi.org/https://doi.org/10.1016/j.jallcom.2004.04.045} {\bibfield
  {journal} {\bibinfo  {journal} {Journal of Alloys and Compounds}\ }\textbf
  {\bibinfo {volume} {383}},\ \bibinfo {pages} {338} (\bibinfo {year}
  {2004})},\ \bibinfo {note} {proceedings of the 14th International Conference
  on Solid Compounds of Transition Elements (SCTE 2003)}\BibitemShut {NoStop}%
\bibitem [{\citenamefont {Nair}\ \emph {et~al.}(2020)\citenamefont {Nair},
  \citenamefont {Maniv}, \citenamefont {John}, \citenamefont {Doyle},
  \citenamefont {Orenstein},\ and\ \citenamefont {Analytis}}]{nair2020}%
  \BibitemOpen
  \bibfield  {author} {\bibinfo {author} {\bibfnamefont {N.~L.}\ \bibnamefont
  {Nair}}, \bibinfo {author} {\bibfnamefont {E.}~\bibnamefont {Maniv}},
  \bibinfo {author} {\bibfnamefont {C.}~\bibnamefont {John}}, \bibinfo {author}
  {\bibfnamefont {S.}~\bibnamefont {Doyle}}, \bibinfo {author} {\bibfnamefont
  {J.}~\bibnamefont {Orenstein}},\ and\ \bibinfo {author} {\bibfnamefont
  {J.~G.}\ \bibnamefont {Analytis}},\ }\bibfield  {title} {\bibinfo {title}
  {Electrical switching in a magnetically intercalated transition metal
  dichalcogenide},\ }\href {https://doi.org/10.1038/s41563-019-0518-x}
  {\bibfield  {journal} {\bibinfo  {journal} {Nature Materials}\ }\textbf
  {\bibinfo {volume} {19}},\ \bibinfo {pages} {153} (\bibinfo {year}
  {2020})}\BibitemShut {NoStop}%
\bibitem [{\citenamefont {Wadley}\ \emph {et~al.}(2016)\citenamefont {Wadley},
  \citenamefont {Howells}, \citenamefont {{\v Z}elezn{\'y}}, \citenamefont
  {Andrews}, \citenamefont {Hills}, \citenamefont {Campion}, \citenamefont
  {Nov{\'a}k}, \citenamefont {Olejn{\'\i}k}, \citenamefont {Maccherozzi},
  \citenamefont {Dhesi}, \citenamefont {Martin}, \citenamefont {Wagner},
  \citenamefont {Wunderlich}, \citenamefont {Freimuth}, \citenamefont
  {Mokrousov}, \citenamefont {Kune{\v s}}, \citenamefont {Chauhan},
  \citenamefont {Grzybowski}, \citenamefont {Rushforth}, \citenamefont
  {Edmonds}, \citenamefont {Gallagher},\ and\ \citenamefont
  {Jungwirth}}]{wadley2016}%
  \BibitemOpen
  \bibfield  {author} {\bibinfo {author} {\bibfnamefont {P.}~\bibnamefont
  {Wadley}}, \bibinfo {author} {\bibfnamefont {B.}~\bibnamefont {Howells}},
  \bibinfo {author} {\bibfnamefont {J.}~\bibnamefont {{\v Z}elezn{\'y}}},
  \bibinfo {author} {\bibfnamefont {C.}~\bibnamefont {Andrews}}, \bibinfo
  {author} {\bibfnamefont {V.}~\bibnamefont {Hills}}, \bibinfo {author}
  {\bibfnamefont {R.~P.}\ \bibnamefont {Campion}}, \bibinfo {author}
  {\bibfnamefont {V.}~\bibnamefont {Nov{\'a}k}}, \bibinfo {author}
  {\bibfnamefont {K.}~\bibnamefont {Olejn{\'\i}k}}, \bibinfo {author}
  {\bibfnamefont {F.}~\bibnamefont {Maccherozzi}}, \bibinfo {author}
  {\bibfnamefont {S.~S.}\ \bibnamefont {Dhesi}}, \bibinfo {author}
  {\bibfnamefont {S.~Y.}\ \bibnamefont {Martin}}, \bibinfo {author}
  {\bibfnamefont {T.}~\bibnamefont {Wagner}}, \bibinfo {author} {\bibfnamefont
  {J.}~\bibnamefont {Wunderlich}}, \bibinfo {author} {\bibfnamefont
  {F.}~\bibnamefont {Freimuth}}, \bibinfo {author} {\bibfnamefont
  {Y.}~\bibnamefont {Mokrousov}}, \bibinfo {author} {\bibfnamefont
  {J.}~\bibnamefont {Kune{\v s}}}, \bibinfo {author} {\bibfnamefont {J.~S.}\
  \bibnamefont {Chauhan}}, \bibinfo {author} {\bibfnamefont {M.~J.}\
  \bibnamefont {Grzybowski}}, \bibinfo {author} {\bibfnamefont {A.~W.}\
  \bibnamefont {Rushforth}}, \bibinfo {author} {\bibfnamefont {K.~W.}\
  \bibnamefont {Edmonds}}, \bibinfo {author} {\bibfnamefont {B.~L.}\
  \bibnamefont {Gallagher}},\ and\ \bibinfo {author} {\bibfnamefont
  {T.}~\bibnamefont {Jungwirth}},\ }\bibfield  {title} {\bibinfo {title}
  {Electrical switching of an antiferromagnet},\ }\href
  {https://doi.org/10.1126/science.aab1031} {\bibfield  {journal} {\bibinfo
  {journal} {Science}\ }\textbf {\bibinfo {volume} {351}},\ \bibinfo {pages}
  {587} (\bibinfo {year} {2016})}\BibitemShut {NoStop}%
\bibitem [{\citenamefont {Bodnar}\ \emph {et~al.}(2018)\citenamefont {Bodnar},
  \citenamefont {{\v S}mejkal}, \citenamefont {Turek}, \citenamefont
  {Jungwirth}, \citenamefont {Gomonay}, \citenamefont {Sinova}, \citenamefont
  {Sapozhnik}, \citenamefont {Elmers}, \citenamefont {Kl{\"a}ui},\ and\
  \citenamefont {Jourdan}}]{bodnar2018}%
  \BibitemOpen
  \bibfield  {author} {\bibinfo {author} {\bibfnamefont {S.~Y.}\ \bibnamefont
  {Bodnar}}, \bibinfo {author} {\bibfnamefont {L.}~\bibnamefont {{\v
  S}mejkal}}, \bibinfo {author} {\bibfnamefont {I.}~\bibnamefont {Turek}},
  \bibinfo {author} {\bibfnamefont {T.}~\bibnamefont {Jungwirth}}, \bibinfo
  {author} {\bibfnamefont {O.}~\bibnamefont {Gomonay}}, \bibinfo {author}
  {\bibfnamefont {J.}~\bibnamefont {Sinova}}, \bibinfo {author} {\bibfnamefont
  {A.~A.}\ \bibnamefont {Sapozhnik}}, \bibinfo {author} {\bibfnamefont {H.~J.}\
  \bibnamefont {Elmers}}, \bibinfo {author} {\bibfnamefont {M.}~\bibnamefont
  {Kl{\"a}ui}},\ and\ \bibinfo {author} {\bibfnamefont {M.}~\bibnamefont
  {Jourdan}},\ }\bibfield  {title} {\bibinfo {title} {Writing and reading
  antiferromagnetic mn2au by n{\'e}el spin-orbit torques and large anisotropic
  magnetoresistance},\ }\href {https://doi.org/10.1038/s41467-017-02780-x}
  {\bibfield  {journal} {\bibinfo  {journal} {Nature Communications}\ }\textbf
  {\bibinfo {volume} {9}},\ \bibinfo {pages} {348} (\bibinfo {year}
  {2018})}\BibitemShut {NoStop}%
\bibitem [{\citenamefont {Manchon}\ and\ \citenamefont
  {Zhang}(2008)}]{prb2008_manchon}%
  \BibitemOpen
  \bibfield  {author} {\bibinfo {author} {\bibfnamefont {A.}~\bibnamefont
  {Manchon}}\ and\ \bibinfo {author} {\bibfnamefont {S.}~\bibnamefont
  {Zhang}},\ }\bibfield  {title} {\bibinfo {title} {Theory of nonequilibrium
  intrinsic spin torque in a single nanomagnet},\ }\href
  {https://doi.org/10.1103/PhysRevB.78.212405} {\bibfield  {journal} {\bibinfo
  {journal} {Phys. Rev. B}\ }\textbf {\bibinfo {volume} {78}},\ \bibinfo
  {pages} {212405} (\bibinfo {year} {2008})}\BibitemShut {NoStop}%
\bibitem [{\citenamefont {Laar}\ \emph {et~al.}(1971)\citenamefont {Laar},
  \citenamefont {Rietveld},\ and\ \citenamefont {Ijdo}}]{laar1970}%
  \BibitemOpen
  \bibfield  {author} {\bibinfo {author} {\bibfnamefont {B.~V.}\ \bibnamefont
  {Laar}}, \bibinfo {author} {\bibfnamefont {H.}~\bibnamefont {Rietveld}},\
  and\ \bibinfo {author} {\bibfnamefont {D.}~\bibnamefont {Ijdo}},\ }\bibfield
  {title} {\bibinfo {title} {Magnetic and crystallographic structures of
  mexnbs2 and mextas2},\ }\href
  {https://doi.org/https://doi.org/10.1016/0022-4596(71)90019-3} {\bibfield
  {journal} {\bibinfo  {journal} {Journal of Solid State Chemistry}\ }\textbf
  {\bibinfo {volume} {3}},\ \bibinfo {pages} {154 } (\bibinfo {year}
  {1971})}\BibitemShut {NoStop}%
\bibitem [{etd()}]{etde_22152337}%
  \BibitemOpen
  \href@noop {} {\bibinfo {title} {Macs-a new high intensity cold neutron
  spectrometer at nist}}\BibitemShut {NoStop}%
\bibitem [{\citenamefont {Weber}\ and\ \citenamefont
  {Neaton}(2021)}]{weber2021}%
  \BibitemOpen
  \bibfield  {author} {\bibinfo {author} {\bibfnamefont {S.~F.}\ \bibnamefont
  {Weber}}\ and\ \bibinfo {author} {\bibfnamefont {J.~B.}\ \bibnamefont
  {Neaton}},\ }\href@noop {} {\bibinfo {title} {Origins of anisotropic
  transport in electrically-switchable antiferromagnet
  $\mathrm{Fe_{1/3}NbS_2}$}} (\bibinfo {year} {2021}),\ \Eprint
  {https://arxiv.org/abs/2104.07591} {arXiv:2104.07591 [cond-mat.mtrl-sci]}
  \BibitemShut {NoStop}%
\bibitem [{\citenamefont {Doi}\ and\ \citenamefont {Tazuke}(1991)}]{doi1991}%
  \BibitemOpen
  \bibfield  {author} {\bibinfo {author} {\bibfnamefont {N.}~\bibnamefont
  {Doi}}\ and\ \bibinfo {author} {\bibfnamefont {Y.}~\bibnamefont {Tazuke}},\
  }\bibfield  {title} {\bibinfo {title} {Spin glass phases in 2h-fexnbs2},\
  }\href {https://doi.org/10.1143/JPSJ.60.3980} {\bibfield  {journal} {\bibinfo
   {journal} {Journal of the Physical Society of Japan}\ }\textbf {\bibinfo
  {volume} {60}},\ \bibinfo {pages} {3980} (\bibinfo {year}
  {1991})}\BibitemShut {NoStop}%
\bibitem [{\citenamefont {Johnston}(2017)}]{dj2017}%
  \BibitemOpen
  \bibfield  {author} {\bibinfo {author} {\bibfnamefont {D.~C.}\ \bibnamefont
  {Johnston}},\ }\bibfield  {title} {\bibinfo {title} {Influence of uniaxial
  single-ion anisotropy on the magnetic and thermal properties of heisenberg
  antiferromagnets within unified molecular field theory},\ }\href
  {https://doi.org/10.1103/PhysRevB.95.094421} {\bibfield  {journal} {\bibinfo
  {journal} {Phys. Rev. B}\ }\textbf {\bibinfo {volume} {95}},\ \bibinfo
  {pages} {094421} (\bibinfo {year} {2017})}\BibitemShut {NoStop}%
\bibitem [{\citenamefont {Birgeneau}\ \emph {et~al.}(1973)\citenamefont
  {Birgeneau}, \citenamefont {Guggenheim},\ and\ \citenamefont
  {Shirane}}]{birgeneau1973}%
  \BibitemOpen
  \bibfield  {author} {\bibinfo {author} {\bibfnamefont {R.~J.}\ \bibnamefont
  {Birgeneau}}, \bibinfo {author} {\bibfnamefont {H.~J.}\ \bibnamefont
  {Guggenheim}},\ and\ \bibinfo {author} {\bibfnamefont {G.}~\bibnamefont
  {Shirane}},\ }\bibfield  {title} {\bibinfo {title} {Spin waves and magnetic
  ordering in ${\mathrm{k}}_{2}$mn${\mathrm{f}}_{4}$},\ }\href
  {https://doi.org/10.1103/PhysRevB.8.304} {\bibfield  {journal} {\bibinfo
  {journal} {Phys. Rev. B}\ }\textbf {\bibinfo {volume} {8}},\ \bibinfo {pages}
  {304} (\bibinfo {year} {1973})}\BibitemShut {NoStop}%
\bibitem [{\citenamefont {Birgeneau}\ \emph {et~al.}(1977)\citenamefont
  {Birgeneau}, \citenamefont {Als-Nielsen},\ and\ \citenamefont
  {Shirane}}]{birgeneau1977}%
  \BibitemOpen
  \bibfield  {author} {\bibinfo {author} {\bibfnamefont {R.~J.}\ \bibnamefont
  {Birgeneau}}, \bibinfo {author} {\bibfnamefont {J.}~\bibnamefont
  {Als-Nielsen}},\ and\ \bibinfo {author} {\bibfnamefont {G.}~\bibnamefont
  {Shirane}},\ }\bibfield  {title} {\bibinfo {title} {Critical behavior of pure
  and site-random two-dimensional antiferromagnets},\ }\href
  {https://doi.org/10.1103/PhysRevB.16.280} {\bibfield  {journal} {\bibinfo
  {journal} {Phys. Rev. B}\ }\textbf {\bibinfo {volume} {16}},\ \bibinfo
  {pages} {280} (\bibinfo {year} {1977})}\BibitemShut {NoStop}%
\bibitem [{\citenamefont {Pelissetto}\ and\ \citenamefont
  {Vicari}(2002)}]{pelissetto2002}%
  \BibitemOpen
  \bibfield  {author} {\bibinfo {author} {\bibfnamefont {A.}~\bibnamefont
  {Pelissetto}}\ and\ \bibinfo {author} {\bibfnamefont {E.}~\bibnamefont
  {Vicari}},\ }\bibfield  {title} {\bibinfo {title} {Critical phenomena and
  renormalization-group theory},\ }\href
  {https://doi.org/https://doi.org/10.1016/S0370-1573(02)00219-3} {\bibfield
  {journal} {\bibinfo  {journal} {Physics Reports}\ }\textbf {\bibinfo {volume}
  {368}},\ \bibinfo {pages} {549} (\bibinfo {year} {2002})}\BibitemShut
  {NoStop}%
\bibitem [{\citenamefont {Onsager}(1944)}]{onsager}%
  \BibitemOpen
  \bibfield  {author} {\bibinfo {author} {\bibfnamefont {L.}~\bibnamefont
  {Onsager}},\ }\bibfield  {title} {\bibinfo {title} {Crystal statistics. i. a
  two-dimensional model with an order-disorder transition},\ }\href
  {https://doi.org/10.1103/PhysRev.65.117} {\bibfield  {journal} {\bibinfo
  {journal} {Phys. Rev.}\ }\textbf {\bibinfo {volume} {65}},\ \bibinfo {pages}
  {117} (\bibinfo {year} {1944})}\BibitemShut {NoStop}%
\bibitem [{\citenamefont {Birgeneau}\ \emph {et~al.}(1983)\citenamefont
  {Birgeneau}, \citenamefont {Yoshizawa}, \citenamefont {Cowley}, \citenamefont
  {Shirane},\ and\ \citenamefont {Ikeda}}]{birgeneau1983}%
  \BibitemOpen
  \bibfield  {author} {\bibinfo {author} {\bibfnamefont {R.~J.}\ \bibnamefont
  {Birgeneau}}, \bibinfo {author} {\bibfnamefont {H.}~\bibnamefont
  {Yoshizawa}}, \bibinfo {author} {\bibfnamefont {R.~A.}\ \bibnamefont
  {Cowley}}, \bibinfo {author} {\bibfnamefont {G.}~\bibnamefont {Shirane}},\
  and\ \bibinfo {author} {\bibfnamefont {H.}~\bibnamefont {Ikeda}},\ }\bibfield
   {title} {\bibinfo {title} {Random-field effects in the diluted
  two-dimensional ising antiferromagnet rb2co0.7mg0.3f4},\ }\href
  {https://doi.org/10.1103/PhysRevB.28.1438} {\bibfield  {journal} {\bibinfo
  {journal} {Phys. Rev. B}\ }\textbf {\bibinfo {volume} {28}},\ \bibinfo
  {pages} {1438} (\bibinfo {year} {1983})}\BibitemShut {NoStop}%
\bibitem [{\citenamefont {Wills}(2000)}]{sarah}%
  \BibitemOpen
  \bibfield  {author} {\bibinfo {author} {\bibfnamefont {A.~S.}\ \bibnamefont
  {Wills}},\ }\bibfield  {title} {\bibinfo {title} {A new protocol for the
  determination of magnetic structures using simulated annealing and
  representational analysis (sarah)},\ }\href@noop {} {\bibfield  {journal}
  {\bibinfo  {journal} {Physica B}\ }\textbf {\bibinfo {volume} {276-278}},\
  \bibinfo {pages} {680} (\bibinfo {year} {2000})}\BibitemShut {NoStop}%
\bibitem [{\citenamefont {Rodriguez-Carvajal}(1990)}]{fullprof}%
  \BibitemOpen
  \bibfield  {author} {\bibinfo {author} {\bibfnamefont {J.}~\bibnamefont
  {Rodriguez-Carvajal}},\ }\bibfield  {title} {\bibinfo {title} {Fullprof: a
  program for rietveld refinement and pattern matching analysis},\ }in\
  \href@noop {} {\emph {\bibinfo {booktitle} {Satellite meeting on powder
  diffraction of the XV congress of the IUCr}}},\ Vol.\ \bibinfo {volume}
  {127}\ (\bibinfo {organization} {Toulouse, France:[sn]},\ \bibinfo {year}
  {1990})\BibitemShut {NoStop}%
\bibitem [{\citenamefont {Squires}(2012)}]{squires_2012}%
  \BibitemOpen
  \bibfield  {author} {\bibinfo {author} {\bibfnamefont {G.~L.}\ \bibnamefont
  {Squires}},\ }\href {https://doi.org/10.1017/CBO9781139107808} {\emph
  {\bibinfo {title} {Introduction to the Theory of Thermal Neutron
  Scattering}}},\ \bibinfo {edition} {3rd}\ ed.\ (\bibinfo  {publisher}
  {Cambridge University Press},\ \bibinfo {year} {2012})\BibitemShut {NoStop}%
\bibitem [{\citenamefont {Mankovsky}\ \emph {et~al.}(2016)\citenamefont
  {Mankovsky}, \citenamefont {Polesya}, \citenamefont {Ebert},\ and\
  \citenamefont {Bensch}}]{mankovsky2016}%
  \BibitemOpen
  \bibfield  {author} {\bibinfo {author} {\bibfnamefont {S.}~\bibnamefont
  {Mankovsky}}, \bibinfo {author} {\bibfnamefont {S.}~\bibnamefont {Polesya}},
  \bibinfo {author} {\bibfnamefont {H.}~\bibnamefont {Ebert}},\ and\ \bibinfo
  {author} {\bibfnamefont {W.}~\bibnamefont {Bensch}},\ }\bibfield  {title}
  {\bibinfo {title} {Electronic and magnetic properties of
  $2h\ensuremath{-}{\mathrm{nbs}}_{2}$ intercalated by $3d$ transition
  metals},\ }\href {https://doi.org/10.1103/PhysRevB.94.184430} {\bibfield
  {journal} {\bibinfo  {journal} {Phys. Rev. B}\ }\textbf {\bibinfo {volume}
  {94}},\ \bibinfo {pages} {184430} (\bibinfo {year} {2016})}\BibitemShut
  {NoStop}%
\bibitem [{\citenamefont {Ruderman}\ and\ \citenamefont
  {Kittel}(1954)}]{ruderman1954}%
  \BibitemOpen
  \bibfield  {author} {\bibinfo {author} {\bibfnamefont {M.~A.}\ \bibnamefont
  {Ruderman}}\ and\ \bibinfo {author} {\bibfnamefont {C.}~\bibnamefont
  {Kittel}},\ }\bibfield  {title} {\bibinfo {title} {Indirect exchange coupling
  of nuclear magnetic moments by conduction electrons},\ }\href
  {https://doi.org/10.1103/PhysRev.96.99} {\bibfield  {journal} {\bibinfo
  {journal} {Phys. Rev.}\ }\textbf {\bibinfo {volume} {96}},\ \bibinfo {pages}
  {99} (\bibinfo {year} {1954})}\BibitemShut {NoStop}%
\bibitem [{\citenamefont {Yosida}(1957)}]{Yosida1957}%
  \BibitemOpen
  \bibfield  {author} {\bibinfo {author} {\bibfnamefont {K.}~\bibnamefont
  {Yosida}},\ }\bibfield  {title} {\bibinfo {title} {Magnetic properties of
  cu-mn alloys},\ }\href {https://doi.org/10.1103/PhysRev.106.893} {\bibfield
  {journal} {\bibinfo  {journal} {Phys. Rev.}\ }\textbf {\bibinfo {volume}
  {106}},\ \bibinfo {pages} {893} (\bibinfo {year} {1957})}\BibitemShut
  {NoStop}%
\bibitem [{\citenamefont {Kasuya}(1956)}]{kasuya1956}%
  \BibitemOpen
  \bibfield  {author} {\bibinfo {author} {\bibfnamefont {T.}~\bibnamefont
  {Kasuya}},\ }\bibfield  {title} {\bibinfo {title} {{A Theory of Metallic
  Ferro- and Antiferromagnetism on Zener's Model}},\ }\href
  {https://doi.org/10.1143/PTP.16.45} {\bibfield  {journal} {\bibinfo
  {journal} {Progress of Theoretical Physics}\ }\textbf {\bibinfo {volume}
  {16}},\ \bibinfo {pages} {45} (\bibinfo {year} {1956})}\BibitemShut {NoStop}%
\bibitem [{\citenamefont {Goodenough}(1955)}]{goodenough1955}%
  \BibitemOpen
  \bibfield  {author} {\bibinfo {author} {\bibfnamefont {J.~B.}\ \bibnamefont
  {Goodenough}},\ }\bibfield  {title} {\bibinfo {title} {Theory of the role of
  covalence in the perovskite-type manganites $[\mathrm{La},
  m(\mathrm{II})]\mathrm{Mn}{\mathrm{o}}_{3}$},\ }\href
  {https://doi.org/10.1103/PhysRev.100.564} {\bibfield  {journal} {\bibinfo
  {journal} {Phys. Rev.}\ }\textbf {\bibinfo {volume} {100}},\ \bibinfo {pages}
  {564} (\bibinfo {year} {1955})}\BibitemShut {NoStop}%
\bibitem [{\citenamefont {Kanamori}(1959)}]{kanamori1959}%
  \BibitemOpen
  \bibfield  {author} {\bibinfo {author} {\bibfnamefont {J.}~\bibnamefont
  {Kanamori}},\ }\bibfield  {title} {\bibinfo {title} {Superexchange
  interaction and symmetry properties of electron orbitals},\ }\href
  {https://doi.org/https://doi.org/10.1016/0022-3697(59)90061-7} {\bibfield
  {journal} {\bibinfo  {journal} {Journal of Physics and Chemistry of Solids}\
  }\textbf {\bibinfo {volume} {10}},\ \bibinfo {pages} {87} (\bibinfo {year}
  {1959})}\BibitemShut {NoStop}%
\bibitem [{\citenamefont {Aristov}(1997)}]{aristov1997}%
  \BibitemOpen
  \bibfield  {author} {\bibinfo {author} {\bibfnamefont {D.~N.}\ \bibnamefont
  {Aristov}},\ }\bibfield  {title} {\bibinfo {title} {Indirect rkky interaction
  in any dimensionality},\ }\href {https://doi.org/10.1103/PhysRevB.55.8064}
  {\bibfield  {journal} {\bibinfo  {journal} {Phys. Rev. B}\ }\textbf {\bibinfo
  {volume} {55}},\ \bibinfo {pages} {8064} (\bibinfo {year}
  {1997})}\BibitemShut {NoStop}%
\bibitem [{\citenamefont {Ko}\ \emph {et~al.}(2011)\citenamefont {Ko},
  \citenamefont {Kim}, \citenamefont {Kim}, \citenamefont {Kim}, \citenamefont
  {Kim}, \citenamefont {Min}, \citenamefont {Park}, \citenamefont {Chang},
  \citenamefont {Lin}, \citenamefont {Tanaka},\ and\ \citenamefont
  {Cheong}}]{Ko2011}%
  \BibitemOpen
  \bibfield  {author} {\bibinfo {author} {\bibfnamefont {K.-T.}\ \bibnamefont
  {Ko}}, \bibinfo {author} {\bibfnamefont {K.}~\bibnamefont {Kim}}, \bibinfo
  {author} {\bibfnamefont {S.~B.}\ \bibnamefont {Kim}}, \bibinfo {author}
  {\bibfnamefont {H.-D.}\ \bibnamefont {Kim}}, \bibinfo {author} {\bibfnamefont
  {J.-Y.}\ \bibnamefont {Kim}}, \bibinfo {author} {\bibfnamefont {B.~I.}\
  \bibnamefont {Min}}, \bibinfo {author} {\bibfnamefont {J.-H.}\ \bibnamefont
  {Park}}, \bibinfo {author} {\bibfnamefont {F.-H.}\ \bibnamefont {Chang}},
  \bibinfo {author} {\bibfnamefont {H.-J.}\ \bibnamefont {Lin}}, \bibinfo
  {author} {\bibfnamefont {A.}~\bibnamefont {Tanaka}},\ and\ \bibinfo {author}
  {\bibfnamefont {S.-W.}\ \bibnamefont {Cheong}},\ }\bibfield  {title}
  {\bibinfo {title} {Rkky ferromagnetism with ising-like spin states in
  intercalated ${\mathrm{fe}}_{1/4}{\mathrm{tas}}_{2}$},\ }\href
  {https://doi.org/10.1103/PhysRevLett.107.247201} {\bibfield  {journal}
  {\bibinfo  {journal} {Phys. Rev. Lett.}\ }\textbf {\bibinfo {volume} {107}},\
  \bibinfo {pages} {247201} (\bibinfo {year} {2011})}\BibitemShut {NoStop}%
\bibitem [{\citenamefont {Parkin}\ \emph {et~al.}(1991)\citenamefont {Parkin},
  \citenamefont {Bhadra},\ and\ \citenamefont {Roche}}]{parkin1991_prl}%
  \BibitemOpen
  \bibfield  {author} {\bibinfo {author} {\bibfnamefont {S.~S.~P.}\
  \bibnamefont {Parkin}}, \bibinfo {author} {\bibfnamefont {R.}~\bibnamefont
  {Bhadra}},\ and\ \bibinfo {author} {\bibfnamefont {K.~P.}\ \bibnamefont
  {Roche}},\ }\bibfield  {title} {\bibinfo {title} {Oscillatory magnetic
  exchange coupling through thin copper layers},\ }\href
  {https://doi.org/10.1103/PhysRevLett.66.2152} {\bibfield  {journal} {\bibinfo
   {journal} {Phys. Rev. Lett.}\ }\textbf {\bibinfo {volume} {66}},\ \bibinfo
  {pages} {2152} (\bibinfo {year} {1991})}\BibitemShut {NoStop}%
\bibitem [{\citenamefont {Parkin}\ and\ \citenamefont
  {Mauri}(1991)}]{parkin1991}%
  \BibitemOpen
  \bibfield  {author} {\bibinfo {author} {\bibfnamefont {S.~S.~P.}\
  \bibnamefont {Parkin}}\ and\ \bibinfo {author} {\bibfnamefont
  {D.}~\bibnamefont {Mauri}},\ }\bibfield  {title} {\bibinfo {title} {Spin
  engineering: Direct determination of the ruderman-kittel-kasuya-yosida
  far-field range function in ruthenium},\ }\href
  {https://doi.org/10.1103/PhysRevB.44.7131} {\bibfield  {journal} {\bibinfo
  {journal} {Phys. Rev. B}\ }\textbf {\bibinfo {volume} {44}},\ \bibinfo
  {pages} {7131} (\bibinfo {year} {1991})}\BibitemShut {NoStop}%
\bibitem [{\citenamefont {Haley}\ \emph {et~al.}(2020)\citenamefont {Haley},
  \citenamefont {Weber}, \citenamefont {Cookmeyer}, \citenamefont {Parker},
  \citenamefont {Maniv}, \citenamefont {Maksimovic}, \citenamefont {John},
  \citenamefont {Doyle}, \citenamefont {Maniv}, \citenamefont {Ramakrishna},
  \citenamefont {Reyes}, \citenamefont {Singleton}, \citenamefont {Moore},
  \citenamefont {Neaton},\ and\ \citenamefont {Analytis}}]{shannon2020}%
  \BibitemOpen
  \bibfield  {author} {\bibinfo {author} {\bibfnamefont {S.~C.}\ \bibnamefont
  {Haley}}, \bibinfo {author} {\bibfnamefont {S.~F.}\ \bibnamefont {Weber}},
  \bibinfo {author} {\bibfnamefont {T.}~\bibnamefont {Cookmeyer}}, \bibinfo
  {author} {\bibfnamefont {D.~E.}\ \bibnamefont {Parker}}, \bibinfo {author}
  {\bibfnamefont {E.}~\bibnamefont {Maniv}}, \bibinfo {author} {\bibfnamefont
  {N.}~\bibnamefont {Maksimovic}}, \bibinfo {author} {\bibfnamefont
  {C.}~\bibnamefont {John}}, \bibinfo {author} {\bibfnamefont {S.}~\bibnamefont
  {Doyle}}, \bibinfo {author} {\bibfnamefont {A.}~\bibnamefont {Maniv}},
  \bibinfo {author} {\bibfnamefont {S.~K.}\ \bibnamefont {Ramakrishna}},
  \bibinfo {author} {\bibfnamefont {A.~P.}\ \bibnamefont {Reyes}}, \bibinfo
  {author} {\bibfnamefont {J.}~\bibnamefont {Singleton}}, \bibinfo {author}
  {\bibfnamefont {J.~E.}\ \bibnamefont {Moore}}, \bibinfo {author}
  {\bibfnamefont {J.~B.}\ \bibnamefont {Neaton}},\ and\ \bibinfo {author}
  {\bibfnamefont {J.~G.}\ \bibnamefont {Analytis}},\ }\bibfield  {title}
  {\bibinfo {title} {Half-magnetization plateau and the origin of threefold
  symmetry breaking in an electrically switchable triangular antiferromagnet},\
  }\href {https://doi.org/10.1103/PhysRevResearch.2.043020} {\bibfield
  {journal} {\bibinfo  {journal} {Phys. Rev. Research}\ }\textbf {\bibinfo
  {volume} {2}},\ \bibinfo {pages} {043020} (\bibinfo {year}
  {2020})}\BibitemShut {NoStop}%
\bibitem [{\citenamefont {Little}\ \emph {et~al.}(2020)\citenamefont {Little},
  \citenamefont {Lee}, \citenamefont {John}, \citenamefont {Doyle},
  \citenamefont {Maniv}, \citenamefont {Nair}, \citenamefont {Chen},
  \citenamefont {Rees}, \citenamefont {Venderbos}, \citenamefont {Fernandes},
  \citenamefont {Analytis},\ and\ \citenamefont {Orenstein}}]{little2020}%
  \BibitemOpen
  \bibfield  {author} {\bibinfo {author} {\bibfnamefont {A.}~\bibnamefont
  {Little}}, \bibinfo {author} {\bibfnamefont {C.}~\bibnamefont {Lee}},
  \bibinfo {author} {\bibfnamefont {C.}~\bibnamefont {John}}, \bibinfo {author}
  {\bibfnamefont {S.}~\bibnamefont {Doyle}}, \bibinfo {author} {\bibfnamefont
  {E.}~\bibnamefont {Maniv}}, \bibinfo {author} {\bibfnamefont {N.~L.}\
  \bibnamefont {Nair}}, \bibinfo {author} {\bibfnamefont {W.}~\bibnamefont
  {Chen}}, \bibinfo {author} {\bibfnamefont {D.}~\bibnamefont {Rees}}, \bibinfo
  {author} {\bibfnamefont {J.~W.~F.}\ \bibnamefont {Venderbos}}, \bibinfo
  {author} {\bibfnamefont {R.~M.}\ \bibnamefont {Fernandes}}, \bibinfo {author}
  {\bibfnamefont {J.~G.}\ \bibnamefont {Analytis}},\ and\ \bibinfo {author}
  {\bibfnamefont {J.}~\bibnamefont {Orenstein}},\ }\bibfield  {title} {\bibinfo
  {title} {Three-state nematicity in the triangular lattice antiferromagnet
  fe1/3nbs2},\ }\href {https://doi.org/10.1038/s41563-020-0681-0} {\bibfield
  {journal} {\bibinfo  {journal} {Nature Materials}\ }\textbf {\bibinfo
  {volume} {19}},\ \bibinfo {pages} {1062} (\bibinfo {year}
  {2020})}\BibitemShut {NoStop}%
\end{thebibliography}%
\end{document}